\documentclass[]{aastex7}

\usepackage{xcolor}

\usepackage{ifthen}
\newcommand{\betapic}[1]{%
    \ifthenelse{\equal{#1}{short}}{$\beta~Pic$}{%
    \ifthenelse{\equal{#1}{full}}{$\beta~$Pictoris}{%
    \textcolor{red!55!red}{Error: Invalid argument}}%
}}
\newcommand{\aumic}[1]{%
    \ifthenelse{\equal{#1}{short}}{AU Mic}{%
    \ifthenelse{\equal{#1}{full}}{AU Microscopii}{%
    \textcolor{red!55!red}{Error: Invalid argument}}%
}}

\received{October 14, 2025}
\revised{Novemeber 28, 2025}
\accepted{December 2, 2025}
\begin{document}

\title{A Catalogue of Interstellar Material Delivery From Nearby Debris Disks}

\author[orcid=0000-0001-8927-7708,gname=Cole R., sname=Gregg]{Cole R. Gregg}
\affiliation{Department of Physics and Astronomy,
The University of Western Ontario,
London, Canada}
\affiliation{Institute for Earth and Space Exploration (IESX),
The University of Western Ontario,
London, Canada}
\email[show]{cgregg2@uwo.ca}  
\author[orcid=0000-0002-1914-5352,gname=Paul A., sname=Wiegert]{Paul A. Wiegert} 
\affiliation{Department of Physics and Astronomy,
The University of Western Ontario,
London, Canada}
\affiliation{Institute for Earth and Space Exploration (IESX),
The University of Western Ontario,
London, Canada}
\email[show]{pwiegert@uwo.ca}

\begin{abstract}

We modeled the trajectories of material ejected from 20 nearby debris disk stars, including $\epsilon$ Eridani (Ran), Vega, Fomalhaut, and $\beta$ Pictoris, within a simulated Milky Way potential in order to quantify their contribution to the population of interstellar material entering the solar system. Our simulations show that material from each of these 20 systems is currently to be expected within our planetary system. We calculate expected fluxes of both macroscopic interstellar objects (ISOs, $\geq100$ m), which could be detected by telescopic surveys, and smaller meteoroids ($\geq200~\mu$m), which could manifest as meteors in Earth’s atmosphere. We estimate that the ISO population originating from these debris disks and currently within the inner solar system is on the order of $\sim2$, only a fraction of the expected total ISO population but nonetheless likely to be discovered by Rubin.
Meteors in Earth's atmosphere from these systems are expected as well, but current methods, both radar and video, might require decades to collect even a single event. Our sample is found to be rich in relatively low excess velocity particles compared to the broader expected ISO population, which might make them harder to distinguish observationally from bound objects in some cases. These results provide a framework for linking detections of interstellar material to their astrophysical origins, offering new opportunities to probe the composition and dynamical history of nearby planetary systems.

\end{abstract}

\keywords{\uat{Interstellar objects}{52} --- \uat{Meteor radiants}{1033} --- \uat{Meteor streams}{1035}}


\section{Introduction\label{sec:intro}}

The discovery of interstellar objects (ISOs) passing through the solar system has reignited questions about the frequency and origin of such encounters. This material provides a rare opportunity to directly probe the building blocks of other astrophysical systems, including planetary systems beyond our own. However, the mechanisms by which this material leaves their origin systems and traverses the Galaxy before entering the solar system remain poorly constrained.

There have now been three discoveries of macroscopic ISOs: 1I/'Oumuamua \citep{Oumuamua_discovery}, 2I/Borisov \citep{Borisov_charac}, and 3I/ATLAS \citep{3IATLAS_discovery}. At dust sizes, in situ detectors on spacecraft have detected an influx of interstellar particles swept up by the solar system as it moves through the Local Interstellar Cloud (LIC) \citep{baguhl1995, Grun1997, altobelli2003}. At intermediate sizes, those which may be observed as meteors if they encounter Earth's atmosphere, no conclusive evidence has been reported, despite controversial claims \citep{vaubaillon_2022, brown_borovicka_2023, hajdukova_et_al_2024}. One difficulty in detecting interstellar meteors is that the only indicator of the interstellar nature of a particle is its hyperbolic excess velocity \citep{hajdukova_et_al_2019}, which is very sensitive to measurement error \citep{hajdukova_et_al_2020}. Because of this, the detection of interstellar meteoroids has been argued for more than a century \citep{stohl_1970}. Studies have searched meteor data for evidence of interstellar meteoroids, including radar observations of potential $\gtrsim100~\mu$m (diameter) interstellar meteors reported by the Advanced Meteor Orbit Radar (AMOR) located in New Zealand \citep{baggaley2000}. Although this study reported positive detections from a discrete source, the Canadian Meteor Orbit Radar (CMOR), located near Tavistock, Canada \citep{CMOR}, was unable to confirm their results \citep{weryk_brown_2004, froncisz_2020} (granting the larger limiting sizes $\gtrsim200~\mu$m). A recent study has mined the Global Meteor Network (GMN; \cite{GMN}) database and found no evidence of visual-sized intertsellar meteors ($\gtrsim5~$ mm) in the highest-quality 57\% of events \citep{wiegert_et_al_GMN_2025}. The continued absence of confirmed interstellar meteors highlights an unresolved gap in our understanding of interstellar material transfer.

Despite progress, both the precise pathways of the interstellar material followed and their original sources remain uncertain. Understanding these pathways is vital, as this material can contribute to planet formation in emerging planetary systems \citep{griperavn19, Moro-Mart_2022}, while also facilitating the exchange of chemical elements, organic compounds, and possibly even precursors of life between stellar systems, an idea central to panspermia \citep{griperavn19, Adams_Napier_2022, osmanov_2025, smith_2024}. 
Comparison of the observed dynamical and other properties of interstellar material with predicted characteristics would provide valuable clues and offer a rare opportunity to directly link an observation made within our solar system to a specific external stellar system, maximizing the scientific return from these detections.

Here, our aim is to build on the previous work of \cite{gregg_wiegert_alphacen_2025}, using the tools from their $\alpha~$Centauri case study.
In particular, we investigate whether there exist dynamical pathways for the transport of material to our solar system from nearby debris disk systems on short ($\sim100~$Myr) timescales. \citet{Murray_2004} explored the ejection conditions that could arise from the interaction of a debris disk with an embedded planet and analytically estimated what the flux may be at Earth. This work provides a useful framework that we follow closely, with adaptations suited for numerical work.  

Debris disks are dust-dominated disks and are observed in stellar systems of various ages. Sometimes known as secondary disks, they represent a continuously replenished population produced through a collisional cascade, serving as evidence of a successful planet formation phase (see \cite{hughes_debrisdisks_review_2018} for a full review). Around main-sequence stars, debris disks provide extrasolar analogs to small-body populations such as the Kuiper Belt and zodiacal dust in the solar system. This renders them especially intriguing for investigations of planetary architectures and evolution.

Our sample of debris disks originates in part from  the \textit{Catalog of Circumstellar Disks} \footnote{This can be found at \url{https://www.circumstellardisks.org/}, crediting Karl Stapelfeldt. It contains 323 resolved disks (220 pre-main-sequence disks and 103 debris disks).} from which we drew the 20 nearest stellar systems with an IR excess (hereafter called debris disks). This sample includes the four best studied debris disk systems: $\epsilon$ Eridani (Ran), Vega, Fomalhaut, and \betapic{full}. We then drew on the \textit{Catalog of Resolved Debris Disks} \footnote{This can be found at \url{https://www.physik.uni-jena.de/21956/catalog-of-resolved-debris-disks}, crediting Nicole Pawellek (Department for Astrophysics, University of Vienna, Austria) and Alexander Krivov (Astrophysical Institute, Friedrich-Schiller-University Jena, Germany). It contains 175 debris disks.}, which provided additional constraints on the system parameters and pointed us toward further resources. Our 20 selected systems' International Celestial Reference System (ICRS) coordinates are provided in Table \ref{Tab:Params} and their ages and IR excess are provided in Table \ref{tab:star_age_summary}. We note that for consistency, throughout this work we adopt the primary SIMBAD designations for each system when reporting our results.

The overall goal of this work is to provide a catalogue of system-by-system predictions for the contribution of nearby debris disks to the ISO population in the solar system and the interstellar meteoroid flux at Earth. By combining numerical simulations with empirical estimates of ejection rates and velocities, we quantify the expected transfer efficiency, arrival characteristics, fluxes, and survivable grain sizes from each of the 20 nearest debris disk systems. These predictions establish a framework for interpreting future detections of both ISOs and interstellar meteoroids, and they address the broader question of whether such objects should be present and/or detectable at all. In doing so, this study offers a means of bridging the current observational gap between interstellar dust grains and the rare, large ISOs already observed.

% %% The "ht!" tells LaTeX to put the figure "here" first, at the "top" next
% %% and to override the normal way of calculating a float position.
% %% The asterisk after "figure" tells the compiler to span multiple columns
% %% if a two column style is selected.
% \begin{figure*}[ht!]
% \plotone{AuthorChargeInfographic.png}
% \caption{The AAS journals are operated as a nonprofit venture, and author charges fairly recapture costs for the services provided in the publishing process. The chart above breaks down the services that author charges go toward. The AAS Journals' Business Model is outlined in a \href{https://aas.org/posts/news/2023/08/aas-open-access-publishing-model-open-transparent-and-fair}{2023 post}.
% \label{fig:general}}
% \end{figure*}
% \latex\ \footnote{\url{http://www.latex-project.org/}} is a document markup language...

\section{Model}\label{sec:model}

This study follows the model described in \cite{gregg_wiegert_alphacen_2025}, which employs a simplified representation of the gravitational field of the Milky Way through a time-independent axisymmetric model of three components introduced by \citet{Miyamoto_1975}. This treats the Milky Way as a smooth potential field, omitting time-dependent effects (such as stellar and giant molecular cloud (GMC) encounters as well as the evolution of the Milky Way itself) and non-gravitational forces (such as interstellar medium (ISM) drag and magnetic interactions). These omissions restrict the validity of the model to particles larger than a trajectory-dependent threshold size, a limitation addressed in Section~\ref{sec:analysis}. Extremely long stellar relaxation times in the Galactic disk ($t_{\text{relax}} \sim 10^7$ Gyr \citep{binney_tremaine_2008galacticDyn}) justify the exclusion of stellar encounters, as the relevant time scales exceed the $\sim$100 Myr integration times used in this work.

They further emphasise that the properties of the Galactic spiral pattern remain poorly constrained, such that adopting a particular spiral-arm model does not necessarily improve encounter predictions.  
Accordingly, neglecting spiral structure primarily affects distant orbits and absolute miss distances, rather than the overall statistics or population of close Solar encounters.

\added{It is also worth noting that the gravitational influence of the spiral arms were not taken into account. \cite{garcia-sanchez_sprialarms_2001} explored the effect of adding a standard spiral-arm perturbation on stellar orbits integrated in a smooth Galactic potential. They found that the majority of close encounters shift by $\lesssim0.5$~pc within $\pm100$~Myr, and only a small fraction change by more than a few parsecs over that interval. They further emphasize that the properties of the Galactic spiral pattern remain poorly constrained, such that adopting a particular spiral-arm model does not necessarily improve encounter predictions. Accordingly, neglecting spiral structure primarily affects distant orbits and absolute miss distances over large time scales, rather than the overall statistics or population of close Solar encounters.}

A summary of the values adopted for the model parameters extracted from \cite{gregg_wiegert_alphacen_2025} is listed in Table~\ref{Tab:Params}. \added{$M_i$ denotes the total mass of a given Galactic component, while $a_i$ and $b_i$ are scale lengths that characterize its geometry. Because the bulge and halo are modeled as spherically symmetric, they each require only a single scale length ($b_{b,h}$). In contrast, the disk is treated as a flattened spheroid, which necessitates two distinct scale lengths ($a_d$ and $b_d$). The solar values $r_\odot$ and $z_\odot$ represent the Sun's distance from the Galactic center and and the midplane of the disk respectively. While $v_\odot$ is the Sun's velocity relative to the local standard of rest and $v_{\odot,circ}$ is circular velocity in the Galaxy at the Sun's radial distance. In addition, the position ($\alpha,~\delta$) of the Galatic centre and North Galactic Pole are also given for the J2000 epoch, these are necessary for initializing the Sun and star systems into a Galactic frame.}

\begin{deluxetable*}{lccccccc}[htb]
\tablecaption{The Adopted Values Used to Initialize Our Simulation, Together With the International Celestial Reference System (ICRS) Coordinates and Primary SIMBAD Designations of the Debris Disk Stars, Obtained from SIMBAD. \label{Tab:Params}}
\tablewidth{0pt}
\tablehead{
\multicolumn{3}{c}{Parameter} & Units & \multicolumn{3}{c}{Value} & Reference
}
\startdata
\multicolumn{3}{l}{$M_i$, $i=d, b, h$} & $10^{10}M_\odot$ & \multicolumn{3}{c}{7.91, 1.40, 69.80} & (3) \\ 
\multicolumn{3}{l}{$a_i$, $i=d, b, h$} & pc & \multicolumn{3}{c}{3500, 0, 0} & (3) \\
\multicolumn{3}{l}{$b_i$, $i=d, b, h$} & pc & \multicolumn{3}{c}{250, 350, 24000} & (3) \\
\multicolumn{3}{l}{$r_\odot$} & kpc & \multicolumn{3}{c}{8.33$\pm 0.35$} & (4) \\
\multicolumn{3}{l}{$z_\odot$} & pc & \multicolumn{3}{c}{27$\pm 4$} & (2) \\
\multicolumn{3}{l}{$v_\odot$ $(U, V, W)$} & km s$^{-1}$ & \multicolumn{3}{c}{($11.1^{+0.69}_{-0.75}$, $12.24^{+0.47}_{-0.47}$, $7.25^{+0.37}_{-0.36}$)} & (6) \\
\multicolumn{3}{l}{$v_{\odot,circ}$} & km s$^{-1}$ & \multicolumn{3}{c}{218$\pm 6$} & (1) \\
\multicolumn{3}{l}{Galactic Centre (J2000: $\alpha$, $\delta$)} & (hr:min:s, deg:':") & \multicolumn{3}{c}{(17:45:37.224, -28:56:10.23)} & (5) \\ 
\multicolumn{3}{l}{North Galactic Pole (J2000: $\alpha$, $\delta$, $\theta$)} & (hr:min:s, deg:':", deg) & \multicolumn{3}{c}{(12:51:26.282, 27:07:42.01, 122.932)} & (5) \\ 
\hline
\multicolumn{8}{c}{\textbf{Star Systems}} \\ \hline
Star System & $\alpha$ & $\mu_{\alpha}$ & $\delta$ & $\mu_{\delta}$ & Parallax & $v_r$  & Reference \\ 
 & (deg) & (mas yr$^{-1}$) & (deg) & (mas yr$^{-1}$) & (mas) & (km s$^{-1}$) & \\ 
\hline 
\protect* eps Eri (Ran) & 53.232685 & -974.758 & -9.458261 & 20.876 & 310.577 & 16.376 & (7) \\
\protect* tau Cet & 26.017013 & -1721.728 & -15.937480 & 854.963 & 273.810 & -16.597 & (7) \\
\protect* e Eri & 49.981879 & 3035.017 & -43.069782 & 726.964 & 165.524 & 87.882 & (7) \\
BD-07 4003 & 229.861779 & -1221.278 & -7.722275 & -97.229 & 158.718 & -9.240 & (7) \\
\protect* alf Lyr (Vega) & 279.234735 & 200.940 & 38.783689 & 286.230 & 130.230 & -13.500 & (7) \\
\protect* alf PsA (Fomalhaut) & 344.412693 & 328.950 & -29.622237 & -164.670 & 129.810 & 6.500 & (7) \\
\protect* 61 Vir & 199.601308 & -1070.202 & -18.311194 & -1063.849 & 117.173 & -7.824 & (7) \\
HD 197481 & 311.289719 & 281.319 & -31.340899 & -360.148 & 102.943 & -4.710 & (7) \\
(\aumic{full}) & & & & & & & \\
\protect* bet Leo & 177.264910 & -497.680 & 14.572058 & -114.670 & 90.910 & -0.200 & (7) \\
HD 166 & 1.653267 & 380.159 & 29.021504 & -177.730 & 72.642 & -6.121 & (7) \\
HD 38858 & 87.145584 & 61.427 & -4.094645 & -229.291 & 65.745 & 31.456 & (7) \\
HD 207129 & 327.065630 & 165.069 & -47.303616 & -295.553 & 64.272 & -7.455 & (7) \\
\protect* b Her & 271.756628 & -100.150 & 30.562107 & 53.998 & 63.394 & -0.170 & (7) \\
HD 23484 & 56.038221 & 209.058 & -38.281770 & 289.316 & 61.844 & 29.616 & (7) \\
\protect* q01 Eri & 25.622144 & 166.041 & -53.740831 & -105.496 & 57.641 & 27.662 & (7) \\
HD 35650 & 81.125706 & 43.159 & -38.969652 & -57.276 & 57.476 & 32.264 & (7) \\
\protect* g Lup & 235.297403 & -169.106 & -44.661206 & -266.391 & 57.476 & -7.080 & (7) \\
\protect* eta Crv & 188.017610 & -424.597 & -16.196005 & -58.241 & 54.814 & 0.690 & (7) \\
HD 53143 & 104.998563 & -161.873 & -61.336181 & 264.836 & 54.525 & 22.040 & (7) \\
\protect* bet Pic (\betapic{full})& 86.821199 & 5.160 & -51.066511 & 84.041 & 50.931 & 16.840 & (7) \\
\enddata
\tablereferences{Table adopted from \cite{gregg_wiegert_alphacen_2025} - (1) \cite{Bovy_2015}; (2) \cite{chen2001}; (3) \cite{Dauphole_1995}; (4) \cite{Gillessen_2009}; (5) \cite{Reid_2004}; (6) \cite{schonrich_2010}; (7) \cite{SIMBAD}}
\tablecomments{The $\theta$ for the North Galactic Pole (NGP) is the position angle of the NGP from the North Celestial Pole.\\
The star system initial parameters are provided for the J2000 epoch, and have not been corrected for light-travel time. This correction would be small: for example, the most distant system considered here is \textit{* bet Pic} at 64 light-years, which would have a light travel time adjustment of 250 au. This is much less than the parallax error, which amounts to a distance uncertainty of ($\pm12000~$au). The largest light-travel time adjustment of 520 au is a much closer system at 20 light-years, \textit{* e Eri}, due to its large relative velocity. Comparable, yet still smaller than the distance uncertainty from the parallax error ($\pm590~$au).  Future work considering systems at larger distances should account for this correction.}
\end{deluxetable*}

\subsection{Ejection Model}\label{sec:ejection}

We do not explicitly model the ejection mechanisms responsible for launching particles from each stellar system, but instead simply assume that material \added{continuously} departs its origin system \added{with directions drawn from an isotropic random distribution and} with some excess velocity. We adopt a velocity distribution expected for the gravitational scattering of residual planetesimals by massive planets which has been widely explored in the context of planetary system evolution \citep{Bailer-Jones_2018, brasser2013, charnoz2003, correa-otto_stability_2019, duncan1987, fernandez1984, Zwart2018}.

Following the work of \cite{Bailer-Jones_2018}, we make use of the ejection speed distribution originally derived for solar system planetesimals interacting with the giant planets (adapted from \cite{brasser2013}). In this case, their results show that Jupiter dominates the ejection statistics, contributing more than 70\% of the total. While the architectures of each exoplanetary system certainly differ and in many cases the planetary content of our chosen systems is unknown, we will assume the presence of at least one Jupiter analog in each system  as the basis for our ejected speed distribution. We adopt the speed distribution represented by the black curve in Figure 6 of \cite{Bailer-Jones_2018} as our fiducial model.

This distribution defines the asymptotic escape speed, $v_{\mathrm{eject}}$, of each particle relative to its host system. To compute the galactocentric velocity, we multiply a unit direction vector drawn randomly from the unit sphere by $v_{\mathrm{eject}}$, and add it to the velocity of the originating star. Ejected particles are initialized at the star’s position with this resultant velocity vector.

During the simulation, particle-solar distances are linearly interpolated between time steps to avoid missing close encounters by ``stepping over them" due to our finite step size. If a particle enters a spherical region around the Sun with radius $100,000~$au \textemdash chosen as a rough proxy for the outer extent of the Oort Cloud (OC) \textemdash it is flagged as a close approach (CA), and the relevant dynamical parameters are recorded for further analysis.

\section{Simulation}\label{sec:Simulation}

Simulations are performed using a variable time step Runge-Kutta-Fehlberg integrator \citep{fehlberg1974}, employing a fifth-order scheme with an adopted local error tolerance of $10^{-6}$.

Each stellar system is initialized using SIMBAD-derived positions and velocities, along with a reference particle representing the solar system (parameters provided in Table~\ref{Tab:Params}).

The systems are first evolved backward from the J2000 epoch ($t=0$) for a duration of $100~$Myr, corresponding to nearly half a typical galactic orbit. For systems with significantly younger ages, such as \betapic{full} and AU Microscopii, both approximately $21~$Myr old \citep{nilsson_et_al_2009}, backward integration is limited accordingly. From their past states (at $t \approx -100~$Myr or $t \approx -21~$Myr), the systems are then propagated forward to a time slightly beyond the present ($t \approx +5~$Myr or $t \approx +4~$Myr), using a nominal timestep of $\sim5,000~$years.

The integration is symmetric in the sense that all systems return within the original positions at $t = 0$, the largest deviation observed being $2.63\times10^{-2}~$au (corresponding to a relative error of $1.53\times10^{-11}$).

During forward integration, particles are continuously added to simulate the continuous ejection of material. Ejecta are introduced every $10,000~$yr, 1000 particles per injection interval for the full-duration runs and 4200 per interval for the shorter ones. This ensures a consistent total particle count across scenarios (10,500,000 in each case). These ejecta represent a wide range of size scales, from km-class bodies such as asteroids and comets to much smaller particles (cm and below) capable of generating meteors observable at Earth.

\section{Analysis} \label{sec:analysis}

\subsection{Grain Sizes}\label{sec:sizes}

Small particles traversing the interstellar medium (ISM) are subject to physical processes that are not accounted for in our simulations. As a result, in some cases where our simulations show that the dynamics, in principle, permit particles to reach the solar system, the survivability of such particles (particularly those in the millimeter size range and smaller, which may be observed as meteors in Earth’s atmosphere) depends critically on factors such as transit time and relative velocity through the ISM.

To estimate the lower bound on the particle sizes capable of surviving this journey, we follow the framework laid out by \cite{Murray_2004}. Their analysis evaluates three key loss mechanisms: (1) deflection by interstellar magnetic fields, which can significantly alter the paths of small, charged grains; (2) deceleration due to ISM drag forces that may effectively halt smaller particles; and (3) destruction via sputtering by energetic gas atoms or catastrophic grain-grain collisions. These criteria allow us to define a conservative minimum particle size threshold for survivability across interstellar distances.

For each of the CAs identified in our simulation, we extracted the corresponding trajectory and computed the minimum grain size required to survive transit through the interstellar medium. This analysis applies the criteria described in \cite{Murray_2004}, specifically their equations 43, 44, 45, and 47, which address the effects of ISM drag (within and outside the local bubble), magnetic deflection (via the gyroradius), and grain destruction mechanisms, respectively. These calculations assume representative ISM and grain parameters: $\rho = 3.5~{\rm g~cm}^{-3}$, $n_H = 1{\rm cm}^{-3}$, $U = 1~$V, $B = 5~\mu$G, and $T_{\rm gas} = 10^6~$K.

\subsection{Flux Calculations}\label{sec:flux}

The number of simulated particles seen to enter our solar system is weighted according to the expected rate of particle ejection by the origin system appropriate to its age. This ``dust luminosity" is assumed by \cite{Murray_2004} to 1) drop as $t_{age}^{-3}$ based on inferred rates of dust grain loss \citep{spangler_2001}, and 2) to be zero prior to some ``critical age" $t_{crit}$, representing the age at which planetary formation has proceeded far enough that gravitational ejection of debris can begin, and we adopt their model with some minor modifications.

Starting with their dust luminosity equation (Equation 91), we too opt for an optimistic approach with $f_{ej}=1$ (the fraction of particles ejected). We adjusted for updated estimates of the parameters of the \betapic{short} system, in particular its age ($t_{age}(\beta~Pic\text{, old})=12~$Myr \citep{zuckerman_et_al_2001} is updated to $t_{age}(\beta~Pic\text{, new})=21~$Myr \citep{binks_jefferies_2013}), and its infrared flux (specific flux $F_{\nu}(800~\mu\text{m, old})=115\pm30~$mJy \citep{zuckerman_becklin_1993} is updated to $F_{\nu}(800~\mu\text{m, new})=94.5\pm25.5~$mJy, which is an average of values drawn from \cite{vandenbussche_et_al_2010} and \cite{ballering_et_al_2016}).
We integrate their expression across all sizes from $a_{min}\rightarrow\infty$ to get

\begin{equation}\label{eject_rate}
    N_{eject}(\geq a_{min}, t_{age}) = \frac{7.8\times10^{18}}{2.5}~\cdot~\bigg(\frac{t_{age}(\beta~Pic\text{, old})}{t_{age}(\beta~Pic\text{, new})}\bigg)~\cdot~\bigg(\frac{F_{\nu}(800~\mu\text{m, new})}{F_{\nu}(800~\mu\text{m, old})}\bigg)~\cdot~\bigg( \frac{21~\text{Myr}}{t_{age}}\bigg)^3 ~\cdot~ \bigg( \frac{a_{min}}{\text{cm}} \bigg) ^{-2.5}.
\end{equation}
Where this applies for ages larger than the critical age $t_{crit}$ which we take to be 10~Myr, and where we omit the speed range fraction, as our simulations naturally account for a range of ejected speeds.

Equation~\ref{eject_rate} provides the number of particles ejected larger than a minimum particle radius ($a_{min}$, (units of cm), noting that we report values in diameter for all other areas of this work) given the age of the system at the time of particle ejection ($t_{age}$). 

An important point: the age used in \cite{Murray_2004} is an IR excess age $t_{IR}$ derived from $\tau$, the ratio of re-emitted light from the disk to the luminosity of the star; this typically differs from the current best estimate for the actual age of the system, sometimes by orders of magnitude (see Table~\ref{tab:star_age_summary}). As the mass loss of debris disks is not well understood, we follow \citep{Murray_2004} in adopting a dust luminosity based on the IR-excess age. We assume that each disk is an exact analogue of \betapic{short} and determine its age from its observed $\tau$, with the assumption that $\tau \propto t^{-2}$ \citep{spangler_2001}:

\begin{equation}\label{IRage}
    t_{IR}(\tau) = t(\beta~Pic)\bigg(\frac{\tau(\beta~Pic)}{\tau}\bigg)^{0.5}.
\end{equation}

Because of the conflicts between the best-estimate ages and IR-excess age, we use both to determine the expected ejection rates. If the time in question is older than the best-estimate age of the system, the ejection rate is taken to be zero. If this is not the case, if the IR-excess age is less than the critical age but not less than the best estimate age of the system, the ejection rate is capped at the value expected for $t_{crit}$. Otherwise, Equation~\ref{eject_rate} is used. Table \ref{tab:star_age_summary} summarizes the estimated age of each system, the adopted age used in this work, its IR excess, and $t_{\mathrm{IR}}$.

\begin{deluxetable*}{cccccc}
\tablecaption{Summary of Star System Age Parameters. References Provided Are for the System Age and IR Excess Columns Respectively.\label{tab:star_age_summary}}
\tablehead{
\colhead{Star System} & \colhead{System Age} & \colhead{Adopted Best-Estimate Age} & \colhead{IR Excess} & \colhead{$t_{IR}$} & \colhead{References} \\
\colhead{{}} & \colhead{(Myr)} & \colhead{(Myr)} & \colhead{{}} & \colhead{(Myr)} & \colhead{{}}
}
\startdata
\protect* eps Eri       & $400-800$       & 400   & 1.10$\times10^{-4}$    & 83   & (9), (1) \\
\protect* tau Cet      & $10000 \pm 500$ & 10000 & 1.20$\times10^{-5}$    & 250  & (8), (6) \\
\protect* e Eri        & $6000 \pm 300$  & 6000  & 4.30$\times10^{-6}$    & 418  & (5), (21) \\
BD-07 4003             & $8900 \pm 1100$ & 8900  & 8.90$\times10^{-5}$    & 92   & (5), (14) \\
\protect* alf Lyr      & $400-800$       & 400   & 1.58$\times10^{-5}$    & 218  & (9), (6) \\
\protect* alf PsA      & $440 \pm 40$    & 440   & 4.60$\times10^{-5}$    & 128  & (9), (6) \\
\protect* 61 Vir       & $5100 \pm 1300$ & 5100  & 2.70$\times10^{-5}$    & 167  & (5), (21) \\
HD 197481              & $23 \pm 3$      & 21    & 3.90$\times10^{-4}$    & 44   & (15), (16) \\
\protect* bet Leo      & 100             & 100   & 1.90$\times10^{-5}$    & 199  & (20), (6) \\
HD 166                 & 160             & 160   & 5.90$\times10^{-5}$    & 113  & (20), (19) \\
HD 38858               & $4300 \pm 1000$ & 4300  & 8.00$\times10^{-5}$    & 97   & (5), (21) \\
HD 207129              & $600-8300$      & 600   & 1.58$\times10^{-4}$    & 69   & (12), (4) \\
\protect* b Her        & $6000-10000$    & 6000  & 1.40$\times10^{-5}$    & 231  & (11) \\
HD 23484               & $760-930$       & 760   & 9.50$\times10^{-5}$    & 89   & (7), (18) \\
\protect* q01 Eri      & $960-1740$      & 960   & 5.37$\times10^{-4}$    & 37   & (7), (4) \\
\protect* g Lup        & $300^{+700}_{-200}$ & 300 & 9.00$\times10^{-5}$    & 91   & (10) \\
HD 35650               & $50-200$        & 100   & 1.75$\times10^{-4}$    & 65   & (3) \\
\protect* eta Crv      & 1400            & 1400  & 3.00$\times10^{-4}$    & 50   & (13) \\
HD 53143               & $1000 \pm 200$  & 1000  & 2.50$\times10^{-4}$    & 55   & (10) \\
\protect* bet Pic      & $21 \pm 4$      & 21    & 1.70$\times10^{-3}$    & 21   & (2), (17) \\
\enddata
\tablereferences{(1) \cite{backman_et_al_2009};
(2) \cite{binks_jefferies_2013};
(3) \cite{choquet_et_al_2016};
(4) \cite{decin_et_al_2003};
(5) \cite{desgrange_et_al_2023};
(6) \cite{di_folco_et_al_2004};
(7) \cite{eiroa_et_al_2013};
(8) \cite{greaves_et_al_2004};
(9) \cite{janson_et_al_2015};
(10) \cite{kalas_et_al_2006};
(11) \cite{kennedy_et_al_2012};
(12) \cite{krist_et_al_2010};
(13) \cite{lebreton_et_al_2016};
(14) \cite{lestrade_et_al_2012};
(15) \cite{mamajek_bell_2014};
(16) \cite{matthews_et_al_2015};
(17) \cite{nilsson_et_al_2009};
(18) \cite{schuppler_et_al_2014};
(19) \cite{trilling_et_al_2008};
(20) \cite{vican_age_2012};
(21) \cite{wyatt_et_al_2012}.}
\end{deluxetable*}

Each of our simulated particles then represents $w_{j}$ real particles $\geq a_{min}$

\begin{equation}\label{fraction_real}
    w_{j} = N_{eject}(\geq a_{min}, t_{age}) \frac{10,000}{n_{eject}}
\end{equation}

\noindent where 10,000 is the time interval between simulated ejections (in yr), and $n_{eject}$ is the number of simulated particles created per ejection interval (1000 or 4200 as described in Section \ref{sec:Simulation}).

The flux and population estimates for different regions of the solar system are then calculated at each time step. For the flux into the solar system $F_{OC}$ at a particular time is

\begin{equation}\label{flux_OC}
    F_{OC} = \frac{1}{dt}\sum{w_j}
\end{equation}

\noindent where $dt$ is our simulation time step and $w_j$ is the weight of each CA arriving at the solar system during that time step.

For particles $\geq100~$m in diameter, we report the population within an observable limit of 10 au as

\begin{equation}\label{Pop_InnerSS}
P_{10au} = \frac{10^{-12}}{dt}\sum{ \big( w_j\cdot t_{cross,j} \cdot f_{lens}(10~\text{au},v_{\infty,j})\big) },
\end{equation}
where the factor of $10^{-12}$ is the relative volume of a 10 au region within a 100,000 au OC, $t_{cross,j}$ is the OC crossing time of each CA arriving at the solar system during that time step, calculated assuming an average chord length of $4R/3$ through the Oort Cloud and the CA's relative solar velocity ($v_{\infty,j}$, \added{the excess velocity}).
The final term in Equation~\ref{Pop_InnerSS} is $f_{lens}(10au,v_{\infty,j})$, the gravitational lensing enhancement due to the Sun's potential well calculated from $v_{\infty,j}$ and the escape velocity at 10 au ($v_{esc,10au}=13.3~\mathrm{km~s}^{-1}$):
\begin{equation}\label{eq:grav_lens}
    f_{lens}(r, v_\infty) = 1 + \bigg(\frac{v_{esc,r}}{v_\infty}\bigg)^2.
\end{equation}
Lastly, the flux of meteors (calculated for particles $\geq200~\mu$m in diameter) as seen at the Earth is
\begin{equation}\label{flux_earth}
    F_{Earth} = \frac{1.8\times10^{-19}}{dt}\sum{ \big( w_j \cdot f_{lens}(1~\text{au},v_{\infty,j}) \big)}
\end{equation}
where $1.8\times10^{-19}$ is the relative cross section of the Earth compared to the OC, the gravitational lensing factor is computed at 1~au and we note $v_{esc,1au}=42.2~\mathrm{km~s}^{-1}$.

For each of these flux and population estimates we report current values around the J2000 epoch ($t=0$) as well as maximum values at the times where each system's transfer rate peaks. In all cases these are computed using a windowing average over 21 time steps ($\pm10$ time steps or $\pm$50,000 yr from the time in question).

Uncertainties are calculated using the bootstrap method with 1,000 samples. In cases where the sample consists of a single value or multiple identical values (i.e., no variability), Poisson errors are adopted.

\subsection{Relationship Between Excess Velocity, Perihelion Distance and Orbital Eccentricity}

In the context of interstellar meteoroids, which become meteors upon entering Earth's atmosphere, it is helpful to calculate their heliocentric velocities as they would be observed from Earth. As their distance from the Sun would be $\sim1$ au when entering Earth's atmosphere, the heliocentric velocity is simply
\begin{equation}\label{eq:helio_vel}
  v_{hel,1au}^2 = v_{\infty}^2 + \frac{2G M_{\odot}}{r_{1au}}.
\end{equation}
However, it is worth noting that the eccentricity of an interstellar particle's orbit depends on both $v_{\infty}$ and $q$ through
\begin{equation}\label{eq:e}
  v_{\infty}^2 = \frac{G M_{\odot} (e-1)}{q},
\end{equation}
which results in larger $e$ at larger $q$ even at the same excess velocity $v_{\infty}$. For the values of $e$  reported in this work, we assume $q=1$ au. The values of $e$ computed for other ISOs crossing the solar system (even with the same $v_{\infty}$) will depend on their perihelion distances.

\section{Results} \label{sec:results}

We find that each of the 20 debris disks analyzed is capable of delivering material to the solar system at some level. The transfer efficiencies, defined as the fraction of simulated particles from a given system that reach the solar system, ranged between 0.01\% and 0.09\% across the sample. This relatively high transfer efficiency arises because these debris disks are particularly close to us at the current time. 

In the sections below, in particular those sections reporting on individual systems and which are based on a simple reporting ``template", significant figures are not strictly maintained and the number of decimal places shown are chosen for simplicity, consistency and to enhance readability. However, 95\% confidence bounds are reported for all quantities, so accurate statistical uncertainties can be derived if needed. We also note that the statistical summaries of the particles that transfer to the solar system are based on the unweighted distribution of simulated particles. Weighting is only applied when calculating flux values, at which point each particle is scaled according to its contribution to the overall flux.

\begin{figure*}
% \digitalasset
\plotone{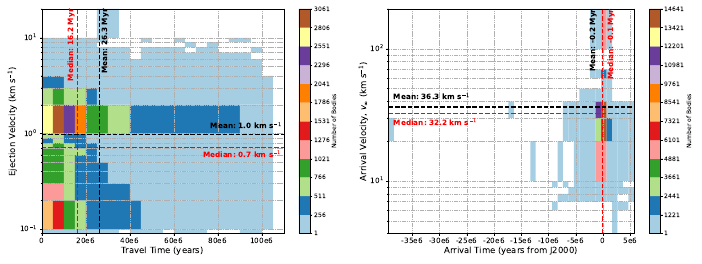}
\caption{The left figure shows the ejection velocities vs the time spent traveling in the ISM of the simulated particles that enter our Oort cloud from all the systems. The right figure shows the arrival velocity vs arrival time of the simulated particles that enter our Oort cloud from all the systems. \\The complete figure set (20 images) displaying the data for each individual system separately is displayed in Appendix~\ref{append:A:heat}.
\label{fig:Total_heatplots}}
\end{figure*}

The 70,022 particles that resulted in a CA were ejected from their origin systems at a median velocity of $\widetilde{v}_{eject}=0.7~\mathrm{km~s}^{-1}$ (95\%: 0.1-3.4 km s$^{-1}$; first panel in Figure \ref{fig:Total_heatplots}). The particles arrived at the solar system from $t_\mathrm{arr}=-3.9\times10^7~$yr to $t_\mathrm{arr}=5.0\times10^6~$yr, the median time $\widetilde{t}_\mathrm{arr}=-1.3\times10^5~$yr (95\%: $-1.3\times10^6$-$1.2\times10^6$; second panel in Figure \ref{fig:Total_heatplots}).

These particles typically traveled for $1.6\times10^7~$yr (95\%: $1.6\times10^{6}-9.1\times10^{7}~$yr) in the Galactic potential (first panel of Figure \ref{fig:Total_heatplots}). The typical distance traveled relative to their origin system is 9.7 pc (95\%: 2.2-973.5 pc). During their transit, particles typically traveled at 10.6 km s$^{-1}$ (95\%: 0.2-47.3 km s$^{-1}$) relative to the circular velocity of the Sun.
The median velocity at the time of arrival \added{the excess velocity} was $\widetilde{v}_{\infty}\pm\sigma_{v_{\infty}}=32.2\pm27.5~\mathrm{km~s}^{-1}$ (95\%: 14.9-125.6 km s$^{-1}$; second panel in Figure \ref{fig:Total_heatplots}) corresponding to a heliocentric velocity at 1 au of $\widetilde{v}_{hel,1au}=53.0~\mathrm{km~s}^{-1}$ (95\%: 44.7-132.5 km s$^{-1}$). Assuming the particles have a perihelion with respect to the Sun of $q=1~$au, the implied eccentricity is $\widetilde{e}_{q=1au}=2.17$ (95\%: 1.25-18.80). The distribution of both $v_{\infty}$ and $e$, weighted by their contribution to the flux at Earth, can be seen in Figure \ref{fig:vel_e}.

We examined the simulated particles that arrived at our solar system under the criteria laid out in Section \ref{sec:sizes} to determine the minimum particle sizes that could survive magnetic deflection, ISM drag, and grain-grain sputtering.
The least demanding trajectory could safely transport a particle as small as $d_{\mathrm{min}}=5.5~\mu$m. The most demanding trajectory required a particle of at least $d_{\mathrm{survive}}=2.4\times10^4~\mu$m (or 24 mm). The smallest particles that can be delivered from the different systems varies considerably; however, any particles above a few cm in size can in principle reach us effectively unperturbed from any of the debris disks considered here.

\begin{figure*}
% \digitalasset
\begin{centering}
\includegraphics[width=\textwidth]{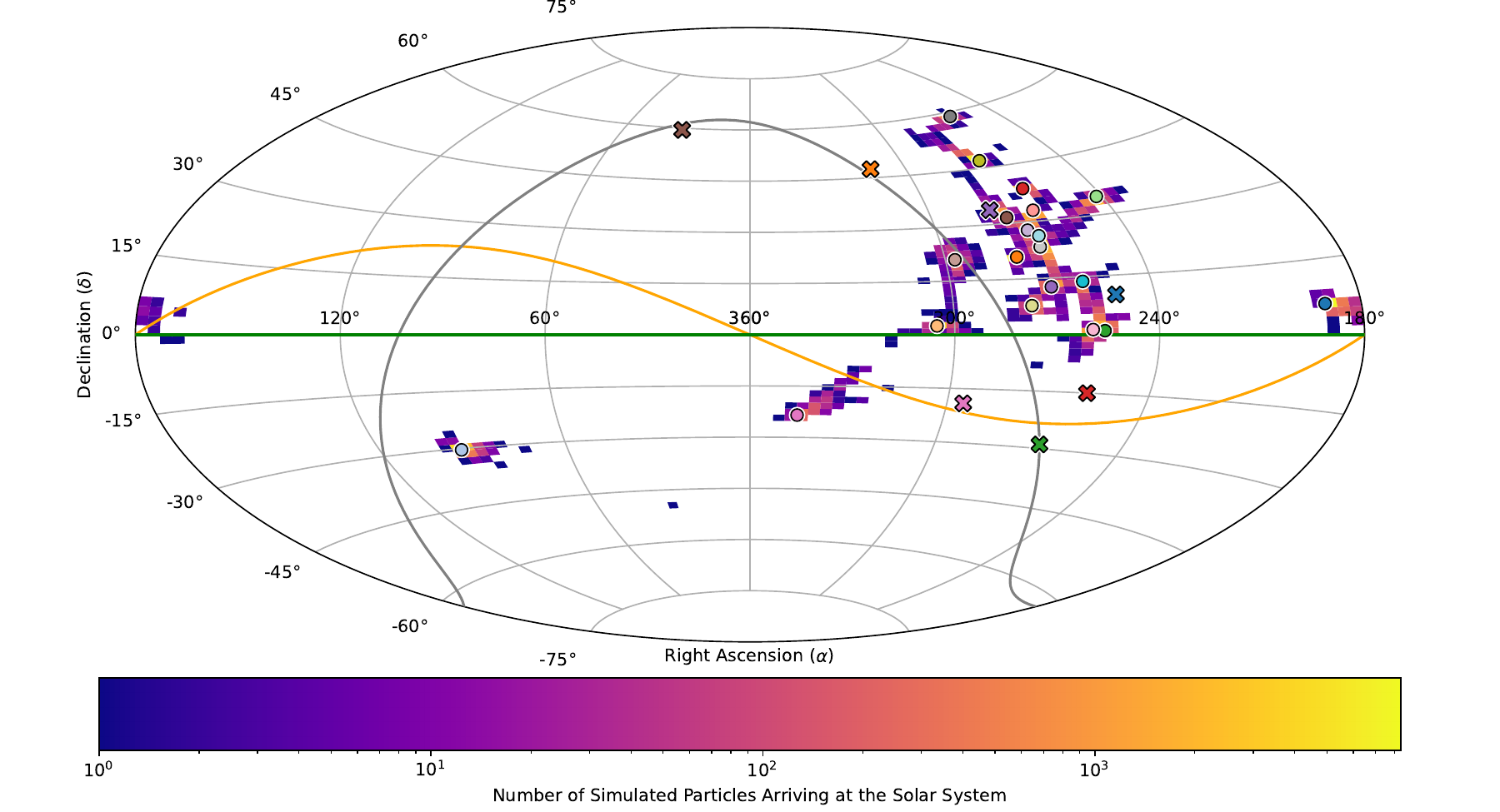}
\includegraphics[width=\textwidth]{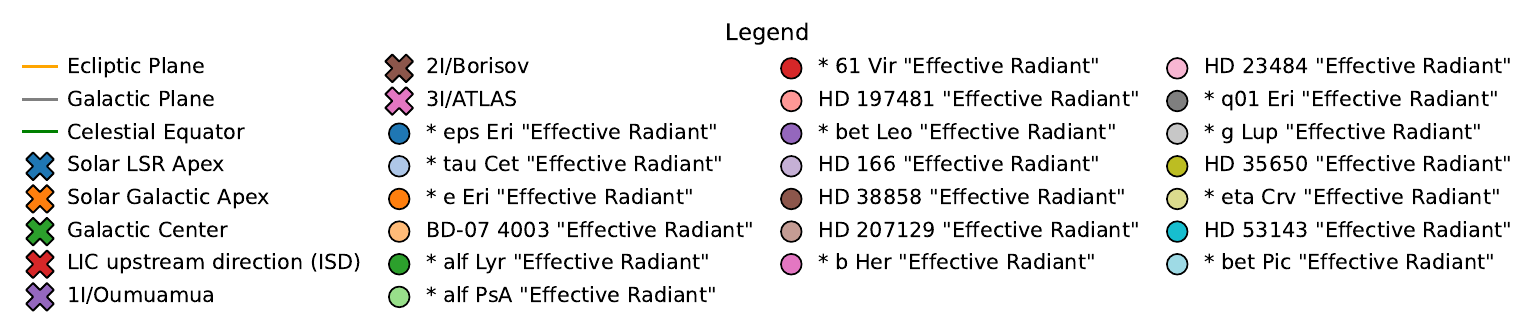}
\caption{The heliocentric equatorial radiant for the close approaches at the time of their closest Solar approach (``Arrival Time"), with the current ``effective radiant" of each debris disk system corresponding to it's apparent velocity relative to the Sun. The arrival directions of the three confirmed km-scale interstellar objects (ISOs), the direction of the local interstellar cloud (LIC) upstream direction of interstellar dust (ISD), the Solar apex with respect to the Local Standard of Rest (LSR) and within the Milky Way are also included, along with the direction towards the Galactic Center (see Table \ref{tab:ISO}). The heat density is in a log-scale for the number of simulated particles transferred to the solar system.\\
The complete figure set (20 images) displaying the data for each individual system separately is displayed in Appendix~\ref{append:A:Rad}. \label{fig:radiant}}
\end{centering}
\end{figure*}

More specifically, 95\% of the particles had critical sizes within 6.3-283.7 $\mu$m. All but 17 of the simulated particles were limited by magnetic forces. The remaining 17 were all from the \textit{* e Eri} system  and were limited by grain-grain sputtering due to their high relative velocity (see Section \ref{*-e-Eri}). The direction in which these particles arrive at the solar system can be seen in Figure \ref{fig:radiant}, where the heliocentric equatorial radiant ($\alpha,~\delta$) is plotted, along with the effective radiant of each star (i.e., the velocity of the star relative to the Sun, reversed to show their arrival direction on the sky).

At the current time\footnote{Note that 10 Myr is the assumed $t_{crit}$ for all values reported here, see Tables within Appendix \ref{append:tcrit3} for $t_{crit}=3~$Myr values, resulting current values scaled up typically by a factor $\sim20$}, the flux of ISOs $\geq100~$m into the OC is $3.0\times10^7\pm3.9\times10^6~\mathrm{yr}^{-1}$. This results in $2.3\pm0.2$ ISOs expected within the inner solar system, that is within 10 au of the Sun. This is $\sim1\%$ of the ISOs predicted to be within 10 au of the Sun from the extrapolation of predicted sporadic ISO number densities \citep{engelhardt_2017}. The flux of $\geq200~\mu\mathrm{m}$ particles into the OC is $1.9\times10^{22}\pm1.3\times10^{21}~\mathrm{yr}^{-1}$, resulting in a flux at Earth of  $1.1\times10^4\pm7.7\times10^2~\mathrm{yr}^{-1}$.

The maximum values observed in our simulations for the debris disks in question is only a factor of a few higher than current values. The maximum flux of $\geq100~$m ISOs into the OC is $4.9\times10^8\pm1.5\times10^7~\mathrm{yr}^{-1}$ at $t=-6.0\times10^5~$yr, resulting in a population of $9.6\pm0.3$ ISOs within the inner solar system at that time. Simultaneously, there is a flux of $8.7\times10^{22}\pm2.6\times10^{21}~\mathrm{yr}^{-1}~\geq200~\mu\mathrm{m}$ particles into the OC, resulting in a flux at Earth of $4.3\times10^4\pm1.2\times10^3~\mathrm{yr}^{-1}$.

\begin{figure*}
\plotone{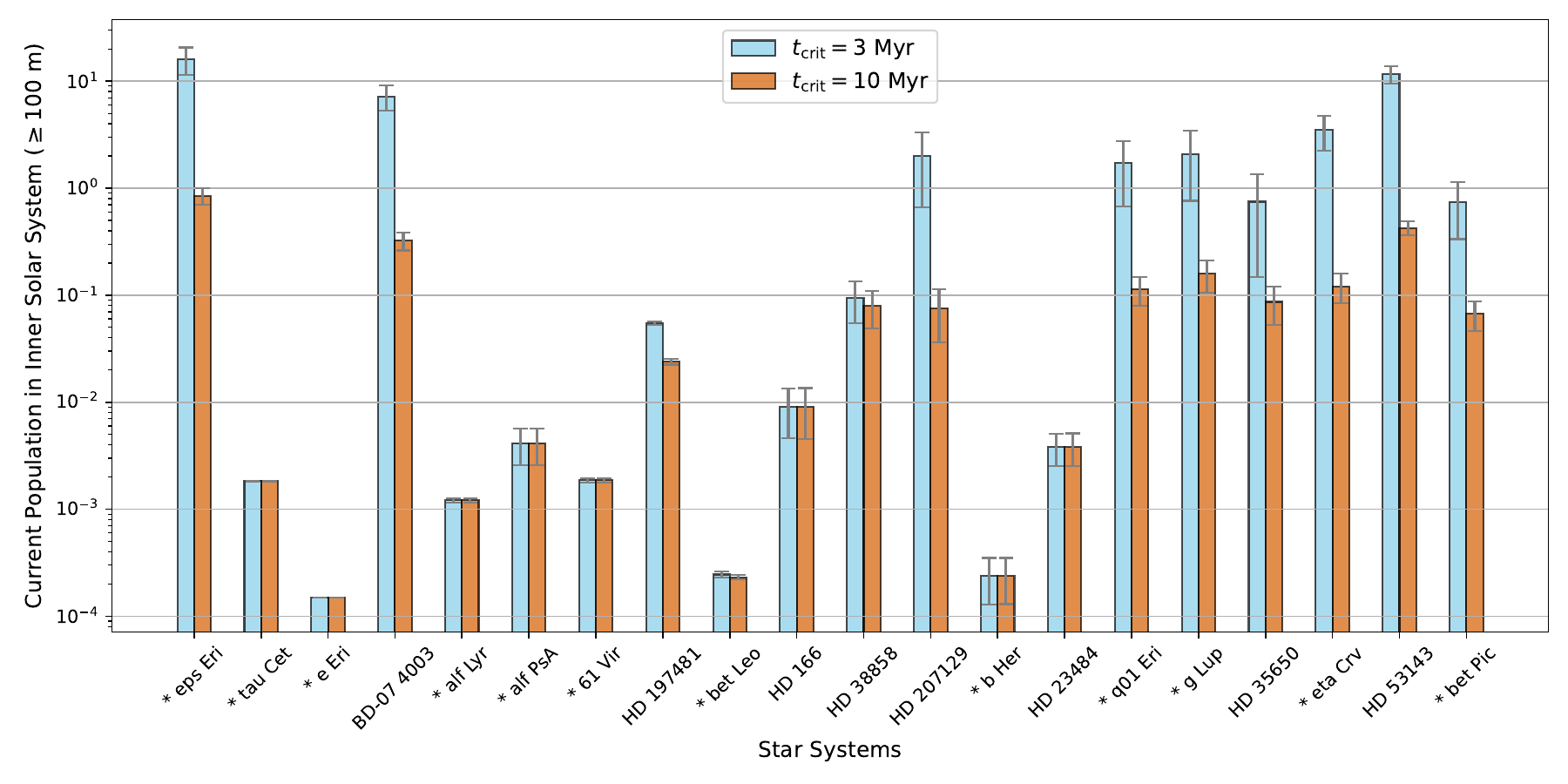}
\plotone{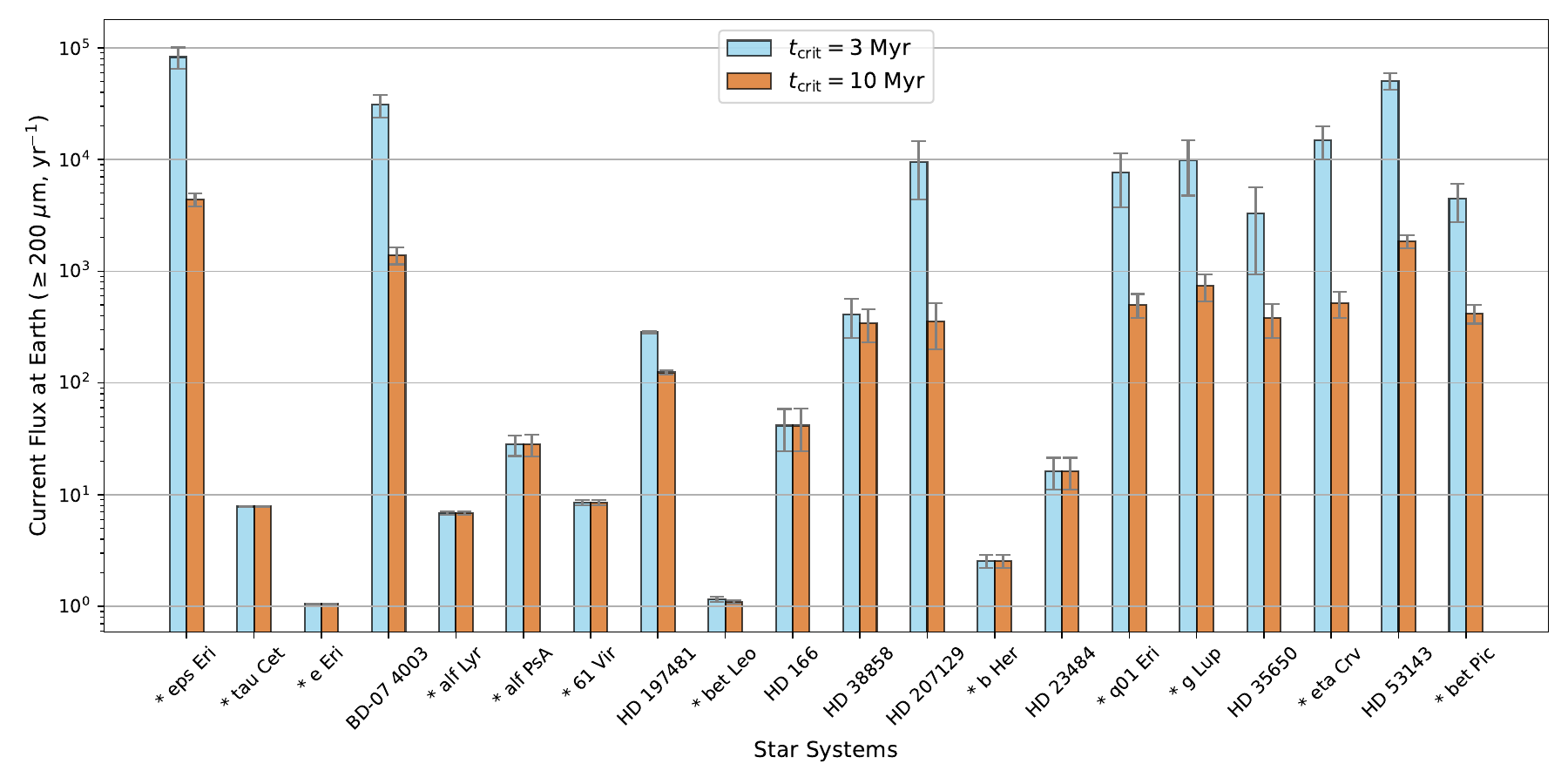}
\caption{The top plot shows the expected current population of $\geq 100~$m particles in the inner solar system. The bottom plot is the current flux of $\geq 200~\mu$m particles at the Earth.
\label{fig:pop&flux}}
\end{figure*}

A detailed system-by-system is provided below. Each subsection follows the same template and can be read selectively for systems of interest. Key transfer characteristics are summarized in Table \ref{tab:transfer_summary}; flux/population estimates for $\geq100~$m ISOs are in Table \ref{tab:100m_ageDepend_tcrit10}, and flux estimates of $\geq200~\mu$m particles at Earth are in Table \ref{tab:200micron_ageDepend_tcrit10}. These are also presented as histograms in Figure \ref{fig:pop&flux}.

\subsection{* eps Eri - Ran}

\textit{* eps Eri}, otherwise known as \textit{Ran}, is a 400-800 Myr old star \citep{janson_et_al_2015} with an observed IR excess of $1.10\times10^{-4}$ \citep{backman_et_al_2009}. From Equation~\ref{IRage}, we derive $t_{IR}=83~$Myr. This star system delivered 0.09\% of the simulated particles to the solar system. The 9,441 particles that reach us were typically ejected at $\widetilde{t}_\mathrm{eject} = -6.3\times10^{6}~\mathrm{yr}$ (95\%: $-6.4\times10^{7}- -9.5\times10^{5}$ yr) with $\widetilde{v}_\mathrm{eject} = 0.6~\mathrm{km~s^{-1}}$ (95\%: $0.1-3.4~\mathrm{km~s^{-1}}$; first panel in Figure~\ref{fig:heatplots:star-eps-Eri}). The particles typically arrived at the solar system at $\widetilde{t}_\mathrm{arr} = -1.1\times10^{5}~\mathrm{yr}$ (95\%: $-5.6\times10^{5}-3.0\times10^{5}$; second panel in Figure~\ref{fig:heatplots:star-eps-Eri}). The median velocity at arrival was $\widetilde{v}_\infty \pm \sigma_{v_\infty} = 22.1 \pm 0.8~\mathrm{km~s^{-1}}$ (95\%: $20.5-23.9~\mathrm{km~s^{-1}}$; second panel in Figure~\ref{fig:heatplots:star-eps-Eri}), corresponding to a heliocentric velocity at 1~au of $\widetilde{v}_\mathrm{hel,1au} = 47.6~\mathrm{km~s^{-1}}$ (95\%: $46.9-48.4~\mathrm{km~s^{-1}}$). Assuming $q=1~\mathrm{au}$, the implied eccentricity is $\widetilde{e}_{q=1au} = 1.55$ (95\%: $1.47-1.65$). The distribution of both $v_{\infty}$ and $e$, weighted by their contribution to the flux at Earth for \textit{* eps Eri}, can be seen in Figure~\ref{fig:velE:star-eps-Eri}. The average direction of arrival is $(\alpha,\delta) = (189.2^\circ\,\pm\,2.1^\circ,\ 6.0^\circ\,\pm\,0.7^\circ)$ (Figure~\ref{fig:radiant:star-eps-Eri}).

These particles typically traveled for $6.2\times10^{6}$~yr (95\%: $8.3\times10^{5}-6.3\times10^{7}$~yr) in the Galactic potential (first panel of Figure~\ref{fig:heatplots:star-eps-Eri}). The typical distance traveled relative to their origin system is 2.9~pc (95\%: $1.8-20.3~$pc). During their transit, the typical velocity relative to the circular velocity of the Sun is $19.2~\mathrm{km~s}^{-1}$ (95\%: $7.2-21.5~\mathrm{km~s}^{-1}$).

The least demanding trajectory in terms of magnetic deflection, ISM drag, and grain-grain sputtering would have allowed a particle of $d_{\min}=5.9~\mu\mathrm{m}$ to survive its passage through the ISM. The most demanding delivery trajectory requires $d_{\mathrm{survive}}=1.4\times10^{3}~\mu\mathrm{m}$ (or $1.4~\mathrm{mm}$). 95\% of the particles had critical sizes within $6.6-30~\mu\mathrm{m}$, all of which are set by magnetic forces.

From \textit{* eps Eri}, the current flux of ISOs $\geq100$~m into the OC is $3.0\times10^{7}\pm3.9\times10^{6}$. This results in $0.8\pm0.1$ ISOs within the inner solar system. The flux of $\geq200~\mu\mathrm{m}$ particles into the OC is $5.2\times10^{21}\pm6.8\times10^{20}$, resulting in a flux at Earth of $4.4\times10^{3}\pm5.8\times10^{2}~\mathrm{yr}^{-1}$.

The maximum flux of $\geq100$~m ISOs into the OC is $4.6\times10^{7}\pm5.1\times10^{6}~\mathrm{yr}^{-1}$ at $t=-6.0\times10^{4}~\mathrm{yr}$, resulting in a population of $1.3\pm0.2$ ISOs within the inner solar system at that time. Simultaneously, there is a flux of $8.1\times10^{21}\pm9.0\times10^{20}~\mathrm{yr}^{-1}\,\geq200~\mu\mathrm{m}$ particles into the OC, resulting in a flux at Earth of $6.7\times10^{3}\pm7.7\times10^{2}~\mathrm{yr}^{-1}$.

The values above assume $t_{\mathrm{crit}}=10~\mathrm{Myr}$, see Tables within Appendix~\ref{append:tcrit3} for flux values with the assumption $t_{\mathrm{crit}}=3~\mathrm{Myr}$.\subsection{* eps Eri - Ran}

\textit{* eps Eri}, otherwise known as \textit{Ran}, is a 400-800 Myr old star \citep{janson_et_al_2015} with an observed IR excess of $1.10\times10^{-4}$ \citep{backman_et_al_2009}. From Equation~\ref{IRage}, we derive $t_{IR}=83~$Myr. This star system delivered 0.09\% of the simulated particles to the solar system. The 9,441 particles that reach us were typically ejected at $\widetilde{t}_\mathrm{eject} = -6.3\times10^{6}~\mathrm{yr}$ (95\%: $-6.4\times10^{7}- -9.5\times10^{5}$ yr) with $\widetilde{v}_\mathrm{eject} = 0.6~\mathrm{km~s^{-1}}$ (95\%: $0.1-3.4~\mathrm{km~s^{-1}}$; first panel in Figure~\ref{fig:heatplots:star-eps-Eri}). The particles typically arrived at the solar system at $\widetilde{t}_\mathrm{arr} = -1.1\times10^{5}~\mathrm{yr}$ (95\%: $-5.6\times10^{5}-3.0\times10^{5}$; second panel in Figure~\ref{fig:heatplots:star-eps-Eri}). The median velocity at arrival was $\widetilde{v}_\infty \pm \sigma_{v_\infty} = 22.1 \pm 0.8~\mathrm{km~s^{-1}}$ (95\%: $20.5-23.9~\mathrm{km~s^{-1}}$; second panel in Figure~\ref{fig:heatplots:star-eps-Eri}), corresponding to a heliocentric velocity at 1~au of $\widetilde{v}_\mathrm{hel,1au} = 47.6~\mathrm{km~s^{-1}}$ (95\%: $46.9-48.4~\mathrm{km~s^{-1}}$). Assuming $q=1~\mathrm{au}$, the implied eccentricity is $\widetilde{e}_{q=1au} = 1.55$ (95\%: $1.47-1.65$). The distribution of both $v_{\infty}$ and $e$, weighted by their contribution to the flux at Earth for \textit{* eps Eri}, can be seen in Figure~\ref{fig:velE:star-eps-Eri}. The average direction of arrival is $(\alpha,\delta) = (189.2^\circ\,\pm\,2.1^\circ,\ 6.0^\circ\,\pm\,0.7^\circ)$ (Figure~\ref{fig:radiant:star-eps-Eri}).

These particles typically traveled for $6.2\times10^{6}$~yr (95\%: $8.3\times10^{5}-6.3\times10^{7}$~yr) in the Galactic potential (first panel of Figure~\ref{fig:heatplots:star-eps-Eri}). The typical distance traveled relative to their origin system is 2.9~pc (95\%: $1.8-20.3~$pc). During their transit, the typical velocity relative to the circular velocity of the Sun is $19.2~\mathrm{km~s}^{-1}$ (95\%: $7.2-21.5~\mathrm{km~s}^{-1}$).

The least demanding trajectory in terms of magnetic deflection, ISM drag, and grain-grain sputtering would have allowed a particle of $d_{\min}=5.9~\mu\mathrm{m}$ to survive its passage through the ISM. The most demanding delivery trajectory requires $d_{\mathrm{survive}}=1.4\times10^{3}~\mu\mathrm{m}$ (or $1.4~\mathrm{mm}$). 95\% of the particles had critical sizes within $6.6-30~\mu\mathrm{m}$, all of which are set by magnetic forces.

From \textit{* eps Eri}, the current flux of ISOs $\geq100$~m into the OC is $3.0\times10^{7}\pm3.9\times10^{6}~\mathrm{yr}^{-1}$. This results in $0.8\pm0.1$ ISOs within the inner solar system. The flux of $\geq200~\mu\mathrm{m}$ particles into the OC is $5.2\times10^{21}\pm6.8\times10^{20}~\mathrm{yr}^{-1}$, resulting in a flux at Earth of $4.4\times10^{3}\pm5.8\times10^{2}~\mathrm{yr}^{-1}$.

The maximum flux of $\geq100$~m ISOs into the OC is $4.6\times10^{7}\pm5.1\times10^{6}~\mathrm{yr}^{-1}$ at $t=-6.0\times10^{4}~\mathrm{yr}$, resulting in a population of $1.3\pm0.2$ ISOs within the inner solar system at that time. Simultaneously, there is a flux of $8.1\times10^{21}\pm9.0\times10^{20}~\mathrm{yr}^{-1}\,\geq200~\mu\mathrm{m}$ particles into the OC, resulting in a flux at Earth of $6.7\times10^{3}\pm7.7\times10^{2}~\mathrm{yr}^{-1}$.

The values above assume $t_{\mathrm{crit}}=10~\mathrm{Myr}$, see Tables within Appendix~\ref{append:tcrit3} for flux values with the assumption $t_{\mathrm{crit}}=3~\mathrm{Myr}$.

\subsection{* tau Cet}

\textit{* tau Cet} is a $10\pm0.5$ Gyr old star \citep{greaves_et_al_2004} with an observed IR excess of $1.20\times10^{-5}$ \citep{di_folco_et_al_2004}. From Equation~\ref{IRage}, we assume $t_{IR}=250~$Myr. This star system delivered 0.07\% of the simulated particles to the solar system. The 7,421 particles that reach us were typically ejected at $\widetilde{t}_\mathrm{eject} = -1.4\times10^{7}~\mathrm{yr}$ (95\%: $-8.8\times10^{7}- -1.3\times10^{6}$ yr) with $\widetilde{v}_\mathrm{eject} = 0.6~\mathrm{km~s^{-1}}$ (95\%: $0.1-3.2~\mathrm{km~s^{-1}}$; first panel in Figure~\ref{fig:heatplots:star-tau-Cet}). The particles typically arrived at the solar system at $\widetilde{t}_\mathrm{arr} = 7.5\times10^{4}~\mathrm{yr}$ (95\%: $-4.9\times10^{5}-8.0\times10^{5}$; second panel in Figure~\ref{fig:heatplots:star-tau-Cet}). The median velocity at arrival was $\widetilde{v}_\infty \pm \sigma_{v_\infty} = 37.1 \pm 0.8~\mathrm{km~s^{-1}}$ (95\%: $35.6-38.9~\mathrm{km~s^{-1}}$; second panel in Figure~\ref{fig:heatplots:star-tau-Cet}), corresponding to a heliocentric velocity at 1~au of $\widetilde{v}_\mathrm{hel,1au} = 56.1~\mathrm{km~s^{-1}}$ (95\%: $55.2-57.3~\mathrm{km~s^{-1}}$). Assuming $q=1~\mathrm{au}$, the implied eccentricity is $\widetilde{e}_{q=1au} = 2.55$ (95\%: $2.43-2.71$). The distribution of both $v_{\infty}$ and $e$, weighted by their contribution to the flux at Earth for \textit{* tau Cet}, can be seen in Figure~\ref{fig:velE:star-tau-Cet}. The average direction of arrival is $(\alpha,\delta) = (93.5^\circ\,\pm\,1.3^\circ,\ -30.4^\circ\,\pm\,0.3^\circ)$ (Figure~\ref{fig:radiant:star-tau-Cet}).

These particles typically traveled for $1.4\times10^{7}$~yr (95\%: $1.4\times10^{6}-8.8\times10^{7}$~yr) in the Galactic potential (first panel of Figure~\ref{fig:heatplots:star-tau-Cet}). The typical distance traveled relative to their origin system is 5.8~pc (95\%: $3.0-49.2~$pc). During their transit, the typical velocity relative to the circular velocity of the Sun is $37.5~\mathrm{km~s}^{-1}$ (95\%: $1.4-43.8~\mathrm{km~s}^{-1}$).

The least demanding trajectory in terms of magnetic deflection, ISM drag, and grain-grain sputtering would have allowed a particle of $d_{\min}=5.6~\mu\mathrm{m}$ to survive its passage through the ISM. The most demanding delivery trajectory requires $d_{\mathrm{survive}}=1.8\times10^{3}~\mu\mathrm{m}$ (or $1.8~\mathrm{mm}$). 95\% of the particles had critical sizes within $5.8-88~\mu\mathrm{m}$, all of which are set by magnetic forces.

From \textit{* tau Cet}, the current flux of ISOs $\geq100$~m into the OC is $1.1\times10^{5}\pm9.1\times10^{2}~\mathrm{yr}^{-1}$. This results in $1.8\times10^{-3}\pm1.9\times10^{-5}$ ISOs within the inner solar system. The flux of $\geq200~\mu\mathrm{m}$ particles into the OC is $1.9\times10^{19}\pm1.6\times10^{17}~\mathrm{yr}^{-1}$, resulting in a flux at Earth of $7.8\pm7.0\times10^{-2}~\mathrm{yr}^{-1}$.

The maximum flux of $\geq100$~m ISOs into the OC is $1.3\times10^{5}\pm1.0\times10^{3}~\mathrm{yr}^{-1}$ at $t=6.5\times10^{4}~\mathrm{yr}$, resulting in a population of $2.2\times10^{-3}\pm1.9\times10^{-5}$ ISOs within the inner solar system at that time. Simultaneously, there is a flux of $2.3\times10^{19}\pm1.8\times10^{17}~\mathrm{yr}^{-1}\,\geq200~\mu\mathrm{m}$ particles into the OC, resulting in a flux at Earth of $9.5\pm7.3\times10^{-2}~\mathrm{yr}^{-1}$.

The values above assume $t_{\mathrm{crit}}=10~\mathrm{Myr}$, see Tables within Appendix~\ref{append:tcrit3} for flux values with the assumption $t_{\mathrm{crit}}=3~\mathrm{Myr}$.

\subsection{* e Eri}\label{*-e-Eri}

\textit{* e Eri} is a $6\pm0.3$ Gyr old star \citep{desgrange_et_al_2023} with an observed IR excess of $4.30\times10^{-5}$ \citep{wyatt_et_al_2012}. From Equation~\ref{IRage}, we assume $t_{IR}=418~$Myr. This star system delivered 0.05\% of the simulated particles to the solar system. The 5,354 particles that reach us were typically ejected at $\widetilde{t}_\mathrm{eject} = -2.3\times10^{7}~\mathrm{yr}$ (95\%: $-9.5\times10^{7}- -1.9\times10^{6}$ yr) with $\widetilde{v}_\mathrm{eject} = 0.6~\mathrm{km~s^{-1}}$ (95\%: $0.2-3.3~\mathrm{km~s^{-1}}$; first panel in Figure~\ref{fig:heatplots:star-e-Eri}). The particles typically arrived at the solar system at $\widetilde{t}_\mathrm{arr} = -5.0\times10^{4}~\mathrm{yr}$ (95\%: $-2.6\times10^{5}-1.3\times10^{5}$; second panel in Figure~\ref{fig:heatplots:star-e-Eri}). The median velocity at arrival was $\widetilde{v}_\infty \pm \sigma_{v_\infty} = 125.5 \pm 0.8~\mathrm{km~s^{-1}}$ (95\%: $123.8-127.1~\mathrm{km~s^{-1}}$; second panel in Figure~\ref{fig:heatplots:star-e-Eri}), corresponding to a heliocentric velocity at 1~au of $\widetilde{v}_\mathrm{hel,1au} = 132.4~\mathrm{km~s^{-1}}$ (95\%: $130.7-133.9~\mathrm{km~s^{-1}}$). Assuming $q=1~\mathrm{au}$, the implied eccentricity is $\widetilde{e}_{q=1au} = 18.76$ (95\%: $18.27-19.20$). The distribution of both $v_{\infty}$ and $e$, weighted by their contribution to the flux at Earth for \textit{* e Eri}, can be seen in Figure~\ref{fig:velE:star-e-Eri}. The average direction of arrival is $(\alpha,\delta) = (278.1^\circ\,\pm\,0.2^\circ,\ 20.8^\circ\,\pm\,0.2^\circ)$ (Figure~\ref{fig:radiant:star-e-Eri}).

These particles typically traveled for $2.3\times10^{7}$~yr (95\%: $1.9\times10^{6}-9.5\times10^{7}$~yr) in the Galactic potential (first panel of Figure~\ref{fig:heatplots:star-e-Eri}). The typical distance traveled relative to their origin system is 10.1~pc (95\%: $4.2-68.2~$pc). During their transit, the typical velocity relative to the circular velocity of the Sun is $42.2~\mathrm{km~s}^{-1}$ (95\%: $4.3-60.6~\mathrm{km~s}^{-1}$).

The least demanding trajectory in terms of magnetic deflection, ISM drag, and grain-grain sputtering would have allowed a particle of $d_{\min}=5.4~\mu\mathrm{m}$ to survive its passage through the ISM. The most demanding delivery trajectory requires $d_{\mathrm{survive}}=3.9\times10^{2}~\mu\mathrm{m}$. 95\% of the particles had critical sizes within $5.8-46~\mu\mathrm{m}$. All but 17 of the simulated particles were limited by magnetic forces, these outliers were limited from grain-grain sputtering due to their high velocity relative to the circular velocity of the Sun.

From \textit{* e Eri}, the current flux of ISOs $\geq100$~m into the OC is $3.0\times10^{4}\pm1.7\times10^{2}~\mathrm{yr}^{-1}$. This results in $1.5\times10^{-4}\pm8.3\times10^{-7}$ ISOs within the inner solar system. The flux of $\geq200~\mu\mathrm{m}$ particles into the OC is $5.2\times10^{18}\pm3.0\times10^{16}~\mathrm{yr}^{-1}$, resulting in a flux at Earth of $1.1\pm6.0\times10^{-3}~\mathrm{yr}^{-1}$.

The maximum flux of $\geq100$~m ISOs into the OC is $4.0\times10^{4}\pm1.9\times10^{2}~\mathrm{yr}^{-1}$ at $t=-5.5\times10^{4}~\mathrm{yr}$, resulting in a population of $2.0\times10^{-4}\pm1.0\times10^{-6}$ ISOs within the inner solar system at that time. Simultaneously, there is a flux of $7.2\times10^{18}\pm3.4\times10^{16}~\mathrm{yr}^{-1}\,\geq200~\mu\mathrm{m}$ particles into the OC, resulting in a flux at Earth of $1.4\pm7.0\times10^{-3}~\mathrm{yr}^{-1}$.

The values above assume $t_{\mathrm{crit}}=10~\mathrm{Myr}$, see Tables within Appendix~\ref{append:tcrit3} for flux values with the assumption $t_{\mathrm{crit}}=3~\mathrm{Myr}$.

\subsection{BD-07 4003}

\textit{BD-07 4003} is a $8.9\pm1.1$ Gyr old star \citep{desgrange_et_al_2023} with an observed IR excess of $8.90\times10^{-5}$ \citep{lestrade_et_al_2012}. From Equation~\ref{IRage}, we assume $t_{IR}=92~$Myr. This star system delivered 0.04\% of the simulated particles to the solar system. The 3,808 particles that reach us were typically ejected at $\widetilde{t}_\mathrm{eject} = -2.2\times10^{7}~\mathrm{yr}$ (95\%: $-9.1\times10^{7}- -2.6\times10^{6}$ yr) with $\widetilde{v}_\mathrm{eject} = 0.6~\mathrm{km~s^{-1}}$ (95\%: $0.2-3.1~\mathrm{km~s^{-1}}$; first panel in Figure~\ref{fig:heatplots:BD-07-4003}). The particles typically arrived at the solar system at $\widetilde{t}_\mathrm{arr} = -6.5\times10^{4}~\mathrm{yr}$ (95\%: $-8.6\times10^{5}-6.0\times10^{5}$; second panel in Figure~\ref{fig:heatplots:BD-07-4003}). The median velocity at arrival was $\widetilde{v}_\infty \pm \sigma_{v_\infty} = 38.0 \pm 0.8~\mathrm{km~s^{-1}}$ (95\%: $36.3-39.4~\mathrm{km~s^{-1}}$; second panel in Figure~\ref{fig:heatplots:BD-07-4003}), corresponding to a heliocentric velocity at 1~au of $\widetilde{v}_\mathrm{hel,1au} = 56.7~\mathrm{km~s^{-1}}$ (95\%: $55.6-57.7~\mathrm{km~s^{-1}}$). Assuming $q=1~\mathrm{au}$, the implied eccentricity is $\widetilde{e}_{q=1au} = 2.63$ (95\%: $2.49-2.75$). The distribution of both $v_{\infty}$ and $e$, weighted by their contribution to the flux at Earth for \textit{* BD 07 4003}, can be seen in Figure~\ref{fig:velE:BD-07-4003}. The average direction of arrival is $(\alpha,\delta) = (304.5^\circ\,\pm\,1.1^\circ,\ 2.3^\circ\,\pm\,0.2^\circ)$ (Figure~\ref{fig:radiant:BD-07-4003}).

These particles typically traveled for $2.2\times10^{7}$~yr (95\%: $2.6\times10^{6}-9.1\times10^{7}$~yr) in the Galactic potential (first panel of Figure~\ref{fig:heatplots:BD-07-4003}). The typical distance traveled relative to their origin system is 11.3~pc (95\%: $6.0-55.6~$pc). During their transit, the typical velocity relative to the circular velocity of the Sun is $8.9~\mathrm{km~s}^{-1}$ (95\%: $0.6-15.9~\mathrm{km~s}^{-1}$).

The least demanding trajectory in terms of magnetic deflection, ISM drag, and grain-grain sputtering would have allowed a particle of $d_{\min}=16~\mu\mathrm{m}$ to survive its passage through the ISM. The most demanding delivery trajectory requires $d_{\mathrm{survive}}=2.4\times10^{3}~\mu\mathrm{m}$ (or $2.4~\mathrm{mm}$). 95\% of the particles had critical sizes within $16-1.2\times10^{2}~\mu\mathrm{m}$, all of which are set by magnetic forces.

From \textit{* BD 07 4003}, the current flux of ISOs $\geq100$~m into the OC is $1.9\times10^{7}\pm3.5\times10^{6}~\mathrm{yr}^{-1}$. This results in $0.3\pm0.1$ ISOs within the inner solar system. The flux of $\geq200~\mu\mathrm{m}$ particles into the OC is $3.4\times10^{21}\pm6.1\times10^{20}~\mathrm{yr}^{-1}$, resulting in a flux at Earth of $1.4\times10^{3}\pm2.4\times10^{2}~\mathrm{yr}^{-1}$.

The maximum flux of $\geq100$~m ISOs into the OC is $6.1\times10^{7}\pm5.3\times10^{6}~\mathrm{yr}^{-1}$ at $t=-2.2\times10^{5}~\mathrm{yr}$, resulting in a population of $1.0\pm0.1$ ISOs within the inner solar system at that time. Simultaneously, there is a flux of $1.1\times10^{22}\pm1.0\times10^{21}~\mathrm{yr}^{-1}\,\geq200~\mu\mathrm{m}$ particles into the OC, resulting in a flux at Earth of $4.3\times10^{3}\pm3.9\times10^{2}~\mathrm{yr}^{-1}$.

The values above assume $t_{\mathrm{crit}}=10~\mathrm{Myr}$, see Tables within Appendix~\ref{append:tcrit3} for flux values with the assumption $t_{\mathrm{crit}}=3~\mathrm{Myr}$.

\subsection{* alf Lyr - Vega}

\textit{* alf Lyr}, also known as \textit{Vega}, is a $400-800$ Myr old star \citep{janson_et_al_2015} with an observed IR excess of $1.58\times10^{-5}$ \citep{di_folco_et_al_2004}. From Equation~\ref{IRage}, we assume $t_{IR}=218~$Myr. This star system delivered 0.05\% of the simulated particles to the solar system. The 5,379 particles that reach us were typically ejected at $\widetilde{t}_\mathrm{eject} = -2.3\times10^{7}~\mathrm{yr}$ (95\%: $-9.1\times10^{7}- -2.1\times10^{6}$ yr) with $\widetilde{v}_\mathrm{eject} = 0.5~\mathrm{km~s^{-1}}$ (95\%: $0.1-3.0~\mathrm{km~s^{-1}}$; first panel in Figure~\ref{fig:heatplots:star-alf-Lyr}). The particles typically arrived at the solar system at $\widetilde{t}_\mathrm{arr} = 3.0\times10^{5}~\mathrm{yr}$ (95\%: $-7.6\times10^{5}-1.3\times10^{6}$; second panel in Figure~\ref{fig:heatplots:star-alf-Lyr}). The median velocity at arrival was $\widetilde{v}_\infty \pm \sigma_{v_\infty} = 18.6 \pm 0.7~\mathrm{km~s^{-1}}$ (95\%: $17.1-20.3~\mathrm{km~s^{-1}}$; second panel in Figure~\ref{fig:heatplots:star-alf-Lyr}), corresponding to a heliocentric velocity at 1~au of $\widetilde{v}_\mathrm{hel,1au} = 46.0~\mathrm{km~s^{-1}}$ (95\%: $45.5-46.8~\mathrm{km~s^{-1}}$). Assuming $q=1~\mathrm{au}$, the implied eccentricity is $\widetilde{e}_{q=1au} = 1.39$ (95\%: $1.33-1.47$). The distribution of both $v_{\infty}$ and $e$, weighted by their contribution to the flux at Earth for \textit{* alf Lyr}, can be seen in Figure~\ref{fig:velE:star-alf-Lyr}. The average direction of arrival is $(\alpha,\delta) = (256.6^\circ\,\pm\,1.2^\circ,\ 2.2^\circ\,\pm\,1.8^\circ)$ (Figure~\ref{fig:radiant:star-alf-Lyr}).

These particles typically traveled for $2.3\times10^{7}$~yr (95\%: $2.4\times10^{6}-9.1\times10^{7}$~yr) in the Galactic potential (first panel of Figure~\ref{fig:heatplots:star-alf-Lyr}). The typical distance traveled relative to their origin system is 7.2~pc (95\%: $5.0-44.3~$pc). During their transit, the typical velocity relative to the circular velocity of the Sun is $7.7~\mathrm{km~s}^{-1}$ (95\%: $4.2-9.8~\mathrm{km~s}^{-1}$).

The least demanding trajectory in terms of magnetic deflection, ISM drag, and grain-grain sputtering would have allowed a particle of $d_{\min}=16~\mu\mathrm{m}$ to survive its passage through the ISM. The most demanding delivery trajectory requires $d_{\mathrm{survive}}=2.0\times10^{2}~\mu\mathrm{m}$. 95\% of the particles had critical sizes within $17-49~\mu\mathrm{m}$, all of which are set by magnetic forces.

From \textit{* alf Lyr}, the current flux of ISOs $\geq100$~m into the OC is $3.6\times10^{4}\pm1.2\times10^{3}~\mathrm{yr}^{-1}$. This results in $1.2\times10^{-3}\pm6.3\times10^{-5}$ ISOs within the inner solar system. The flux of $\geq200~\mu\mathrm{m}$ particles into the OC is $6.4\times10^{18}\pm2.2\times10^{17}~\mathrm{yr}^{-1}$, resulting in a flux at Earth of $6.9\pm2.3\times10^{-1}~\mathrm{yr}^{-1}$.

The maximum flux of $\geq100$~m ISOs into the OC is $1.6\times10^{5}\pm2.5\times10^{3}~\mathrm{yr}^{-1}$ at $t=3.3\times10^{5}~\mathrm{yr}$, resulting in a population of $5.6\times10^{-3}\pm1.2\times10^{-4}$ ISOs within the inner solar system at that time. Simultaneously, there is a flux of $2.9\times10^{19}\pm4.4\times10^{17}~\mathrm{yr}^{-1}\,\geq200~\mu\mathrm{m}$ particles into the OC, resulting in a flux at Earth of $3.2\times10^{1}\pm4.7\times10^{-1}~\mathrm{yr}^{-1}$.

The values above assume $t_{\mathrm{crit}}=10~\mathrm{Myr}$, see Tables within Appendix~\ref{append:tcrit3} for flux values with the assumption $t_{\mathrm{crit}}=3~\mathrm{Myr}$.

\subsection{* alf PsA - Fomalhaut}

\textit{* alf PsA}, also known as \textit{Fomalhaut}, is a $440\pm40$ Myr old star \citep{janson_et_al_2015} with an observed IR excess of $4.60\times10^{-4}$ \citep{di_folco_et_al_2004}. From Equation~\ref{IRage}, we assume $t_{IR}=128~$Myr. This star system delivered 0.03\% of the simulated particles to the solar system. The 3,018 particles that reach us were typically ejected at $\widetilde{t}_\mathrm{eject} = -2.1\times10^{7}~\mathrm{yr}$ (95\%: $-9.3\times10^{7}- -2.7\times10^{6}$ yr) with $\widetilde{v}_\mathrm{eject} = 0.6~\mathrm{km~s^{-1}}$ (95\%: $0.1-3.3~\mathrm{km~s^{-1}}$; first panel in Figure~\ref{fig:heatplots:star-alf-PsA}). The particles typically arrived at the solar system at $\widetilde{t}_\mathrm{arr} = -5.0\times10^{5}~\mathrm{yr}$ (95\%: $-1.4\times10^{6}-6.7\times10^{5}$; second panel in Figure~\ref{fig:heatplots:star-alf-PsA}). The median velocity at arrival was $\widetilde{v}_\infty \pm \sigma_{v_\infty} = 15.1 \pm 0.7~\mathrm{km~s^{-1}}$ (95\%: $13.8-16.6~\mathrm{km~s^{-1}}$; second panel in Figure~\ref{fig:heatplots:star-alf-PsA}), corresponding to a heliocentric velocity at 1~au of $\widetilde{v}_\mathrm{hel,1au} = 44.8~\mathrm{km~s^{-1}}$ (95\%: $44.3-45.3~\mathrm{km~s^{-1}}$). Assuming $q=1~\mathrm{au}$, the implied eccentricity is $\widetilde{e}_{q=1au} = 1.26$ (95\%: $1.21-1.31$). The distribution of both $v_{\infty}$ and $e$, weighted by their contribution to the flux at Earth for \textit{* alf PsA}, can be seen in Figure~\ref{fig:velE:star-alf-PsA}. The average direction of arrival is $(\alpha,\delta) = (244.1^\circ\,\pm\,3.1^\circ,\ 33.5^\circ\,\pm\,1.1^\circ)$ (Figure~\ref{fig:radiant:star-alf-PsA}).

These particles typically traveled for $2.1\times10^{7}$~yr (95\%: $2.6\times10^{6}-9.2\times10^{7}$~yr) in the Galactic potential (first panel of Figure~\ref{fig:heatplots:star-alf-PsA}). The typical distance traveled relative to their origin system is 9.7~pc (95\%: $6.8-42.4~$pc). During their transit, the typical velocity relative to the circular velocity of the Sun is $3.0~\mathrm{km~s}^{-1}$ (95\%: $2.0-6.1~\mathrm{km~s}^{-1}$).

The least demanding trajectory in terms of magnetic deflection, ISM drag, and grain-grain sputtering would have allowed a particle of $d_{\min}=26~\mu\mathrm{m}$ to survive its passage through the ISM. The most demanding delivery trajectory requires $d_{\mathrm{survive}}=8.5\times10^{2}~\mu\mathrm{m}$. 95\% of the particles had critical sizes within $30-78~\mu\mathrm{m}$, all of which are set by magnetic forces.

From \textit{* alf PsA}, the current flux of ISOs $\geq100$~m into the OC is $9.7\times10^{4}\pm2.1\times10^{4}~\mathrm{yr}^{-1}$. This results in $4.1\times10^{-3}\pm1.6\times10^{-3}$ ISOs within the inner solar system. The flux of $\geq200~\mu\mathrm{m}$ particles into the OC is $1.7\times10^{19}\pm4.0\times10^{18}~\mathrm{yr}^{-1}$, resulting in a flux at Earth of $2.8\times10^{1}\pm6.2~\mathrm{yr}^{-1}$.

The maximum flux of $\geq100$~m ISOs into the OC is $1.7\times10^{6}\pm1.1\times10^{5}~\mathrm{yr}^{-1}$ at $t=-6.2\times10^{5}~\mathrm{yr}$, resulting in a population of $0.1\pm8.1\times10^{-3}$ ISOs within the inner solar system at that time. Simultaneously, there is a flux of $3.0\times10^{20}\pm1.9\times10^{19}~\mathrm{yr}^{-1}\,\geq200~\mu\mathrm{m}$ particles into the OC, resulting in a flux at Earth of $4.8\times10^{2}\pm3.1\times10^{1}~\mathrm{yr}^{-1}$.

The values above assume $t_{\mathrm{crit}}=10~\mathrm{Myr}$, see Tables within Appendix~\ref{append:tcrit3} for flux values with the assumption $t_{\mathrm{crit}}=3~\mathrm{Myr}$.

\subsection{* 61 Vir}

\textit{* 61 Vir} is a $5.1\pm1.3$ Gyr old star \citep{desgrange_et_al_2023} with an observed IR excess of $2.70\times10^{-5}$ \citep{wyatt_et_al_2012}. From Equation~\ref{IRage}, we assume $t_{IR}=167~$Myr. This star system delivered 0.02\% of the simulated particles to the solar system. The 2,261 particles that reach us were typically ejected at $\widetilde{t}_\mathrm{eject} = -2.3\times10^{7}~\mathrm{yr}$ (95\%: $-9.5\times10^{7}- -3.5\times10^{6}$ yr) with $\widetilde{v}_\mathrm{eject} = 0.7~\mathrm{km~s^{-1}}$ (95\%: $0.2-3.3~\mathrm{km~s^{-1}}$; first panel in Figure~\ref{fig:heatplots:star-61-Vir}). The particles typically arrived at the solar system at $\widetilde{t}_\mathrm{arr} = 7.5\times10^{4}~\mathrm{yr}$ (95\%: $-3.9\times10^{5}-5.9\times10^{5}$; second panel in Figure~\ref{fig:heatplots:star-61-Vir}). The median velocity at arrival was $\widetilde{v}_\infty \pm \sigma_{v_\infty} = 61.5 \pm 0.7~\mathrm{km~s^{-1}}$ (95\%: $59.9-63.2~\mathrm{km~s^{-1}}$; second panel in Figure~\ref{fig:heatplots:star-61-Vir}), corresponding to a heliocentric velocity at 1~au of $\widetilde{v}_\mathrm{hel,1au} = 74.5~\mathrm{km~s^{-1}}$ (95\%: $73.2-75.9~\mathrm{km~s^{-1}}$). Assuming $q=1~\mathrm{au}$, the implied eccentricity is $\widetilde{e}_{q=1au} = 5.26$ (95\%: $5.04-5.50$). The distribution of both $v_{\infty}$ and $e$, weighted by their contribution to the flux at Earth for \textit{* 61 Vir}, can be seen in Figure~\ref{fig:velE:star-61-Vir}. The average direction of arrival is $(\alpha,\delta) = (263.3^\circ\,\pm\,0.8^\circ,\ 38.4^\circ\,\pm\,0.4^\circ)$ (Figure~\ref{fig:radiant:star-61-Vir}).

These particles typically traveled for $2.3\times10^{7}$~yr (95\%: $3.5\times10^{6}-9.6\times10^{7}$~yr) in the Galactic potential (first panel of Figure~\ref{fig:heatplots:star-61-Vir}). The typical distance traveled relative to their origin system is 13.2~pc (95\%: $8.3-63.5~$pc). During their transit, the typical velocity relative to the circular velocity of the Sun is $23.4~\mathrm{km~s}^{-1}$ (95\%: $2.1-31.2~\mathrm{km~s}^{-1}$).

The least demanding trajectory in terms of magnetic deflection, ISM drag, and grain-grain sputtering would have allowed a particle of $d_{\min}=11~\mu\mathrm{m}$ to survive its passage through the ISM. The most demanding delivery trajectory requires $d_{\mathrm{survive}}=4.1\times10^{2}~\mu\mathrm{m}$. 95\% of the particles had critical sizes within $11-70~\mu\mathrm{m}$, all of which are set by magnetic forces.

From \textit{* 61 Vir}, the current flux of ISOs $\geq100$~m into the OC is $1.8\times10^{5}\pm8.4\times10^{3}~\mathrm{yr}^{-1}$. This results in $1.9\times10^{-3}\pm9.5\times10^{-5}$ ISOs within the inner solar system. The flux of $\geq200~\mu\mathrm{m}$ particles into the OC is $3.2\times10^{19}\pm1.5\times10^{18}~\mathrm{yr}^{-1}$, resulting in a flux at Earth of $8.5\pm4.3\times10^{-1}~\mathrm{yr}^{-1}$.

The maximum flux of $\geq100$~m ISOs into the OC is $3.4\times10^{5}\pm1.3\times10^{4}~\mathrm{yr}^{-1}$ at $t=1.6\times10^{5}~\mathrm{yr}$, resulting in a population of $3.5\times10^{-3}\pm1.4\times10^{-4}$ ISOs within the inner solar system at that time. Simultaneously, there is a flux of $6.0\times10^{19}\pm2.3\times10^{18}~\mathrm{yr}^{-1}\,\geq200~\mu\mathrm{m}$ particles into the OC, resulting in a flux at Earth of $1.6\times10^{1}\pm5.9\times10^{-1}~\mathrm{yr}^{-1}$.

The values above assume $t_{\mathrm{crit}}=10~\mathrm{Myr}$, see Tables within Appendix~\ref{append:tcrit3} for flux values with the assumption $t_{\mathrm{crit}}=3~\mathrm{Myr}$.

\subsection{HD 197481 - AU Microscopii}

\textit{HD 197481}, also known as \textit{AU Microscopii}, is a $23\pm3$ Myr old star \citep{mamajek_bell_2014} with an observed IR excess of $3.90\times10^{-4}$ \citep{matthews_et_al_2015}. From Equation~\ref{IRage}, we assume $t_{IR}=44~$Myr. This star system delivered 0.04\% of the simulated particles to the solar system. The 3,782 particles that reach us were typically ejected at $\widetilde{t}_\mathrm{eject} = -1.1\times10^{7}~\mathrm{yr}$ (95\%: $-2.0\times10^{7}- -2.7\times10^{6}$ yr) with $\widetilde{v}_\mathrm{eject} = 1.2~\mathrm{km~s^{-1}}$ (95\%: $0.6-4.4~\mathrm{km~s^{-1}}$; first panel in Figure~\ref{fig:heatplots:HD-197481}). The particles typically arrived at the solar system at $\widetilde{t}_\mathrm{arr} = 2.5\times10^{4}~\mathrm{yr}$ (95\%: $-1.1\times10^{6}-1.1\times10^{6}$; second panel in Figure~\ref{fig:heatplots:HD-197481}). The median velocity at arrival was $\widetilde{v}_\infty \pm \sigma_{v_\infty} = 21.8 \pm 1.1~\mathrm{km~s^{-1}}$ (95\%: $19.7-24.2~\mathrm{km~s^{-1}}$; second panel in Figure~\ref{fig:heatplots:HD-197481}), corresponding to a heliocentric velocity at 1~au of $\widetilde{v}_\mathrm{hel,1au} = 47.5~\mathrm{km~s^{-1}}$ (95\%: $46.5-48.6~\mathrm{km~s^{-1}}$). Assuming $q=1~\mathrm{au}$, the implied eccentricity is $\widetilde{e}_{q=1au} = 1.54$ (95\%: $1.44-1.66$). The distribution of both $v_{\infty}$ and $e$, weighted by their contribution to the flux at Earth for \textit{* HD 197481}, can be seen in Figure~\ref{fig:velE:HD-197481}. The average direction of arrival is $(\alpha,\delta) = (267.2^\circ\,\pm\,1.7^\circ,\ 30.7^\circ\,\pm\,1.8^\circ)$ (Figure~\ref{fig:radiant:HD-197481}).

These particles typically traveled for $1.1\times10^{7}$~yr (95\%: $2.8\times10^{6}-2.0\times10^{7}$~yr) in the Galactic potential (first panel of Figure~\ref{fig:heatplots:HD-197481}). The typical distance traveled relative to their origin system is 11.0~pc (95\%: $9.3-30.9~$pc). During their transit, the typical velocity relative to the circular velocity of the Sun is $3.5~\mathrm{km~s}^{-1}$ (95\%: $2.2-5.5~\mathrm{km~s}^{-1}$).

The least demanding trajectory in terms of magnetic deflection, ISM drag, and grain-grain sputtering would have allowed a particle of $d_{\min}=28~\mu\mathrm{m}$ to survive its passage through the ISM. The most demanding delivery trajectory requires $d_{\mathrm{survive}}=4.2\times10^{2}~\mu\mathrm{m}$. 95\% of the particles had critical sizes within $33-72~\mu\mathrm{m}$, all of which are set by magnetic forces.

From \textit{* HD 197481}, the current flux of ISOs $\geq100$~m into the OC is $8.4\times10^{5}\pm3.7\times10^{4}~\mathrm{yr}^{-1}$. This results in $0.0\pm1.5\times10^{-3}$ ISOs within the inner solar system. The flux of $\geq200~\mu\mathrm{m}$ particles into the OC is $1.5\times10^{20}\pm6.6\times10^{18}~\mathrm{yr}^{-1}$, resulting in a flux at Earth of $1.2\times10^{2}\pm5.8~\mathrm{yr}^{-1}$.

The maximum flux of $\geq100$~m ISOs into the OC is $1.0\times10^{6}\pm3.8\times10^{4}~\mathrm{yr}^{-1}$ at $t=7.0\times10^{4}~\mathrm{yr}$, resulting in a population of $0.0\pm1.5\times10^{-3}$ ISOs within the inner solar system at that time. Simultaneously, there is a flux of $1.8\times10^{20}\pm7.0\times10^{18}~\mathrm{yr}^{-1}\,\geq200~\mu\mathrm{m}$ particles into the OC, resulting in a flux at Earth of $1.5\times10^{2}\pm5.7~\mathrm{yr}^{-1}$.

The values above assume $t_{\mathrm{crit}}=10~\mathrm{Myr}$, see Tables within Appendix~\ref{append:tcrit3} for flux values with the assumption $t_{\mathrm{crit}}=3~\mathrm{Myr}$.

\subsection{* bet Leo}

\textit{* bet Leo} is a $100$ Myr old star \citep{vican_age_2012} with an observed IR excess of $1.90\times10^{-5}$ \citep{di_folco_et_al_2004}. From Equation~\ref{IRage}, we assume $t_{IR}=199~$Myr. This star system delivered 0.02\% of the simulated particles to the solar system. The 1,625 particles that reach us were typically ejected at $\widetilde{t}_\mathrm{eject} = -3.3\times10^{7}~\mathrm{yr}$ (95\%: $-9.5\times10^{7}- -4.3\times10^{6}$ yr) with $\widetilde{v}_\mathrm{eject} = 1.0~\mathrm{km~s^{-1}}$ (95\%: $0.4-3.3~\mathrm{km~s^{-1}}$; first panel in Figure~\ref{fig:heatplots:star-bet-Leo}). The particles typically arrived at the solar system at $\widetilde{t}_\mathrm{arr} = 7.1\times10^{5}~\mathrm{yr}$ (95\%: $-8.1\times10^{5}-2.2\times10^{6}$; second panel in Figure~\ref{fig:heatplots:star-bet-Leo}). The median velocity at arrival was $\widetilde{v}_\infty \pm \sigma_{v_\infty} = 26.0 \pm 0.9~\mathrm{km~s^{-1}}$ (95\%: $24.7-28.2~\mathrm{km~s^{-1}}$; second panel in Figure~\ref{fig:heatplots:star-bet-Leo}), corresponding to a heliocentric velocity at 1~au of $\widetilde{v}_\mathrm{hel,1au} = 49.5~\mathrm{km~s^{-1}}$ (95\%: $48.8-50.7~\mathrm{km~s^{-1}}$). Assuming $q=1~\mathrm{au}$, the implied eccentricity is $\widetilde{e}_{q=1au} = 1.76$ (95\%: $1.69-1.89$). The distribution of both $v_{\infty}$ and $e$, weighted by their contribution to the flux at Earth for \textit{* bet Leo}, can be seen in Figure~\ref{fig:velE:star-bet-Leo}. The average direction of arrival is $(\alpha,\delta) = (269.3^\circ\,\pm\,1.7^\circ,\ 13.1^\circ\,\pm\,0.7^\circ)$ (Figure~\ref{fig:radiant:star-bet-Leo}).

These particles typically traveled for $3.4\times10^{7}$~yr (95\%: $4.3\times10^{6}-9.6\times10^{7}$~yr) in the Galactic potential (first panel of Figure~\ref{fig:heatplots:star-bet-Leo}). The typical distance traveled relative to their origin system is 31.4~pc (95\%: $11.0-92.9~$pc). During their transit, the typical velocity relative to the circular velocity of the Sun is $3.2~\mathrm{km~s}^{-1}$ (95\%: $0.2-12.3~\mathrm{km~s}^{-1}$).

The least demanding trajectory in terms of magnetic deflection, ISM drag, and grain-grain sputtering would have allowed a particle of $d_{\min}=35~\mu\mathrm{m}$ to survive its passage through the ISM. The most demanding delivery trajectory requires $d_{\mathrm{survive}}=3.7\times10^{3}~\mu\mathrm{m}$ (or $3.7~\mathrm{mm}$). 95\% of the particles had critical sizes within $38-2.4\times10^{2}~\mu\mathrm{m}$, all of which are set by magnetic forces.

From \textit{* bet Leo}, the current flux of ISOs $\geq100$~m into the OC is $9.7\times10^{3}\pm3.8\times10^{2}~\mathrm{yr}^{-1}$. This results in $2.3\times10^{-4}\pm1.1\times10^{-5}$ ISOs within the inner solar system. The flux of $\geq200~\mu\mathrm{m}$ particles into the OC is $1.7\times10^{18}\pm6.6\times10^{16}~\mathrm{yr}^{-1}$, resulting in a flux at Earth of $1.1\pm4.2\times10^{-2}~\mathrm{yr}^{-1}$.

The maximum flux of $\geq100$~m ISOs into the OC is $2.5\times10^{4}\pm1.2\times10^{3}~\mathrm{yr}^{-1}$ at $t=1.3\times10^{6}~\mathrm{yr}$, resulting in a population of $6.2\times10^{-4}\pm3.8\times10^{-5}$ ISOs within the inner solar system at that time. Simultaneously, there is a flux of $4.5\times10^{18}\pm2.2\times10^{17}~\mathrm{yr}^{-1}\,\geq200~\mu\mathrm{m}$ particles into the OC, resulting in a flux at Earth of $3.0\pm1.4\times10^{-1}~\mathrm{yr}^{-1}$.

The values above assume $t_{\mathrm{crit}}=10~\mathrm{Myr}$, see Tables within Appendix~\ref{append:tcrit3} for flux values with the assumption $t_{\mathrm{crit}}=3~\mathrm{Myr}$.

\subsection{HD 166}

\textit{HD 166} is a $160$ Myr old star \citep{vican_age_2012} with an observed IR excess of $5.90\times10^{-5}$ \citep{trilling_et_al_2008}. From Equation~\ref{IRage}, we assume $t_{IR}=113~$Myr. This star system delivered 0.01\% of the simulated particles to the solar system. The 1,499 particles that reach us were typically ejected at $\widetilde{t}_\mathrm{eject} = -4.1\times10^{7}~\mathrm{yr}$ (95\%: $-9.7\times10^{7}- -5.5\times10^{6}$ yr) with $\widetilde{v}_\mathrm{eject} = 0.9~\mathrm{km~s^{-1}}$ (95\%: $0.2-3.2~\mathrm{km~s^{-1}}$; first panel in Figure~\ref{fig:heatplots:HD-166}). The particles typically arrived at the solar system at $\widetilde{t}_\mathrm{arr} = -5.0\times10^{5}~\mathrm{yr}$ (95\%: $-1.4\times10^{6}-7.0\times10^{5}$; second panel in Figure~\ref{fig:heatplots:HD-166}). The median velocity at arrival was $\widetilde{v}_\infty \pm \sigma_{v_\infty} = 28.4 \pm 0.7~\mathrm{km~s^{-1}}$ (95\%: $27.0-29.9~\mathrm{km~s^{-1}}$; second panel in Figure~\ref{fig:heatplots:HD-166}), corresponding to a heliocentric velocity at 1~au of $\widetilde{v}_\mathrm{hel,1au} = 50.8~\mathrm{km~s^{-1}}$ (95\%: $50.0-51.6~\mathrm{km~s^{-1}}$). Assuming $q=1~\mathrm{au}$, the implied eccentricity is $\widetilde{e}_{q=1au} = 1.91$ (95\%: $1.82-2.01$). The distribution of both $v_{\infty}$ and $e$, weighted by their contribution to the flux at Earth for \textit{* HD 166}, can be seen in Figure~\ref{fig:velE:HD-166}. The average direction of arrival is $(\alpha,\delta) = (271.8^\circ\,\pm\,1.6^\circ,\ 28.0^\circ\,\pm\,1.0^\circ)$ (Figure~\ref{fig:radiant:HD-166}).

These particles typically traveled for $4.0\times10^{7}$~yr (95\%: $5.6\times10^{6}-9.7\times10^{7}$~yr) in the Galactic potential (first panel of Figure~\ref{fig:heatplots:HD-166}). The typical distance traveled relative to their origin system is 28.7~pc (95\%: $13.5-70.7~$pc). During their transit, the typical velocity relative to the circular velocity of the Sun is $7.4~\mathrm{km~s}^{-1}$ (95\%: $0.4-12.4~\mathrm{km~s}^{-1}$).

The least demanding trajectory in terms of magnetic deflection, ISM drag, and grain-grain sputtering would have allowed a particle of $d_{\min}=22~\mu\mathrm{m}$ to survive its passage through the ISM. The most demanding delivery trajectory requires $d_{\mathrm{survive}}=1.7\times10^{3}~\mu\mathrm{m}$ (or $1.7~\mathrm{mm}$). 95\% of the particles had critical sizes within $26-1.8\times10^{2}~\mu\mathrm{m}$, all of which are set by magnetic forces.

From \textit{* HD 166}, the current flux of ISOs $\geq100$~m into the OC is $4.0\times10^{5}\pm1.5\times10^{5}~\mathrm{yr}^{-1}$. This results in $9.0\times10^{-3}\pm4.5\times10^{-3}$ ISOs within the inner solar system. The flux of $\geq200~\mu\mathrm{m}$ particles into the OC is $7.2\times10^{19}\pm2.8\times10^{19}~\mathrm{yr}^{-1}$, resulting in a flux at Earth of $4.2\times10^{1}\pm1.7\times10^{1}~\mathrm{yr}^{-1}$.

The maximum flux of $\geq100$~m ISOs into the OC is $1.3\times10^{7}\pm1.3\times10^{6}~\mathrm{yr}^{-1}$ at $t=-6.6\times10^{5}~\mathrm{yr}$, resulting in a population of $0.3\pm0.0$ ISOs within the inner solar system at that time. Simultaneously, there is a flux of $2.2\times10^{21}\pm2.3\times10^{20}~\mathrm{yr}^{-1}\,\geq200~\mu\mathrm{m}$ particles into the OC, resulting in a flux at Earth of $1.3\times10^{3}\pm1.4\times10^{2}~\mathrm{yr}^{-1}$.

The values above assume $t_{\mathrm{crit}}=10~\mathrm{Myr}$, see Tables within Appendix~\ref{append:tcrit3} for flux values with the assumption $t_{\mathrm{crit}}=3~\mathrm{Myr}$.

\subsection{HD 38858}

\textit{HD 38858} is a $4.3\pm1.0$ Gyr old star \citep{desgrange_et_al_2023} with an observed IR excess of $8.00\times10^{-5}$ \citep{di_folco_et_al_2004}. From Equation~\ref{IRage}, we assume $t_{IR}=97~$Myr. This star system delivered 0.04\% of the simulated particles to the solar system. The 4,061 particles that reach us were typically ejected at $\widetilde{t}_\mathrm{eject} = -2.8\times10^{7}~\mathrm{yr}$ (95\%: $-9.2\times10^{7}- -3.9\times10^{6}$ yr) with $\widetilde{v}_\mathrm{eject} = 0.6~\mathrm{km~s^{-1}}$ (95\%: $0.2-3.0~\mathrm{km~s^{-1}}$; first panel in Figure~\ref{fig:heatplots:HD-38858}). The particles typically arrived at the solar system at $\widetilde{t}_\mathrm{arr} = -3.4\times10^{5}~\mathrm{yr}$ (95\%: $-9.9\times10^{5}-2.5\times10^{5}$; second panel in Figure~\ref{fig:heatplots:HD-38858}). The median velocity at arrival was $\widetilde{v}_\infty \pm \sigma_{v_\infty} = 35.7 \pm 0.7~\mathrm{km~s^{-1}}$ (95\%: $34.4-37.3~\mathrm{km~s^{-1}}$; second panel in Figure~\ref{fig:heatplots:HD-38858}), corresponding to a heliocentric velocity at 1~au of $\widetilde{v}_\mathrm{hel,1au} = 55.2~\mathrm{km~s^{-1}}$ (95\%: $54.4-56.2~\mathrm{km~s^{-1}}$). Assuming $q=1~\mathrm{au}$, the implied eccentricity is $\widetilde{e}_{q=1au} = 2.44$ (95\%: $2.33-2.57$). The distribution of both $v_{\infty}$ and $e$, weighted by their contribution to the flux at Earth for \textit{* HD 38858}, can be seen in Figure~\ref{fig:velE:HD-38858}. The average direction of arrival is $(\alpha,\delta) = (275.7^\circ\,\pm\,0.7^\circ,\ 32.2^\circ\,\pm\,0.9^\circ)$ (Figure~\ref{fig:radiant:HD-38858}).

These particles typically traveled for $2.8\times10^{7}$~yr (95\%: $3.5\times10^{6}-9.2\times10^{7}$~yr) in the Galactic potential (first panel of Figure~\ref{fig:heatplots:HD-38858}). The typical distance traveled relative to their origin system is 10.0~pc (95\%: $7.1-52.8~$pc). During their transit, the typical velocity relative to the circular velocity of the Sun is $11.5~\mathrm{km~s}^{-1}$ (95\%: $0.7-17.7~\mathrm{km~s}^{-1}$).

The least demanding trajectory in terms of magnetic deflection, ISM drag, and grain-grain sputtering would have allowed a particle of $d_{\min}=13~\mu\mathrm{m}$ to survive its passage through the ISM. The most demanding delivery trajectory requires $d_{\mathrm{survive}}=2.4\times10^{3}~\mu\mathrm{m}$ (or $2.4~\mathrm{mm}$). 95\% of the particles had critical sizes within $14-1.1\times10^{2}~\mu\mathrm{m}$, all of which are set by magnetic forces.

From \textit{* HD 38858}, the current flux of ISOs $\geq100$~m into the OC is $4.5\times10^{6}\pm1.5\times10^{6}~\mathrm{yr}^{-1}$. This results in $0.1\pm0.0$ ISOs within the inner solar system. The flux of $\geq200~\mu\mathrm{m}$ particles into the OC is $7.9\times10^{20}\pm2.6\times10^{20}~\mathrm{yr}^{-1}$, resulting in a flux at Earth of $3.4\times10^{2}\pm1.1\times10^{2}~\mathrm{yr}^{-1}$.

The maximum flux of $\geq100$~m ISOs into the OC is $8.5\times10^{7}\pm7.2\times10^{6}~\mathrm{yr}^{-1}$ at $t=-3.2\times10^{5}~\mathrm{yr}$, resulting in a population of $1.5\pm0.1$ ISOs within the inner solar system at that time. Simultaneously, there is a flux of $1.5\times10^{22}\pm1.2\times10^{21}~\mathrm{yr}^{-1}\,\geq200~\mu\mathrm{m}$ particles into the OC, resulting in a flux at Earth of $6.5\times10^{3}\pm5.3\times10^{2}~\mathrm{yr}^{-1}$.

The values above assume $t_{\mathrm{crit}}=10~\mathrm{Myr}$, see Tables within Appendix~\ref{append:tcrit3} for flux values with the assumption $t_{\mathrm{crit}}=3~\mathrm{Myr}$.

\subsection{HD 207129}

\textit{HD 207129} is a $600-8000$ Myr old star \citep{krist_et_al_2010} with an observed IR excess of $1.58\times10^{-4}$ \citep{decin_et_al_2003}. From Equation~\ref{IRage}, we assume $t_{IR}=69~$Myr. This star system delivered 0.01\% of the simulated particles to the solar system. The 858 particles that reach us were typically ejected at $\widetilde{t}_\mathrm{eject} = -2.7\times10^{7}~\mathrm{yr}$ (95\%: $-9.5\times10^{7}- -5.0\times10^{6}$ yr) with $\widetilde{v}_\mathrm{eject} = 1.4~\mathrm{km~s^{-1}}$ (95\%: $0.8-4.2~\mathrm{km~s^{-1}}$; first panel in Figure~\ref{fig:heatplots:HD-207129}). The particles typically arrived at the solar system at $\widetilde{t}_\mathrm{arr} = 4.2\times10^{5}~\mathrm{yr}$ (95\%: $-3.0\times10^{6}-4.7\times10^{6}$; second panel in Figure~\ref{fig:heatplots:HD-207129}). The median velocity at arrival was $\widetilde{v}_\infty \pm \sigma_{v_\infty} = 25.8 \pm 1.3~\mathrm{km~s^{-1}}$ (95\%: $23.9-28.7~\mathrm{km~s^{-1}}$; second panel in Figure~\ref{fig:heatplots:HD-207129}), corresponding to a heliocentric velocity at 1~au of $\widetilde{v}_\mathrm{hel,1au} = 49.4~\mathrm{km~s^{-1}}$ (95\%: $48.4-51.0~\mathrm{km~s^{-1}}$). Assuming $q=1~\mathrm{au}$, the implied eccentricity is $\widetilde{e}_{q=1au} = 1.75$ (95\%: $1.64-1.93$). The distribution of both $v_{\infty}$ and $e$, weighted by their contribution to the flux at Earth for \textit{* HD 207129}, can be seen in Figure~\ref{fig:velE:HD-207129}. The average direction of arrival is $(\alpha,\delta) = (297.0^\circ\,\pm\,1.9^\circ,\ 20.5^\circ\,\pm\,2.5^\circ)$ (Figure~\ref{fig:radiant:HD-207129}).

These particles typically traveled for $2.7\times10^{7}$~yr (95\%: $5.2\times10^{6}-9.8\times10^{7}$~yr) in the Galactic potential (first panel of Figure~\ref{fig:heatplots:HD-207129}). The typical distance traveled relative to their origin system is 39.2~pc (95\%: $14.9-158.2~$pc). During their transit, the typical velocity relative to the circular velocity of the Sun is $7.3~\mathrm{km~s}^{-1}$ (95\%: $1.1-11.4~\mathrm{km~s}^{-1}$).

The least demanding trajectory in terms of magnetic deflection, ISM drag, and grain-grain sputtering would have allowed a particle of $d_{\min}=29~\mu\mathrm{m}$ to survive its passage through the ISM. The most demanding delivery trajectory requires $d_{\mathrm{survive}}=1.9\times10^{3}~\mu\mathrm{m}$ (or $1.9~\mathrm{mm}$). 95\% of the particles had critical sizes within $29-1.8\times10^{2}~\mu\mathrm{m}$, all of which are set by magnetic forces.

From \textit{* HD 207129}, the current flux of ISOs $\geq100$~m into the OC is $3.1\times10^{6}\pm1.3\times10^{6}~\mathrm{yr}^{-1}$. This results in $0.1\pm0.0$ ISOs within the inner solar system. The flux of $\geq200~\mu\mathrm{m}$ particles into the OC is $5.6\times10^{20}\pm2.4\times10^{20}~\mathrm{yr}^{-1}$, resulting in a flux at Earth of $3.6\times10^{2}\pm1.6\times10^{2}~\mathrm{yr}^{-1}$.

The maximum flux of $\geq100$~m ISOs into the OC is $1.0\times10^{7}\pm6.9\times10^{5}~\mathrm{yr}^{-1}$ at $t=3.6\times10^{6}~\mathrm{yr}$, resulting in a population of $0.3\pm0.0$ ISOs within the inner solar system at that time. Simultaneously, there is a flux of $1.8\times10^{21}\pm1.2\times10^{20}~\mathrm{yr}^{-1}\,\geq200~\mu\mathrm{m}$ particles into the OC, resulting in a flux at Earth of $1.3\times10^{3}\pm8.7\times10^{1}~\mathrm{yr}^{-1}$.

The values above assume $t_{\mathrm{crit}}=10~\mathrm{Myr}$, see Tables within Appendix~\ref{append:tcrit3} for flux values with the assumption $t_{\mathrm{crit}}=3~\mathrm{Myr}$.

\subsection{* b Her}

\textit{* b Her}, or \textit{99 Herculis}, is a $6-10$ Gyr old star with an observed IR excess of $1.40\times10^{-5}$ \citep{kennedy_et_al_2012}. From Equation~\ref{IRage}, we assume $t_{IR}=231~$Myr. This star system delivered 0.01\% of the simulated particles to the solar system. The 1,089 particles that reach us were typically ejected at $\widetilde{t}_\mathrm{eject} = -3.6\times10^{7}~\mathrm{yr}$ (95\%: $-9.7\times10^{7}- -6.0\times10^{6}$ yr) with $\widetilde{v}_\mathrm{eject} = 0.7~\mathrm{km~s^{-1}}$ (95\%: $0.1-3.6~\mathrm{km~s^{-1}}$; first panel in Figure~\ref{fig:heatplots:star-b-Her}). The particles typically arrived at the solar system at $\widetilde{t}_\mathrm{arr} = -5.9\times10^{5}~\mathrm{yr}$ (95\%: $-3.5\times10^{6}-1.8\times10^{6}$; second panel in Figure~\ref{fig:heatplots:star-b-Her}). The median velocity at arrival was $\widetilde{v}_\infty \pm \sigma_{v_\infty} = 8.8 \pm 0.6~\mathrm{km~s^{-1}}$ (95\%: $7.6-10.1~\mathrm{km~s^{-1}}$; second panel in Figure~\ref{fig:heatplots:star-b-Her}), corresponding to a heliocentric velocity at 1~au of $\widetilde{v}_\mathrm{hel,1au} = 43.0~\mathrm{km~s^{-1}}$ (95\%: $42.8-43.3~\mathrm{km~s^{-1}}$). Assuming $q=1~\mathrm{au}$, the implied eccentricity is $\widetilde{e}_{q=1au} = 1.09$ (95\%: $1.07-1.12$). The distribution of both $v_{\infty}$ and $e$, weighted by their contribution to the flux at Earth for \textit{* b Her}, can be seen in Figure~\ref{fig:velE:star-b-Her}. The average direction of arrival is $(\alpha,\delta) = (340.8^\circ\,\pm\,4.5^\circ,\ -21.5^\circ\,\pm\,3.4^\circ)$ (Figure~\ref{fig:radiant:star-b-Her}).

These particles typically traveled for $3.5\times10^{7}$~yr (95\%: $5.7\times10^{6}-9.6\times10^{7}$~yr) in the Galactic potential (first panel of Figure~\ref{fig:heatplots:star-b-Her}). The typical distance traveled relative to their origin system is 18.9~pc (95\%: $15.7-61.2~$pc). During their transit, the typical velocity relative to the circular velocity of the Sun is $6.1~\mathrm{km~s}^{-1}$ (95\%: $3.9-8.8~\mathrm{km~s}^{-1}$).

The least demanding trajectory in terms of magnetic deflection, ISM drag, and grain-grain sputtering would have allowed a particle of $d_{\min}=29~\mu\mathrm{m}$ to survive its passage through the ISM. The most demanding delivery trajectory requires $d_{\mathrm{survive}}=1.2\times10^{2}~\mu\mathrm{m}$. 95\% of the particles had critical sizes within $30-77~\mu\mathrm{m}$, all of which are set by magnetic forces.

From \textit{* b Her}, the current flux of ISOs $\geq100$~m into the OC is $3.4\times10^{3}\pm4.2\times10^{2}~\mathrm{yr}^{-1}$. This results in $2.4\times10^{-4}\pm1.1\times10^{-4}$ ISOs within the inner solar system. The flux of $\geq200~\mu\mathrm{m}$ particles into the OC is $5.9\times10^{17}\pm7.4\times10^{16}~\mathrm{yr}^{-1}$, resulting in a flux at Earth of $2.6\pm3.4\times10^{-1}~\mathrm{yr}^{-1}$.

The maximum flux of $\geq100$~m ISOs into the OC is $2.3\times10^{4}\pm8.8\times10^{2}~\mathrm{yr}^{-1}$ at $t=-6.2\times10^{5}~\mathrm{yr}$, resulting in a population of $1.7\times10^{-3}\pm2.2\times10^{-4}$ ISOs within the inner solar system at that time. Simultaneously, there is a flux of $4.0\times10^{18}\pm1.6\times10^{17}~\mathrm{yr}^{-1}\,\geq200~\mu\mathrm{m}$ particles into the OC, resulting in a flux at Earth of $1.8\times10^{1}\pm7.2\times10^{-1}~\mathrm{yr}^{-1}$.

The values above assume $t_{\mathrm{crit}}=10~\mathrm{Myr}$, see Tables within Appendix~\ref{append:tcrit3} for flux values with the assumption $t_{\mathrm{crit}}=3~\mathrm{Myr}$.

\subsection{HD 23484}

\textit{HD 23484} is a $760-930$ Myr old star \citep{eiroa_et_al_2013} with an observed IR excess of $9.50\times10^{-5}$ \citep{schuppler_et_al_2014}. From Equation~\ref{IRage}, we assume $t_{IR}=89~$Myr. This star system delivered 0.02\% of the simulated particles to the solar system. The 1,665 particles that reach us were typically ejected at $\widetilde{t}_\mathrm{eject} = -2.9\times10^{7}~\mathrm{yr}$ (95\%: $-9.2\times10^{7}- -4.7\times10^{6}$ yr) with $\widetilde{v}_\mathrm{eject} = 0.9~\mathrm{km~s^{-1}}$ (95\%: $0.5-3.5~\mathrm{km~s^{-1}}$; first panel in Figure~\ref{fig:heatplots:HD-23484}). The particles typically arrived at the solar system at $\widetilde{t}_\mathrm{arr} = -5.8\times10^{5}~\mathrm{yr}$ (95\%: $-1.4\times10^{6}-2.4\times10^{5}$; second panel in Figure~\ref{fig:heatplots:HD-23484}). The median velocity at arrival was $\widetilde{v}_\infty \pm \sigma_{v_\infty} = 40.9 \pm 0.8~\mathrm{km~s^{-1}}$ (95\%: $39.0-42.3~\mathrm{km~s^{-1}}$; second panel in Figure~\ref{fig:heatplots:HD-23484}), corresponding to a heliocentric velocity at 1~au of $\widetilde{v}_\mathrm{hel,1au} = 58.7~\mathrm{km~s^{-1}}$ (95\%: $57.4-59.7~\mathrm{km~s^{-1}}$). Assuming $q=1~\mathrm{au}$, the implied eccentricity is $\widetilde{e}_{q=1au} = 2.89$ (95\%: $2.71-3.01$). The distribution of both $v_{\infty}$ and $e$, weighted by their contribution to the flux at Earth for \textit{* HD 23484}, can be seen in Figure~\ref{fig:velE:HD-23484}. The average direction of arrival is $(\alpha,\delta) = (259.7^\circ\,\pm\,0.7^\circ,\ 0.7^\circ\,\pm\,0.9^\circ)$ (Figure~\ref{fig:radiant:HD-23484}).

These particles typically traveled for $2.8\times10^{7}$~yr (95\%: $4.5\times10^{6}-9.2\times10^{7}$~yr) in the Galactic potential (first panel of Figure~\ref{fig:heatplots:HD-23484}). The typical distance traveled relative to their origin system is 22.8~pc (95\%: $10.9-75.2~$pc). During their transit, the typical velocity relative to the circular velocity of the Sun is $10.1~\mathrm{km~s}^{-1}$ (95\%: $1.9-19.0~\mathrm{km~s}^{-1}$).

The least demanding trajectory in terms of magnetic deflection, ISM drag, and grain-grain sputtering would have allowed a particle of $d_{\min}=23~\mu\mathrm{m}$ to survive its passage through the ISM. The most demanding delivery trajectory requires $d_{\mathrm{survive}}=91~\mu\mathrm{m}$. 95\% of the particles had critical sizes within $25-61~\mu\mathrm{m}$, all of which are set by magnetic forces.

From \textit{* HD 23484}, the current flux of ISOs $\geq100$~m into the OC is $2.4\times10^{5}\pm7.9\times10^{4}~\mathrm{yr}^{-1}$. This results in $3.8\times10^{-3}\pm1.3\times10^{-3}$ ISOs within the inner solar system. The flux of $\geq200~\mu\mathrm{m}$ particles into the OC is $4.3\times10^{19}\pm1.4\times10^{19}~\mathrm{yr}^{-1}$, resulting in a flux at Earth of $1.6\times10^{1}\pm5.2~\mathrm{yr}^{-1}$.

The maximum flux of $\geq100$~m ISOs into the OC is $1.8\times10^{7}\pm2.9\times10^{6}~\mathrm{yr}^{-1}$ at $t=-8.8\times10^{5}~\mathrm{yr}$, resulting in a population of $0.3\pm0.1$ ISOs within the inner solar system at that time. Simultaneously, there is a flux of $3.2\times10^{21}\pm5.1\times10^{20}~\mathrm{yr}^{-1}\,\geq200~\mu\mathrm{m}$ particles into the OC, resulting in a flux at Earth of $1.2\times10^{3}\pm1.9\times10^{2}~\mathrm{yr}^{-1}$.

The values above assume $t_{\mathrm{crit}}=10~\mathrm{Myr}$, see Tables within Appendix~\ref{append:tcrit3} for flux values with the assumption $t_{\mathrm{crit}}=3~\mathrm{Myr}$.

\subsection{* q01 Eri}

\textit{* q01 Eri} is a $960-1740$ Myr old star \citep{eiroa_et_al_2013} with an observed IR excess of $5.37\times10^{-4}$ \citep{decin_et_al_2003}. From Equation~\ref{IRage}, we assume $t_{IR}=37~$Myr. This star system delivered 0.02\% of the simulated particles to the solar system. The 2,091 particles that reach us were typically ejected at $\widetilde{t}_\mathrm{eject} = -2.7\times10^{7}~\mathrm{yr}$ (95\%: $-9.6\times10^{7}- -3.9\times10^{6}$ yr) with $\widetilde{v}_\mathrm{eject} = 0.7~\mathrm{km~s^{-1}}$ (95\%: $0.1-3.5~\mathrm{km~s^{-1}}$; first panel in Figure~\ref{fig:heatplots:star-q01-Eri}). The particles typically arrived at the solar system at $\widetilde{t}_\mathrm{arr} = -6.8\times10^{5}~\mathrm{yr}$ (95\%: $-1.3\times10^{6}-7.4\times10^{4}$; second panel in Figure~\ref{fig:heatplots:star-q01-Eri}). The median velocity at arrival was $\widetilde{v}_\infty \pm \sigma_{v_\infty} = 32.2 \pm 0.7~\mathrm{km~s^{-1}}$ (95\%: $30.8-33.8~\mathrm{km~s^{-1}}$; second panel in Figure~\ref{fig:heatplots:star-q01-Eri}), corresponding to a heliocentric velocity at 1~au of $\widetilde{v}_\mathrm{hel,1au} = 53.0~\mathrm{km~s^{-1}}$ (95\%: $52.2-54.0~\mathrm{km~s^{-1}}$). Assuming $q=1~\mathrm{au}$, the implied eccentricity is $\widetilde{e}_{q=1au} = 2.17$ (95\%: $2.07-2.29$). The distribution of both $v_{\infty}$ and $e$, weighted by their contribution to the flux at Earth for \textit{* q01 Eri}, can be seen in Figure~\ref{fig:velE:star-q01-Eri}. The average direction of arrival is $(\alpha,\delta) = (262.5^\circ\,\pm\,2.4^\circ,\ 58.4^\circ\,\pm\,0.4^\circ)$ (Figure~\ref{fig:radiant:star-q01-Eri}).

These particles typically traveled for $2.6\times10^{7}$~yr (95\%: $3.4\times10^{6}-9.5\times10^{7}$~yr) in the Galactic potential (first panel of Figure~\ref{fig:heatplots:star-q01-Eri}). The typical distance traveled relative to their origin system is 3240.6~pc (95\%: $358.0-1.9\times10^{4}~$pc). During their transit, the typical velocity relative to the circular velocity of the Sun is $13.8~\mathrm{km~s}^{-1}$ (95\%: $1.1-16.9~\mathrm{km~s}^{-1}$).

The least demanding trajectory in terms of magnetic deflection, ISM drag, and grain-grain sputtering would have allowed a particle of $d_{\min}=57~\mu\mathrm{m}$ to survive its passage through the ISM. The most demanding delivery trajectory requires $d_{\mathrm{survive}}=2.4\times10^{4}~\mu\mathrm{m}$ (or $24~\mathrm{mm}$). 95\% of the particles had critical sizes within $1.0\times10^{2}-2.2\times10^{3}~\mu\mathrm{m}$, all of which are set by magnetic forces.

From \textit{* q01 Eri}, the current flux of ISOs $\geq100$~m into the OC is $5.8\times10^{6}\pm1.4\times10^{6}~\mathrm{yr}^{-1}$. This results in $0.1\pm0.0$ ISOs within the inner solar system. The flux of $\geq200~\mu\mathrm{m}$ particles into the OC is $1.0\times10^{21}\pm2.6\times10^{20}~\mathrm{yr}^{-1}$, resulting in a flux at Earth of $5.0\times10^{2}\pm1.2\times10^{2}~\mathrm{yr}^{-1}$.

The maximum flux of $\geq100$~m ISOs into the OC is $2.8\times10^{8}\pm4.9\times10^{6}~\mathrm{yr}^{-1}$ at $t=-7.8\times10^{5}~\mathrm{yr}$, resulting in a population of $5.5\pm0.1$ ISOs within the inner solar system at that time. Simultaneously, there is a flux of $4.9\times10^{22}\pm8.8\times10^{20}~\mathrm{yr}^{-1}\,\geq200~\mu\mathrm{m}$ particles into the OC, resulting in a flux at Earth of $2.4\times10^{4}\pm4.7\times10^{2}~\mathrm{yr}^{-1}$.

The values above assume $t_{\mathrm{crit}}=10~\mathrm{Myr}$, see Tables within Appendix~\ref{append:tcrit3} for flux values with the assumption $t_{\mathrm{crit}}=3~\mathrm{Myr}$.

\subsection{* g Lup}

\textit{* g Lup} is a $300^{+700}_{-200}$ Myr old star with an observed IR excess of $9.00\times10^{-5}$ \citep{kalas_et_al_2006}. From Equation~\ref{IRage}, we assume $t_{IR}=91~$Myr. This star system delivered 0.01\% of the simulated particles to the solar system. The 1,484 particles that reach us were typically ejected at $\widetilde{t}_\mathrm{eject} = -3.6\times10^{7}~\mathrm{yr}$ (95\%: $-9.4\times10^{7}- -6.9\times10^{6}$ yr) with $\widetilde{v}_\mathrm{eject} = 0.8~\mathrm{km~s^{-1}}$ (95\%: $0.4-3.2~\mathrm{km~s^{-1}}$; first panel in Figure~\ref{fig:heatplots:star-g-Lup}). The particles typically arrived at the solar system at $\widetilde{t}_\mathrm{arr} = 2.3\times10^{5}~\mathrm{yr}$ (95\%: $-9.6\times10^{5}-1.4\times10^{6}$; second panel in Figure~\ref{fig:heatplots:star-g-Lup}). The median velocity at arrival was $\widetilde{v}_\infty \pm \sigma_{v_\infty} = 27.0 \pm 0.7~\mathrm{km~s^{-1}}$ (95\%: $25.7-28.5~\mathrm{km~s^{-1}}$; second panel in Figure~\ref{fig:heatplots:star-g-Lup}), corresponding to a heliocentric velocity at 1~au of $\widetilde{v}_\mathrm{hel,1au} = 50.1~\mathrm{km~s^{-1}}$ (95\%: $49.4-50.9~\mathrm{km~s^{-1}}$). Assuming $q=1~\mathrm{au}$, the implied eccentricity is $\widetilde{e}_{q=1au} = 1.83$ (95\%: $1.75-1.91$). The distribution of both $v_{\infty}$ and $e$, weighted by their contribution to the flux at Earth for \textit{* g Lup}, can be seen in Figure~\ref{fig:velE:star-g-Lup}. The average direction of arrival is $(\alpha,\delta) = (269.2^\circ\,\pm\,1.2^\circ,\ 22.2^\circ\,\pm\,1.3^\circ)$ (Figure~\ref{fig:radiant:star-g-Lup}).

These particles typically traveled for $3.6\times10^{7}$~yr (95\%: $7.2\times10^{6}-9.4\times10^{7}$~yr) in the Galactic potential (first panel of Figure~\ref{fig:heatplots:star-g-Lup}). The typical distance traveled relative to their origin system is 21.8~pc (95\%: $16.8-79.6~$pc). During their transit, the typical velocity relative to the circular velocity of the Sun is $3.3~\mathrm{km~s}^{-1}$ (95\%: $0.2-12.0~\mathrm{km~s}^{-1}$).

The least demanding trajectory in terms of magnetic deflection, ISM drag, and grain-grain sputtering would have allowed a particle of $d_{\min}=31~\mu\mathrm{m}$ to survive its passage through the ISM. The most demanding delivery trajectory requires $d_{\mathrm{survive}}=1.2\times10^{3}~\mu\mathrm{m}$ (or $1.2~\mathrm{mm}$). 95\% of the particles had critical sizes within $33-2.4\times10^{2}~\mu\mathrm{m}$, all of which are set by magnetic forces.

From \textit{* g Lup}, the current flux of ISOs $\geq100$~m into the OC is $6.8\times10^{6}\pm1.9\times10^{6}~\mathrm{yr}^{-1}$. This results in $0.2\pm0.1$ ISOs within the inner solar system. The flux of $\geq200~\mu\mathrm{m}$ particles into the OC is $1.2\times10^{21}\pm3.4\times10^{20}~\mathrm{yr}^{-1}$, resulting in a flux at Earth of $7.4\times10^{2}\pm2.0\times10^{2}~\mathrm{yr}^{-1}$.

The maximum flux of $\geq100$~m ISOs into the OC is $1.7\times10^{7}\pm2.9\times10^{6}~\mathrm{yr}^{-1}$ at $t=2.6\times10^{5}~\mathrm{yr}$, resulting in a population of $0.4\pm0.1$ ISOs within the inner solar system at that time. Simultaneously, there is a flux of $3.0\times10^{21}\pm5.3\times10^{20}~\mathrm{yr}^{-1}\,\geq200~\mu\mathrm{m}$ particles into the OC, resulting in a flux at Earth of $1.9\times10^{3}\pm3.3\times10^{2}~\mathrm{yr}^{-1}$.

The values above assume $t_{\mathrm{crit}}=10~\mathrm{Myr}$, see Tables within Appendix~\ref{append:tcrit3} for flux values with the assumption $t_{\mathrm{crit}}=3~\mathrm{Myr}$.

\subsection{HD 35650}

\textit{HD 35650} is a $50-200$ Myr old star with an observed IR excess of $1.75\times10^{-4}$ \citep{choquet_et_al_2016}. From Equation~\ref{IRage}, we assume $t_{IR}=65~$Myr. This star system delivered 0.07\% of the simulated particles to the solar system. The 7,657 particles that reach us were typically ejected at $\widetilde{t}_\mathrm{eject} = -1.4\times10^{7}~\mathrm{yr}$ (95\%: $-8.8\times10^{7}- -1.9\times10^{6}$ yr) with $\widetilde{v}_\mathrm{eject} = 0.5~\mathrm{km~s^{-1}}$ (95\%: $0.1-3.0~\mathrm{km~s^{-1}}$; first panel in Figure~\ref{fig:heatplots:HD-35650}). The particles typically arrived at the solar system at $\widetilde{t}_\mathrm{arr} = -5.4\times10^{5}~\mathrm{yr}$ (95\%: $-9.7\times10^{5}--1.1\times10^{5}$; second panel in Figure~\ref{fig:heatplots:HD-35650}). The median velocity at arrival was $\widetilde{v}_\infty \pm \sigma_{v_\infty} = 32.8 \pm 0.7~\mathrm{km~s^{-1}}$ (95\%: $31.3-34.4~\mathrm{km~s^{-1}}$; second panel in Figure~\ref{fig:heatplots:HD-35650}), corresponding to a heliocentric velocity at 1~au of $\widetilde{v}_\mathrm{hel,1au} = 53.4~\mathrm{km~s^{-1}}$ (95\%: $52.5-54.4~\mathrm{km~s^{-1}}$). Assuming $q=1~\mathrm{au}$, the implied eccentricity is $\widetilde{e}_{q=1au} = 2.22$ (95\%: $2.10-2.33$). The distribution of both $v_{\infty}$ and $e$, weighted by their contribution to the flux at Earth for \textit{* HD 35650}, can be seen in Figure~\ref{fig:velE:HD-35650}. The average direction of arrival is $(\alpha,\delta) = (271.0^\circ\,\pm\,1.3^\circ,\ 47.6^\circ\,\pm\,0.8^\circ)$ (Figure~\ref{fig:radiant:HD-35650}).

These particles typically traveled for $1.3\times10^{7}$~yr (95\%: $1.4\times10^{6}-8.7\times10^{7}$~yr) in the Galactic potential (first panel of Figure~\ref{fig:heatplots:HD-35650}). The typical distance traveled relative to their origin system is 4.7~pc (95\%: $2.9-31.4~$pc). During their transit, the typical velocity relative to the circular velocity of the Sun is $15.0~\mathrm{km~s}^{-1}$ (95\%: $1.5-17.6~\mathrm{km~s}^{-1}$).

The least demanding trajectory in terms of magnetic deflection, ISM drag, and grain-grain sputtering would have allowed a particle of $d_{\min}=8.4~\mu\mathrm{m}$ to survive its passage through the ISM. The most demanding delivery trajectory requires $d_{\mathrm{survive}}=3.0\times10^{3}~\mu\mathrm{m}$ (or $3~\mathrm{mm}$). 95\% of the particles had critical sizes within $9.2-65~\mu\mathrm{m}$, all of which are set by magnetic forces.

From \textit{* HD 35650}, the current flux of ISOs $\geq100$~m into the OC is $4.5\times10^{6}\pm1.5\times10^{6}~\mathrm{yr}^{-1}$. This results in $0.1\pm0.0$ ISOs within the inner solar system. The flux of $\geq200~\mu\mathrm{m}$ particles into the OC is $7.9\times10^{20}\pm2.7\times10^{20}~\mathrm{yr}^{-1}$, resulting in a flux at Earth of $3.8\times10^{2}\pm1.3\times10^{2}~\mathrm{yr}^{-1}$.

The maximum flux of $\geq100$~m ISOs into the OC is $3.2\times10^{8}\pm1.3\times10^{7}~\mathrm{yr}^{-1}$ at $t=-5.9\times10^{5}~\mathrm{yr}$, resulting in a population of $6.1\pm0.3$ ISOs within the inner solar system at that time. Simultaneously, there is a flux of $5.6\times10^{22}\pm2.2\times10^{21}~\mathrm{yr}^{-1}\,\geq200~\mu\mathrm{m}$ particles into the OC, resulting in a flux at Earth of $2.7\times10^{4}\pm1.1\times10^{3}~\mathrm{yr}^{-1}$.

The values above assume $t_{\mathrm{crit}}=10~\mathrm{Myr}$, see Tables within Appendix~\ref{append:tcrit3} for flux values with the assumption $t_{\mathrm{crit}}=3~\mathrm{Myr}$.

\subsection{* eta Crv}

\textit{* eta Crv} is a $1.4$ Gyr old star with an observed IR excess of $3.00\times10^{-4}$ \citep{lebreton_et_al_2016}. From Equation~\ref{IRage}, we assume $t_{IR}=50~$Myr. This star system delivered 0.01\% of the simulated particles to the solar system. The 852 particles that reach us were typically ejected at $\widetilde{t}_\mathrm{eject} = -4.8\times10^{7}~\mathrm{yr}$ (95\%: $-9.4\times10^{7}- -7.2\times10^{6}$ yr) with $\widetilde{v}_\mathrm{eject} = 1.4~\mathrm{km~s^{-1}}$ (95\%: $0.7-3.8~\mathrm{km~s^{-1}}$; first panel in Figure~\ref{fig:heatplots:star-eta-Crv}). The particles typically arrived at the solar system at $\widetilde{t}_\mathrm{arr} = 1.2\times10^{6}~\mathrm{yr}$ (95\%: $-8.2\times10^{5}-3.4\times10^{6}$; second panel in Figure~\ref{fig:heatplots:star-eta-Crv}). The median velocity at arrival was $\widetilde{v}_\infty \pm \sigma_{v_\infty} = 35.8 \pm 1.3~\mathrm{km~s^{-1}}$ (95\%: $33.7-38.7~\mathrm{km~s^{-1}}$; second panel in Figure~\ref{fig:heatplots:star-eta-Crv}), corresponding to a heliocentric velocity at 1~au of $\widetilde{v}_\mathrm{hel,1au} = 55.3~\mathrm{km~s^{-1}}$ (95\%: $54.0-57.2~\mathrm{km~s^{-1}}$). Assuming $q=1~\mathrm{au}$, the implied eccentricity is $\widetilde{e}_{q=1au} = 2.45$ (95\%: $2.28-2.69$). The distribution of both $v_{\infty}$ and $e$, weighted by their contribution to the flux at Earth for \textit{* eta Crv}, can be seen in Figure~\ref{fig:velE:star-eta-Crv}. The average direction of arrival is $(\alpha,\delta) = (276.2^\circ\,\pm\,1.4^\circ,\ 8.0^\circ\,\pm\,1.0^\circ)$ (Figure~\ref{fig:radiant:star-eta-Crv}).

These particles typically traveled for $4.9\times10^{7}$~yr (95\%: $7.2\times10^{6}-9.6\times10^{7}$~yr) in the Galactic potential (first panel of Figure~\ref{fig:heatplots:star-eta-Crv}). The typical distance traveled relative to their origin system is 65.1~pc (95\%: $18.4-155.2~$pc). During their transit, the typical velocity relative to the circular velocity of the Sun is $9.3~\mathrm{km~s}^{-1}$ (95\%: $0.4-16.6~\mathrm{km~s}^{-1}$).

The least demanding trajectory in terms of magnetic deflection, ISM drag, and grain-grain sputtering would have allowed a particle of $d_{\min}=39~\mu\mathrm{m}$ to survive its passage through the ISM. The most demanding delivery trajectory requires $d_{\mathrm{survive}}=1.1\times10^{3}~\mu\mathrm{m}$ (or $1.1~\mathrm{mm}$). 95\% of the particles had critical sizes within $41-2.2\times10^{2}~\mu\mathrm{m}$, all of which are set by magnetic forces.

From \textit{* eta Crv}, the current flux of ISOs $\geq100$~m into the OC is $7.1\times10^{6}\pm1.9\times10^{6}~\mathrm{yr}^{-1}$. This results in $0.1\pm0.0$ ISOs within the inner solar system. The flux of $\geq200~\mu\mathrm{m}$ particles into the OC is $1.3\times10^{21}\pm3.4\times10^{20}~\mathrm{yr}^{-1}$, resulting in a flux at Earth of $5.2\times10^{2}\pm1.4\times10^{2}~\mathrm{yr}^{-1}$.

The maximum flux of $\geq100$~m ISOs into the OC is $2.7\times10^{7}\pm1.1\times10^{6}~\mathrm{yr}^{-1}$ at $t=2.0\times10^{6}~\mathrm{yr}$, resulting in a population of $0.5\pm0.0$ ISOs within the inner solar system at that time. Simultaneously, there is a flux of $4.9\times10^{21}\pm1.8\times10^{20}~\mathrm{yr}^{-1}\,\geq200~\mu\mathrm{m}$ particles into the OC, resulting in a flux at Earth of $2.1\times10^{3}\pm8.4\times10^{1}~\mathrm{yr}^{-1}$.

The values above assume $t_{\mathrm{crit}}=10~\mathrm{Myr}$, see Tables within Appendix~\ref{append:tcrit3} for flux values with the assumption $t_{\mathrm{crit}}=3~\mathrm{Myr}$.

\subsection{HD 53143}

\textit{HD 53143} is a $1\pm0.2$ Gyr old star with an observed IR excess of $2.50\times10^{-4}$ \citep{kalas_et_al_2006}. From Equation~\ref{IRage}, we assume $t_{IR}=55~$Myr. This star system delivered 0.01\% of the simulated particles to the solar system. The 1,573 particles that reach us were typically ejected at $\widetilde{t}_\mathrm{eject} = -3.2\times10^{7}~\mathrm{yr}$ (95\%: $-9.2\times10^{7}- -6.0\times10^{6}$ yr) with $\widetilde{v}_\mathrm{eject} = 0.7~\mathrm{km~s^{-1}}$ (95\%: $0.4-3.3~\mathrm{km~s^{-1}}$; first panel in Figure~\ref{fig:heatplots:HD-53143}). The particles typically arrived at the solar system at $\widetilde{t}_\mathrm{arr} = -4.1\times10^{5}~\mathrm{yr}$ (95\%: $-1.2\times10^{6}-4.4\times10^{5}$; second panel in Figure~\ref{fig:heatplots:HD-53143}). The median velocity at arrival was $\widetilde{v}_\infty \pm \sigma_{v_\infty} = 35.0 \pm 0.7~\mathrm{km~s^{-1}}$ (95\%: $33.6-36.4~\mathrm{km~s^{-1}}$; second panel in Figure~\ref{fig:heatplots:HD-53143}), corresponding to a heliocentric velocity at 1~au of $\widetilde{v}_\mathrm{hel,1au} = 54.8~\mathrm{km~s^{-1}}$ (95\%: $53.9-55.7~\mathrm{km~s^{-1}}$). Assuming $q=1~\mathrm{au}$, the implied eccentricity is $\widetilde{e}_{q=1au} = 2.38$ (95\%: $2.27-2.50$). The distribution of both $v_{\infty}$ and $e$, weighted by their contribution to the flux at Earth for \textit{* HD 53143}, can be seen in Figure~\ref{fig:velE:HD-53143}. The average direction of arrival is $(\alpha,\delta) = (260.2^\circ\,\pm\,0.8^\circ,\ 12.9^\circ\,\pm\,1.1^\circ)$ (Figure~\ref{fig:radiant:HD-53143}).

These particles typically traveled for $3.2\times10^{7}$~yr (95\%: $5.8\times10^{6}-9.2\times10^{7}$~yr) in the Galactic potential (first panel of Figure~\ref{fig:heatplots:HD-53143}). The typical distance traveled relative to their origin system is 18.4~pc (95\%: $14.1-70.5~$pc). During their transit, the typical velocity relative to the circular velocity of the Sun is $4.4~\mathrm{km~s}^{-1}$ (95\%: $0.1-15.0~\mathrm{km~s}^{-1}$).

The least demanding trajectory in terms of magnetic deflection, ISM drag, and grain-grain sputtering would have allowed a particle of $d_{\min}=24~\mu\mathrm{m}$ to survive its passage through the ISM. The most demanding delivery trajectory requires $d_{\mathrm{survive}}=2.2\times10^{3}~\mu\mathrm{m}$ (or $2.2~\mathrm{mm}$). 95\% of the particles had critical sizes within $25-2.5\times10^{2}~\mu\mathrm{m}$, all of which are set by magnetic forces.

From \textit{* HD 53143}, the current flux of ISOs $\geq100$~m into the OC is $2.3\times10^{7}\pm3.1\times10^{6}~\mathrm{yr}^{-1}$. This results in $0.4\pm0.1$ ISOs within the inner solar system. The flux of $\geq200~\mu\mathrm{m}$ particles into the OC is $4.1\times10^{21}\pm5.5\times10^{20}~\mathrm{yr}^{-1}$, resulting in a flux at Earth of $1.8\times10^{3}\pm2.5\times10^{2}~\mathrm{yr}^{-1}$.

The maximum flux of $\geq100$~m ISOs into the OC is $8.3\times10^{7}\pm5.2\times10^{6}~\mathrm{yr}^{-1}$ at $t=-3.9\times10^{5}~\mathrm{yr}$, resulting in a population of $1.5\pm0.1$ ISOs within the inner solar system at that time. Simultaneously, there is a flux of $1.5\times10^{22}\pm9.5\times10^{20}~\mathrm{yr}^{-1}\,\geq200~\mu\mathrm{m}$ particles into the OC, resulting in a flux at Earth of $6.5\times10^{3}\pm4.1\times10^{2}~\mathrm{yr}^{-1}$.

The values above assume $t_{\mathrm{crit}}=10~\mathrm{Myr}$, see Tables within Appendix~\ref{append:tcrit3} for flux values with the assumption $t_{\mathrm{crit}}=3~\mathrm{Myr}$.

\subsection{* bet Pic}

\textit{* bet Pic} is a $21\pm4$ Myr old star \citep{binks_jefferies_2013} with an observed IR excess of $1.70\times10^{-3}$ \citep{nilsson_et_al_2009}. This is our reference system on which $t_{IR}$ is based; therefore, from Equation~\ref{IRage}, we assume $t_{IR}=21~$Myr. This star system delivered 0.05\% of the simulated particles to the solar system. The 5,104 particles that reach us were typically ejected at $\widetilde{t}_\mathrm{eject} = -1.2\times10^{7}~\mathrm{yr}$ (95\%: $-2.0\times10^{7}- -3.6\times10^{6}$ yr) with $\widetilde{v}_\mathrm{eject} = 1.1~\mathrm{km~s^{-1}}$ (95\%: $0.5-4.0~\mathrm{km~s^{-1}}$; first panel in Figure~\ref{fig:heatplots:star-bet-Pic}). The particles typically arrived at the solar system at $\widetilde{t}_\mathrm{arr} = -9.8\times10^{5}~\mathrm{yr}$ (95\%: $-2.1\times10^{6}-2.0\times10^{5}$; second panel in Figure~\ref{fig:heatplots:star-bet-Pic}). The median velocity at arrival was $\widetilde{v}_\infty \pm \sigma_{v_\infty} = 18.8 \pm 1.1~\mathrm{km~s^{-1}}$ (95\%: $16.7-21.2~\mathrm{km~s^{-1}}$; second panel in Figure~\ref{fig:heatplots:star-bet-Pic}), corresponding to a heliocentric velocity at 1~au of $\widetilde{v}_\mathrm{hel,1au} = 46.1~\mathrm{km~s^{-1}}$ (95\%: $45.3-47.2~\mathrm{km~s^{-1}}$). Assuming $q=1~\mathrm{au}$, the implied eccentricity is $\widetilde{e}_{q=1au} = 1.40$ (95\%: $1.31-1.51$). The distribution of both $v_{\infty}$ and $e$, weighted by their contribution to the flux at Earth for \textit{* bet Pic}, can be seen in Figure~\ref{fig:velE:star-bet-Pic}. The average direction of arrival is $(\alpha,\delta) = (268.3^\circ\,\pm\,0.5^\circ,\ 23.1^\circ\,\pm\,2.3^\circ)$ (Figure~\ref{fig:radiant:star-bet-Pic}).

These particles typically traveled for $1.1\times10^{7}$~yr (95\%: $2.6\times10^{6}-2.0\times10^{7}$~yr) in the Galactic potential (first panel of Figure~\ref{fig:heatplots:star-bet-Pic}). The typical distance traveled relative to their origin system is 9.7~pc (95\%: $8.0-29.2~$pc). During their transit, the typical velocity relative to the circular velocity of the Sun is $0.4~\mathrm{km~s}^{-1}$ (95\%: $0.0-2.0~\mathrm{km~s}^{-1}$). 

The least demanding trajectory in terms of magnetic deflection, ISM drag, and grain-grain sputtering would have allowed a particle of $d_{\min}=29~\mu\mathrm{m}$ to survive its passage through the ISM. The most demanding delivery trajectory requires $d_{\mathrm{survive}}=8.4\times10^{3}~\mu\mathrm{m}$ (or $8.4~\mathrm{mm}$). 95\% of the particles had critical sizes within $61-4.5\times10^{2}~\mu\mathrm{m}$, all of which are set by magnetic forces.

From \textit{* bet Pic}, the current flux of ISOs $\geq100$~m into the OC is $1.8\times10^{6}\pm3.5\times10^{5}~\mathrm{yr}^{-1}$. This results in $0.1\pm0.0$ ISOs within the inner solar system. The flux of $\geq200~\mu\mathrm{m}$ particles into the OC is $3.1\times10^{20}\pm6.2\times10^{19}~\mathrm{yr}^{-1}$, resulting in a flux at Earth of $4.2\times10^{2}\pm7.9\times10^{1}~\mathrm{yr}^{-1}$.

The maximum flux of $\geq100$~m ISOs into the OC is $2.5\times10^{7}\pm1.1\times10^{6}~\mathrm{yr}^{-1}$ at $t=-9.8\times10^{5}~\mathrm{yr}$, resulting in a population of $0.8\pm0.1$ ISOs within the inner solar system at that time. Simultaneously, there is a flux of $4.4\times10^{21}\pm2.1\times10^{20}~\mathrm{yr}^{-1}\,\geq200~\mu\mathrm{m}$ particles into the OC, resulting in a flux at Earth of $4.8\times10^{3}\pm2.2\times10^{2}~\mathrm{yr}^{-1}$.

The values above assume $t_{\mathrm{crit}}=10~\mathrm{Myr}$, see Tables within Appendix~\ref{append:tcrit3} for flux values with the assumption $t_{\mathrm{crit}}=3~\mathrm{Myr}$.

\begin{deluxetable*}{cccccccc}
\tablecaption{Summary of Star System Transfer Characteristics\label{tab:transfer_summary}}
\tablehead{
\colhead{Star System} & \colhead{Effective Radiant} & \colhead{Transfer} & \colhead{$\widetilde{v}_{\infty}\pm\sigma_{v_{\infty}}$} & \colhead{$\widetilde{v}_{hel,1au}$} & \colhead{$\widetilde{e}$} & \colhead{Radiant} & \colhead{Critical Size Limits} \\
\colhead{} & \colhead{($\alpha$,$\delta$)} & \colhead{Percentage} & \colhead{} & \colhead{} & \colhead{} & \colhead{($\alpha\pm\sigma_\alpha$, $\delta\pm\sigma_\delta$)} & \colhead{[$d_{\mathrm{min}}$, $d_{\mathrm{survive}}$]} \\
\colhead{{}} & \colhead{(deg, deg)} & \colhead{($\%$)} & \colhead{(km s$^{-1}$)} & \colhead{(km s$^{-1}$)} & \colhead{($q=1~$au)} & \colhead{(deg, deg)} & \colhead{($\mu m$)}
}
\startdata
\protect* eps Eri & (190, 6) & 0.09\% & 22.1$\pm$0.8 & 47.6 & 1.55 & (189.8$\pm$2.1, 6.1$\pm$0.7) & [5.9, 1.4e+03] \\
\protect* tau Cet & (94, -30) & 0.07\% & 37.1$\pm$0.8 & 56.1 & 2.55 & (93.9$\pm$1.3, -30.4$\pm$0.3) & [5.6, 1.8e+03] \\
\protect* e Eri & (277, 20) & 0.05\% & 125.5$\pm$0.8 & 132.4 & 18.76 & (278.0$\pm$0.2, 20.9$\pm$0.2) & [5.5, 3.9e+02] \\
BD-07 4003 & (305, 2) & 0.04\% & 38.0$\pm$0.8 & 56.7 & 2.63 & (304.7$\pm$1.1, 2.3$\pm$0.2) & [16, 2.4e+03] \\
\protect* alf Lyr & (256, 1) & 0.05\% & 18.6$\pm$0.7 & 46.0 & 1.39 & (256.5$\pm$1.2, 1.7$\pm$1.8) & [16, 2e+02] \\
\protect* alf PsA & (241, 34) & 0.03\% & 15.1$\pm$0.7 & 44.8 & 1.26 & (243.3$\pm$3.1, 33.8$\pm$1.1) & [26, 8.5e+02] \\
\protect* 61 Vir & (263, 38) & 0.02\% & 61.5$\pm$0.7 & 74.5 & 5.26 & (263.5$\pm$0.8, 38.4$\pm$0.4) & [11, 4.1e+02] \\
HD 197481 & (265, 32) & 0.04\% & 21.8$\pm$1.1 & 47.5 & 1.54 & (266.9$\pm$1.7, 31.2$\pm$1.8) & [28, 4.2e+02] \\
\protect* bet Leo & (270, 12) & 0.02\% & 26.0$\pm$0.9 & 49.5 & 1.76 & (269.5$\pm$1.7, 13.2$\pm$0.7) & [35, 3.7e+03] \\
HD 166 & (271, 27) & 0.01\% & 28.4$\pm$0.7 & 50.8 & 1.91 & (271.4$\pm$1.6, 27.7$\pm$1.0) & [22, 1.7e+03] \\
HD 38858 & (275, 31) & 0.04\% & 35.7$\pm$0.7 & 55.2 & 2.44 & (275.7$\pm$0.7, 32.0$\pm$0.9) & [13, 2.4e+03] \\
HD 207129 & (297, 20) & 0.01\% & 25.8$\pm$1.3 & 49.4 & 1.75 & (297.0$\pm$1.9, 20.4$\pm$2.5) & [29, 1.9e+03] \\
\protect* b Her & (345, -23) & 0.01\% & 8.8$\pm$0.6 & 43.0 & 1.09 & (342.2$\pm$4.5, -22.5$\pm$3.4) & [29, 1.2e+02] \\
HD 23484 & (259, 1) & 0.02\% & 40.9$\pm$0.8 & 58.7 & 2.89 & (259.7$\pm$0.7, 0.9$\pm$0.9) & [23, 91] \\
\protect* q01 Eri & (261, 58) & 0.02\% & 32.2$\pm$0.7 & 53.0 & 2.17 & (261.9$\pm$2.4, 58.5$\pm$0.4) & [57, 2.4e+04] \\
\protect* g Lup & (269, 23) & 0.01\% & 27.0$\pm$0.7 & 50.1 & 1.83 & (269.3$\pm$1.2, 22.5$\pm$1.3) & [31, 1.2e+03] \\
HD 35650 & (270, 46) & 0.07\% & 32.8$\pm$0.7 & 53.4 & 2.22 & (270.7$\pm$1.3, 47.3$\pm$0.8) & [8.4, 3e+03] \\
\protect* eta Crv & (276, 7) & 0.01\% & 35.8$\pm$1.3 & 55.3 & 2.45 & (276.4$\pm$1.4, 8.2$\pm$1.0) & [39, 1.1e+03] \\
HD 53143 & (260, 13) & 0.01\% & 35.0$\pm$0.7 & 54.8 & 2.38 & (260.3$\pm$0.8, 13.2$\pm$1.1) & [24, 2.2e+03] \\
\protect* bet Pic & (268, 26) & 0.05\% & 18.8$\pm$1.1 & 46.1 & 1.40 & (268.4$\pm$0.5, 23.9$\pm$2.3) & [29, 8.4e+03] \\
\enddata
\tablecomments{The average radiant provided for each system is derived from the unweighted simulated particles, and is a heliocentric J2000 RA/Dec.\\The ``Critical Size Limits" column provides the minimum ($d_{min}$) and maximum ($d_{survive}$) sizes (diameter) that were necessary for the simulated particles to survive magnetic deflection, ISM drag, and destruction based on their individual trajectories (see Section \ref{sec:sizes}). Therefore, all particles $\geq d_{survive}$ (the critical size) are expected to survive the journey.}
\end{deluxetable*}

\begin{deluxetable*}{cccccccccc}
\tablecaption{Maximum and Current Flux into the Oort Cloud and the Resulting Population Within the Inner Solar System for Particles $\geq 100$ m (assuming $t_{crit}=10$ Myr)\label{tab:100m_ageDepend_tcrit10}}
\tablehead{
\colhead{Star Name} & \colhead{Time at Max} & \colhead{Max Flux} & \colhead{Error} & \colhead{Max Population} & \colhead{Error} & \colhead{Current Flux} & \colhead{Error} & \colhead{Current Population} & \colhead{Error} \\
\colhead{} & \colhead{} & \colhead{Oort Cloud} & \colhead{} & \colhead{Inner Solar System} & \colhead{} & \colhead{Oort Cloud} & \colhead{} & \colhead{Inner Solar System} & \colhead{} \\
\colhead{} & \colhead{(yr)} & \colhead{(yr$^{-1}$)} & \colhead{} & \colhead{} & \colhead{} & \colhead{(yr$^{-1}$)} & \colhead{} & \colhead{} & \colhead{}
}
\startdata
\protect* eps Eri & -6.0e+04 & 4.6e+07 & 5.1e+06 & 1.3e+00 & 1.9e-01 & 3.0e+07 & 3.9e+06 & 8.5e-01 & 1.5e-01 \\
\protect* tau Cet & 6.5e+04 & 1.3e+05 & 1.0e+03 & 2.2e-03 & 1.9e-05 & 1.1e+05 & 9.1e+02 & 1.8e-03 & 1.9e-05 \\
\protect* e Eri & -5.5e+04 & 4.0e+04 & 1.9e+02 & 2.0e-04 & 1.0e-06 & 3.0e+04 & 1.7e+02 & 1.5e-04 & 8.3e-07 \\
BD-07 4003 & -2.2e+05 & 6.1e+07 & 5.3e+06 & 1.0e+00 & 1.0e-01 & 1.9e+07 & 3.5e+06 & 3.2e-01 & 6.3e-02 \\
\protect* alf Lyr & 3.3e+05 & 1.6e+05 & 2.5e+03 & 5.6e-03 & 1.2e-04 & 3.6e+04 & 1.2e+03 & 1.2e-03 & 6.3e-05 \\
\protect* alf PsA & -6.2e+05 & 1.7e+06 & 1.1e+05 & 7.2e-02 & 8.1e-03 & 9.7e+04 & 2.1e+04 & 4.1e-03 & 1.6e-03 \\
\protect* 61 Vir & 1.6e+05 & 3.4e+05 & 1.3e+04 & 3.5e-03 & 1.4e-04 & 1.8e+05 & 8.4e+03 & 1.9e-03 & 9.5e-05 \\
HD 197481 & 7.0e+04 & 1.0e+06 & 3.8e+04 & 2.9e-02 & 1.5e-03 & 8.4e+05 & 3.7e+04 & 2.4e-02 & 1.5e-03 \\
\protect* bet Leo & 1.3e+06 & 2.5e+04 & 1.2e+03 & 6.2e-04 & 3.8e-05 & 9.7e+03 & 3.8e+02 & 2.3e-04 & 1.1e-05 \\
HD 166 & -6.6e+05 & 1.3e+07 & 1.3e+06 & 2.8e-01 & 3.7e-02 & 4.0e+05 & 1.5e+05 & 9.0e-03 & 4.5e-03 \\
HD 38858 & -3.2e+05 & 8.5e+07 & 7.2e+06 & 1.5e+00 & 1.4e-01 & 4.5e+06 & 1.5e+06 & 7.9e-02 & 3.0e-02 \\
HD 207129 & 3.6e+06 & 1.0e+07 & 6.9e+05 & 2.6e-01 & 2.2e-02 & 3.1e+06 & 1.3e+06 & 7.6e-02 & 3.9e-02 \\
\protect* b Her & -6.2e+05 & 2.3e+04 & 8.8e+02 & 1.7e-03 & 2.2e-04 & 3.4e+03 & 4.2e+02 & 2.4e-04 & 1.1e-04 \\
HD 23484 & -8.8e+05 & 1.8e+07 & 2.9e+06 & 2.8e-01 & 5.0e-02 & 2.4e+05 & 7.9e+04 & 3.8e-03 & 1.3e-03 \\
\protect* q01 Eri & -7.8e+05 & 2.8e+08 & 4.9e+06 & 5.5e+00 & 1.2e-01 & 5.8e+06 & 1.4e+06 & 1.1e-01 & 3.4e-02 \\
\protect* g Lup & 2.6e+05 & 1.7e+07 & 2.9e+06 & 4.0e-01 & 8.6e-02 & 6.8e+06 & 1.9e+06 & 1.6e-01 & 5.4e-02 \\
HD 35650 & -5.9e+05 & 3.2e+08 & 1.3e+07 & 6.1e+00 & 2.8e-01 & 4.5e+06 & 1.5e+06 & 8.7e-02 & 3.4e-02 \\
\protect* eta Crv & 2.0e+06 & 2.7e+07 & 1.1e+06 & 4.9e-01 & 2.0e-02 & 7.1e+06 & 1.9e+06 & 1.2e-01 & 3.7e-02 \\
HD 53143 & -3.9e+05 & 8.3e+07 & 5.2e+06 & 1.5e+00 & 1.1e-01 & 2.3e+07 & 3.1e+06 & 4.3e-01 & 6.4e-02 \\
\protect* bet Pic & -9.8e+05 & 2.5e+07 & 1.1e+06 & 8.4e-01 & 5.6e-02 & 1.8e+06 & 3.5e+05 & 6.7e-02 & 2.1e-02 \\
\tableline
\textbf{Total} & -6.0e+05 & 4.9e+08 & 1.5e+07 & 9.6e+00 & 3.3e-01 & 1.1e+08 & 7.2e+06 & 2.3e+00 & 2.0e-01 \\
\enddata
\end{deluxetable*}

\begin{deluxetable*}{cccccc}
\tablecaption{Maximum and Current Flux into Earth's Atmosphere for Particles $\geq 200~\mu$m (assuming $t_{crit}=10$ Myr)\label{tab:200micron_ageDepend_tcrit10}}
\tablehead{
\colhead{Star Name} & \colhead{Time at Max} & \colhead{Max Flux} & \colhead{Error} & \colhead{Current Flux} & \colhead{Error} \\
\colhead{} & \colhead{} & \colhead{Earth} & \colhead{} & \colhead{Earth} & \colhead{} \\
\colhead{} & \colhead{(yr)} & \colhead{(yr$^{-1}$)} & \colhead{} & \colhead{(yr$^{-1}$)} & \colhead{}
}
\startdata
\protect* eps Eri & -6.0e+04 & 6.7e+03 & 7.7e+02 & 4.4e+03 & 5.8e+02 \\
\protect* tau Cet & 6.5e+04 & 9.5e+00 & 7.3e-02 & 7.8e+00 & 7.0e-02 \\
\protect* e Eri & -5.5e+04 & 1.4e+00 & 7.0e-03 & 1.1e+00 & 6.0e-03 \\
BD-07 4003 & -2.2e+05 & 4.3e+03 & 3.9e+02 & 1.4e+03 & 2.4e+02 \\
\protect* alf Lyr & 3.3e+05 & 3.2e+01 & 4.7e-01 & 6.9e+00 & 2.3e-01 \\
\protect* alf PsA & -6.2e+05 & 4.8e+02 & 3.1e+01 & 2.8e+01 & 6.2e+00 \\
\protect* 61 Vir & 1.6e+05 & 1.6e+01 & 5.9e-01 & 8.5e+00 & 4.3e-01 \\
HD 197481 & 7.0e+04 & 1.5e+02 & 5.7e+00 & 1.2e+02 & 5.8e+00 \\
\protect* bet Leo & 1.3e+06 & 3.0e+00 & 1.4e-01 & 1.1e+00 & 4.2e-02 \\
HD 166 & -6.6e+05 & 1.3e+03 & 1.4e+02 & 4.2e+01 & 1.7e+01 \\
HD 38858 & -3.2e+05 & 6.5e+03 & 5.3e+02 & 3.4e+02 & 1.1e+02 \\
HD 207129 & 3.6e+06 & 1.3e+03 & 8.7e+01 & 3.6e+02 & 1.6e+02 \\
\protect* b Her & -6.2e+05 & 1.8e+01 & 7.2e-01 & 2.6e+00 & 3.4e-01 \\
HD 23484 & -8.8e+05 & 1.2e+03 & 1.9e+02 & 1.6e+01 & 5.2e+00 \\
\protect* q01 Eri & -7.8e+05 & 2.4e+04 & 4.7e+02 & 5.0e+02 & 1.2e+02 \\
\protect* g Lup & 2.6e+05 & 1.9e+03 & 3.3e+02 & 7.4e+02 & 2.0e+02 \\
HD 35650 & -5.9e+05 & 2.7e+04 & 1.1e+03 & 3.8e+02 & 1.3e+02 \\
\protect* eta Crv & 2.0e+06 & 2.1e+03 & 8.4e+01 & 5.2e+02 & 1.4e+02 \\
HD 53143 & -3.9e+05 & 6.5e+03 & 4.1e+02 & 1.8e+03 & 2.5e+02 \\
\protect* bet Pic & -9.8e+05 & 4.8e+03 & 2.2e+02 & 4.2e+02 & 7.9e+01 \\
\tableline
\textbf{Total} & -6.0e+05 & 4.3e+04 & 1.2e+03 & 1.1e+04 & 7.7e+02 \\
\enddata
\end{deluxetable*}

\clearpage

\section{Discussion} \label{sec:disc}

\subsection{Connection to Known Interstellar Objects}

An intriguing question is whether any of the three known macroscopic ISOs (1I/'Oumuamua, 2I/Borisov, and 3I/ATLAS) could have originated from these nearby debris disk systems? From Figure~\ref{fig:radiant}, we see that the radiants of 2I and 3I are far from those expected from our sample of debris disks. However, the arrival direction of 1I/'Oumuamua appears to be roughly coincident with the expected radiants from HD 38858 and HD 166. 

\begin{figure}
\plotone{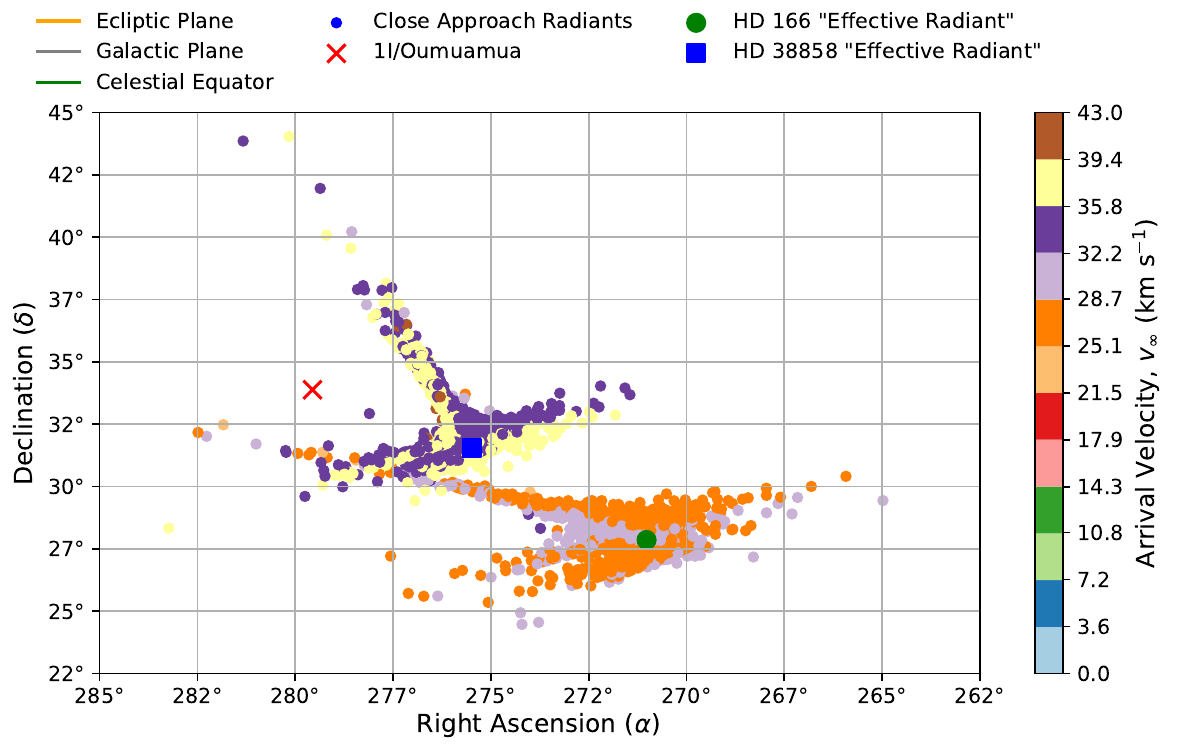}
\caption{This is a zoomed-in version of the radiant plot in Figure \ref{fig:radiant}, focusing on the direction in which 1I/'Oumuamua arrived. HD 38858 and HD 166 are the only systems plotted, to show the comparison in the radiant of their particles arrival directions. These particles are colored by their arrival velocity $v_{\infty}$.
\label{fig:Oumuamua_radiant}}
\end{figure}

Figure \ref{fig:Oumuamua_radiant} provides a more detailed look at the relevant radiant directions. Neither system provides an exact match to 1I/'Oumuamua's incoming direction and velocity, though the match is tantalizingly close. HD 166 in particular has expected radiants within a few degrees of 1I/'Oumuamua's at comparable arrival speeds ( 1I's $v_\infty=26~\mathrm{km~s}^{-1}$). 

However, the trajectories of 1I/'Oumuamua and the two debris disks in question are sufficiently well known to preclude a simple origin at either one. The debris disk radiants most similar to 1I's all were ejected within the past 50 Myr. Conducting a dynamical trace back simulation over that time interval, the closest 1I/'Oumuamua gets to HD 38858 is 5.4 pc with a relative velocity of 9.6 km s$^{-1}$, and to HD 166, 9.8 pc with 4.4 km s$^{-1}$. 
If 1I had originated at either of these disks, we would expect backward integration to take it much closer to these systems.

This is confirmed by the fact that previous studies of 1I's origins did not identify these systems as possible progenitors. Both HD 166 and HD 38858 are in the SIMBAD as well as the Gaia DR2 and DR3 catalogues, which form the basis of most of the previous searches for 'Oumuamua's point of origin \citep{Bailer-Jones_2018, Zwart2018, dybczynski_2018, Hallat_2020}. This result should therefore be viewed as a proof of concept: while firm associations are more difficult to establish, quick radiant and velocity comparisons using Table \ref{tab:transfer_summary} can help flag nearby systems that merit further investigation. 

\begin{deluxetable*}{lcccc}
\tablecaption{Confirmed interstellar objects, including LIC upstream direction of interstellar dust and significant directional values in ICRS coordinates.\label{tab:ISO}}
\tablewidth{0pt}
\tablehead{
Designation & Right Ascension ($\alpha$) & Declination ($\delta$) & Category & References \\
         & (deg) & (deg) &  &  
}
\startdata
1I/'Oumuamua & 279.6 & 33.9 & ISO & [2] \\
2I/Borisov & 32.6 & 59.5 & ISO & [2] \\
3I/ATLAS & 295.0 & -19.1 & ISO & [4]  \\
Local interstellar cloud (LIC) upstream direction& 259 & 8 & ISD & [5] \\
\hline \\
Solar apex relative to Local Standard of Rest (LSR) & 258.7 & -15.0 &  & [3] \\
Solar Galactic Apex & 313.3 & 47.5 &  & [1] \\
Galactic Centre (GC) -  ($\lambda$,$b$)=($0^{\circ}$,$0^{\circ}$) & 266.4 & -28.9 & & [6] \\
\enddata
\tablereferences{
% [1] \cite{baggaley2000},
    [1] \cite{gregg_wiegert_alphacen_2025}
    [2] \cite{Hallat_2020},
    [3] \cite{jaschek_solar_1992},
    [4] \cite{marcos_3I_2025}
    [5] \cite{sterken_flow_2012},
    [6] \cite{Reid_2004}
    }
\end{deluxetable*}

\subsection{Comparison With Previous Work}

Of our 20 systems, 6 have been previously examined in the aforementioned work of \cite{Murray_2004}: \textit{* eps Eri} (Ran), \textit{* tau Cet}, \textit{* alf Lyr} (Vega), \textit{* alf PsA} (Fomalhaut), and * bet Pic (\betapic{full}). Of these, \betapic{full} was by far the largest contributor of small grains in Earth's atmosphere. They found the flux to be $1.5\times10^{-4}~\mathrm{yr}^{-1}~\mathrm{km}^{-2}$ for particles $\geq20~\mu$m assuming $t_{crit}=3~$ Myr. Adjusting this to $\geq200~\mu$m particles hitting the entire cross-sectional area of the Earth yields $\sim600~\mathrm{yr}^{-1}$. This is within an order of magnitude of our value of $4.4\times10^{3}\pm1.7\times10^{3}~\mathrm{yr}^{-1}$ (Table \ref{tab:200micron_ageDepend_tcrit3} of Appendix \ref{append:tcrit3}), and we would argue that our results are essentially consistent with \cite{Murray_2004}'s. We find a similar order of magnitude difference for the other systems, with the exception of \textit{*eps Eri} (Ran). For this system,  our value is larger by a factor of $\sim 10^3$. The study of \cite{Murray_2004} also reported a critical grain size for each system which we compare to our minimum grain size required to survive transport, and find that our values always matched within a factor of 3. Overall our results are highly consistent with those of \cite{Murray_2004}.

\subsection{Why Don't We See Interstellar Meteoroids?}\label{sec:meteors}

Our results are consistent with the conclusion of \cite{Murray_2004} that \cite{baggaley2000} did not detect a significant population of meteoroids from \betapic{full} with the AMOR radar. Although our predicted fluxes are higher than those derived by \cite{Murray_2004}, the expected arrival direction of particles from \betapic{full} do not match the proposed observations of \cite{baggaley2000}.

To investigate this probability of an interstellar meteoroid detection from our sample of debris disks, we consider the fluxes and collecting areas of radar and video meteor networks. 

For the smallest particles observable through radar ($\gtrsim200~\mu$m), CMOR has an optimistic collecting area of 400 km$^{2}$ \citep{froncisz_2020}, only about $10^{-6}$ of Earth's total area. The probability of a single interstellar meteor being detected each year from these nearby debris disks is only 1\%. Assuming Poisson statistics, a time interval of 80 years is needed before reaching a 50\% confidence of observing a single true event. CMOR has been in multi-station operation since January 2002 \citep{CMOR}\footnote{\url{https://aquarid.physics.uwo.ca/research/radar/cmor_intro.html}}, almost a quarter of a century. As a result, there is only a 20\% chance that a meteoroid from one of these nearby debris disk systems would have been observed by CMOR.

Large collecting areas improve ones chances of observing interstellar meteors, making global networks like GMN very appealing for this type of search. GMN currently has a collecting area of 17.5 million km$^2$, however, the limiting size for these video detectors is much larger ($\geq5$ mm; \citep{wiegert_et_al_GMN_2025}). Correcting the expected fluxes from those given at $\geq200~\mu$m sizes, a straightforward correction for particle size can be applied as $(d\mathrm{(\mu m)}/200)^{-2.5}$. A factor of $\sim3\times10^{-4}$ accounts for the difference in size at 5 mm, closer to the detection limit of GMN. This results in 3 meteors at this size or larger across the Earth each year. But if we account for the loss of 50\% of potential events due to the necessity of observing at night, as well as the loss of 55\% to cloud cover \citep{cloud_cover}, it would take 25 years to be 50\% confident of making one real observation. Making the crude assumption that the networks coverage has been the same since its inception in December 2018, we can be 80\% confident that an interstellar meteoroid from these nearby debris disk systems has not yet been observed by GMN.

\subsubsection{The Eccentricities of Meteoroids From Nearby Debris Disks}

The background population of sporadic interstellar particles arriving at the solar system is expected to have a velocity distribution similar to the velocity dispersion of stars in the solar neighbourhood $\approx 20$~km~s$^{-1}$ \citep{mihalas_1981}. These values of $v_{\infty}$ produce a heliocentric velocities at 1~au of 46.6~km~s$^{-1}$ (Equation~\ref{eq:helio_vel}), well above the escape velocity of 42.1~km~s$^{-1}$ and with an eccentricity of 1.45 (Equation~\ref{eq:e}; assuming $q=1$ au). 

\begin{figure*}
% \digitalasset
\plotone{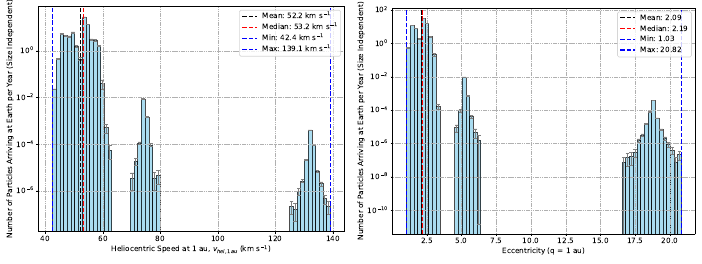}
\caption{These are histograms of all simulated particles arriving at the solar system, weighted according to their contribution to the flux at Earth (see Section \ref{sec:flux}) but with no size dependence. Heliocentric velocity is computed at 1 au ($v_{hel,1au}$) and eccentricities ($e$) with an assumed perihelion at 1 au. Note the log scale on the y-axis.\\The complete figure set (20 images) displaying the data for each individual system separately is displayed in Appendix~\ref{append:A:vel_e}.
\label{fig:vel_e}}
\end{figure*}

We find with this sample set that meteors with more extreme values are to be expected. Figure~\ref{fig:vel_e} shows that heliocentric speeds of up to 140 km s$^{-1}$ at the Earth can be expected, with these highest values expected for \textit{* e Eri}. There is a risk that meteor events determined to be traveling at these speeds might be discarded as bad measurements based on these seemingly impossible values: this suggests caution when using arbitrary cut-offs to eliminate ``bad" events. We also estimate 20\% of the particles entering the solar system from these debris disks are arriving with $v_{\infty}\leq20~\mathrm{km~s}^{-1}$, corresponding to heliocentric velocities at 1 au of $42.4\leq v_{hel,1au}\leq46.5~\mathrm{km~s}^{-1}$ (Figure \ref{fig:vel_e}). Assuming $q=1$ au, these particles would have eccentricities of $1.03\leq e\leq1.45$ (Figure \ref{fig:vel_e}).

Slower arrivals are concentrated by gravitational focusing (Equation~\ref{eq:grav_lens}), which increases the effective cross section of the Earth for low-velocity particles. Slower meteoroids experience stronger deflection due to the gravity of the Sun, making them more likely to be directed into the inner solar system and enhancing their flux at Earth. Therefore, an enhanced population of interstellar meteors is expected near the parabolic boundary. Thus, the near-parabolic region may contain more interstellar particles than simple assumptions about the velocity distribution of ISOs may imply.

If our results are correct, then meteor networks with large collection areas offer the best opportunity to identify elusive interstellar meteors. We also find that the lack of detections to date is not surprising given the current survey limitations, but are encouraging as we realize that discoveries are imminent as coverage and precision improve. Dedicated searches for hyperbolic candidates in archives, improved velocity calibration, and future high-sensitivity meteor networks may soon reveal the first robust detections of interstellar meteors and provide an empirical test of the predictions presented here.

\subsection{Implications}

As observational capabilities improve, it is likely that more ISOs will be discovered, and it is only a matter of time before interstellar meteors are conclusively identified. The key question will then be whether they can be securely associated with a specific source system. The results presented here provide a first step toward that goal: for each of the 20 nearest debris disk systems we now have estimates for the efficiency of delivery, expected arrival speeds, and asymptotic radiants. These predictions can serve as a guide for targeted searches in current meteor datasets and a first check for identifying the home system of newly discovered ISOs.

\section{Conclusion} \label{sec:conc}

In this work, we set out to answer the following questions. If nearby debris disk systems are ejecting material into the Galaxy, how much of it should we expect to reach the solar system? Will there be a population of observable macroscopic ISOs in the inner solar system?  Should we expect to be able to detect interstellar meteors and in what numbers?

This work combines numerical integrations of particle trajectories from the 20 nearest debris disk systems with empirical estimates of their ejection rates and velocities from the literature. This allowed us to compute size-dependent fluxes, expected arrival radiants, and encounter velocities at Earth, for particle ejections taking place over the last 100 Myr. We also provide system-by-system predictions of arrival speeds and radiants that can serve as templates for targeted searches.

Our simulations predict that nearby debris disks collectively inject millions of $\geq100~$m ISOs into the solar system each year and there may currently be a few within the inner solar system that could be detected by current telescopic surveys. For small particles, only observable as meteors, we find impact rates at Earth that should produce thousands of detectable radar ($\geq200~\mu$m) events annually. At visual sizes ($\geq5$ mm), this drops to a few each year. These values are consistent with the current lack of detections of interstellar meteors by meteor networks.

Our flux and population values all make some assumptions about the origin systems ejection rates. The results are perhaps most sensitive to the choice of the critical age (the age after planet formation where ejections could likely begin). Our assumption is that $t_{crit}$ is 10 Myr. If the actual critical age is 3 Myr, the flux and population values within our solar system increase roughly by a factor of $\sim20$ (Appendix \ref{append:tcrit3}). 

Improvements in global meteor systems, as well as targeted re-analyses of archival data with refined velocity calibration, offer the opportunity to test and refine these predictions. Future work combining detailed debris disk modeling and improved dynamical back-tracing will be crucial to establish source associations for any detected ISOs or interstellar meteors.

This work represents the first numerical attempt to link specific nearby debris disk systems to the expected ISO population in the inner solar system and meteoroid flux at Earth. By providing concrete predictions of flux, speed, and radiant distributions, we hope to bridge the gap between theory and observation and enable the first unambiguous detections of interstellar material from known extrasolar systems, a milestone that would open a new window on the exchange of material between planetary systems.

%% Please use the acknowledgment and contribution environments. This will 
%% be anonomyized when the "anonymous" style option is used. 
\begin{acknowledgments}
This work was supported in part by the
NASA Meteoroid Environment Office under Cooperative Agreement No. 80NSSC24M0060 and by the Natural Sciences and Engineering Research Council of Canada (NSERC) Discovery Grant program (grant No. RGPIN-2024-05200).

This research has made use of the SIMBAD database, operated at CDS, Strasbourg, France.

This work has made use of data from the European Space Agency (ESA) mission
{\it Gaia} (\url{https://www.cosmos.esa.int/gaia}), processed by the {\it Gaia}
Data Processing and Analysis Consortium (DPAC,
\url{https://www.cosmos.esa.int/web/gaia/dpac/consortium}). Funding for the DPAC
has been provided by national institutions, in particular the institutions
participating in the {\it Gaia} Multilateral Agreement.
\end{acknowledgments}

\begin{contribution}
%%This section gives authors the space to recognize author contributions. The text inside this environment is NOT counted towards the total word quanta. At a minimum, manuscripts are expected to include this text:
All authors contributed equally.

%% But authors are expected to provide more specific details, e.g. 
%%
%%SC was responsible for writing and submitting the manuscript.
%%WWM came up with the initial research concept and edited the manuscript.
%%OTS obtained the funding and edited the manuscript.
%%EBF provided the formal analysis and validation. He also edited the manuscript.
%%GEH Supervised the undergraduates, wrote the software and administers the project github and Zenodo repositories.
%%
%% Authors can use the Contributor Role Taxonomy (CRediT) at
%% for ideas on how write a good statement tailored to their needs.

\end{contribution}

% \facilities{HST(STIS), Swift(XRT and UVOT), AAVSO, CTIO:1.3m, CTIO:1.5m, CXO}

% \software{astropy \citep{2013A&A...558A..33A,2018AJ....156..123A,2022ApJ...935..167A},  
%           Cloudy \citep{2013RMxAA..49..137F}, 
%           Source Extractor \citep{1996A&AS..117..393B}
%           }


\bibliography{DebrisDisks}{}
\bibliographystyle{aasjournal}


% \clearpage

\appendix

\section{Star-by-Star Plots}\label{append:A}

This appendix presents the individual star-by-star plots corresponding to Figures \ref{fig:Total_heatplots} \ref{fig:radiant} \& \ref{fig:vel_e} that will appear as Figure Sets in the final publication. While the figures in the main body of the paper provide a view of all the systems combined, the plots here illustrate the same quantities on a system-by-system basis for a clear and comprehensive record of the analysis for readers who wish to examine each system in isolation.

% \makeappendixsection{Arrival and Travel Heatplots for All Systems}{append:A:heat}

\subsection{Arrival and Travel Heatplots for All Systems}\label{append:A:heat}

% Helper macro for consistent formatting
\newcommand{\appendixHeatplot}[3]{%
  \begin{figure*}[htb]
    \centering
    \includegraphics[width=\textwidth]{figs_Catalogue/fig01set/#1}
    \caption{Two-panel plot for #2: The left figure shows the ejection velocities vs.\ the time spent traveling in the ISM of the simulated particles that enter our Oort Cloud. The right figure shows the arrival velocity vs.\ arrival time of the simulated particles that enter our Oort Cloud.}
    \label{fig:heatplots:#3}
  \end{figure*}
}

% ------- Individual systems -------
\appendixHeatplot{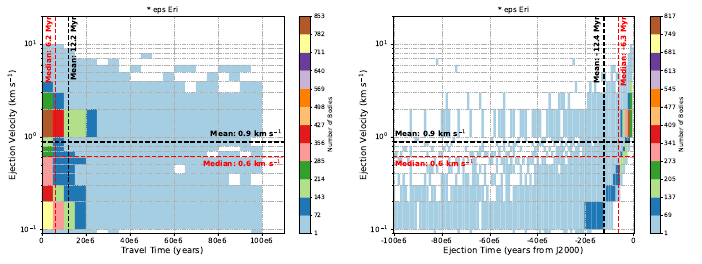}{\textit{* eps Eri}}{star-eps-Eri}
\appendixHeatplot{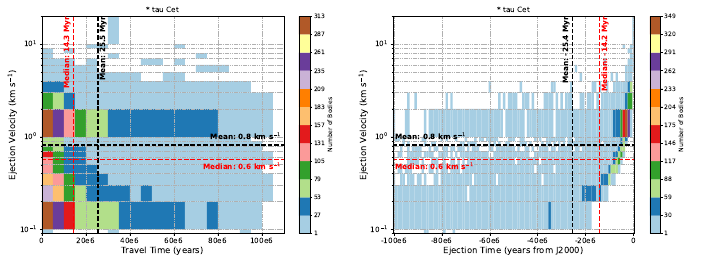}{\textit{* tau Cet}}{star-tau-Cet}
\appendixHeatplot{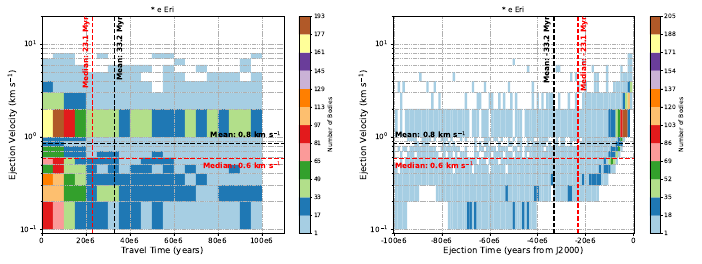}{\textit{* e Eri}}{star-e-Eri}
\appendixHeatplot{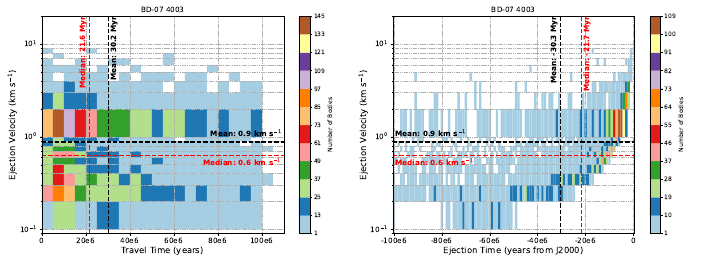}{\textit{BD-07 4003}}{BD-07-4003}
\appendixHeatplot{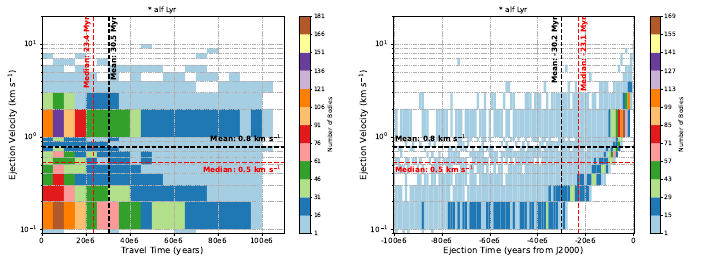}{\textit{* alf Lyr}}{star-alf-Lyr}
\appendixHeatplot{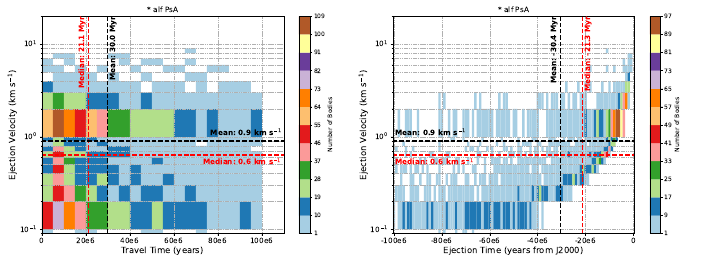}{\textit{* alf PsA}}{star-alf-PsA}
\appendixHeatplot{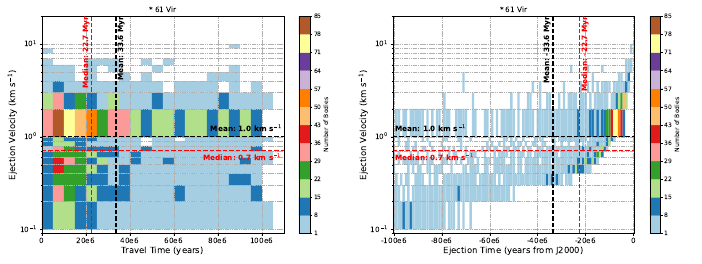}{\textit{* 61 Vir}}{star-61-Vir}
\appendixHeatplot{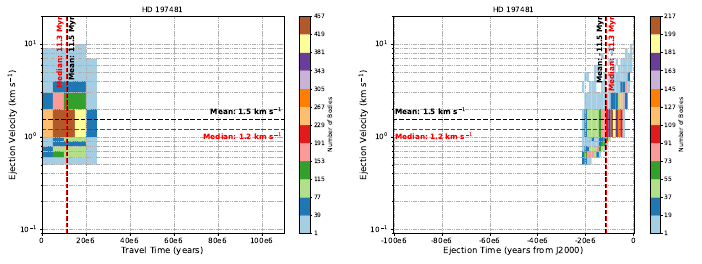}{\textit{HD 197481}}{HD-197481}
\appendixHeatplot{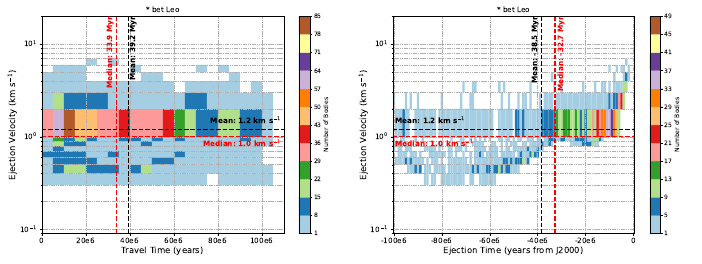}{\textit{* bet Leo}}{star-bet-Leo}
\appendixHeatplot{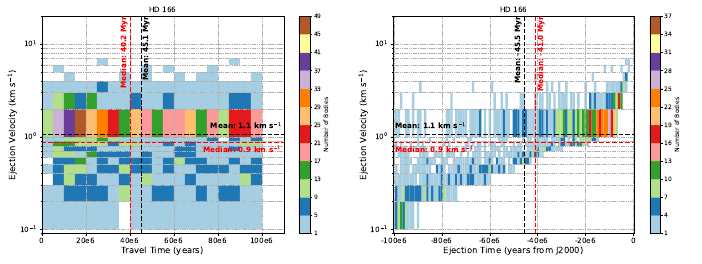}{\textit{HD 166}}{HD-166}
\appendixHeatplot{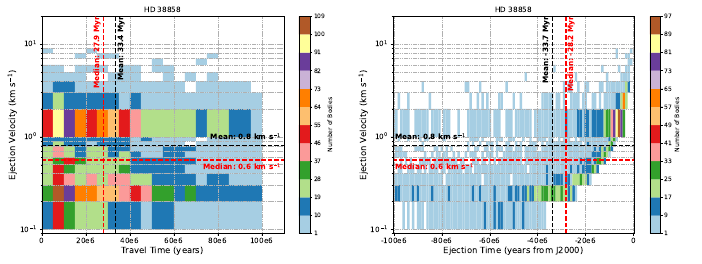}{\textit{HD 38858}}{HD-38858}
\appendixHeatplot{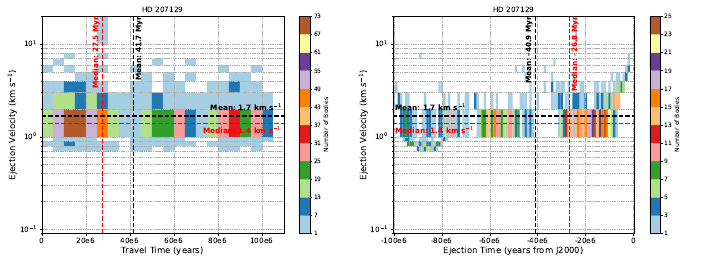}{\textit{HD 207129}}{HD-207129}
\appendixHeatplot{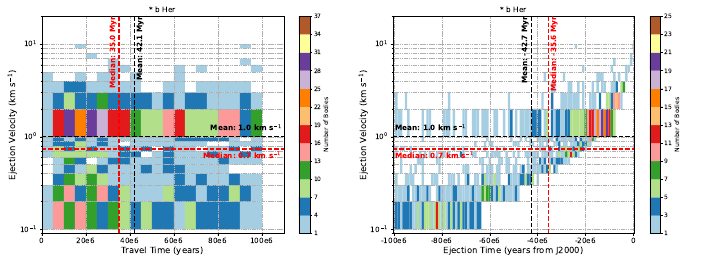}{\textit{* b Her}}{star-b-Her}
\appendixHeatplot{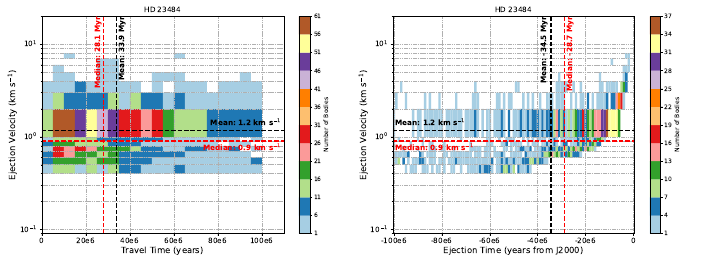}{\textit{HD 23484}}{HD-23484}
\appendixHeatplot{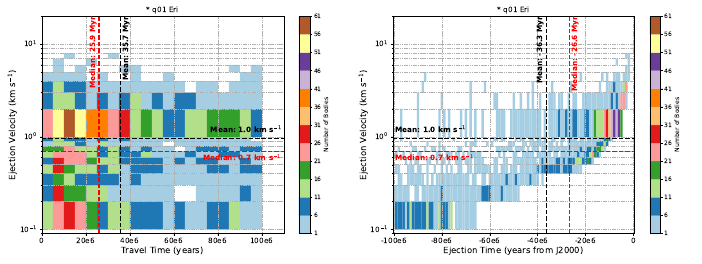}{\textit{* q01 Eri}}{star-q01-Eri}
\appendixHeatplot{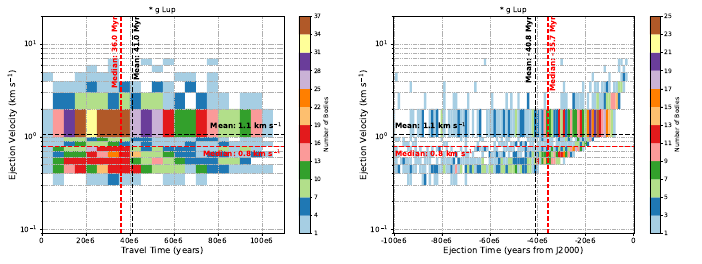}{\textit{* g Lup}}{star-g-Lup}
\appendixHeatplot{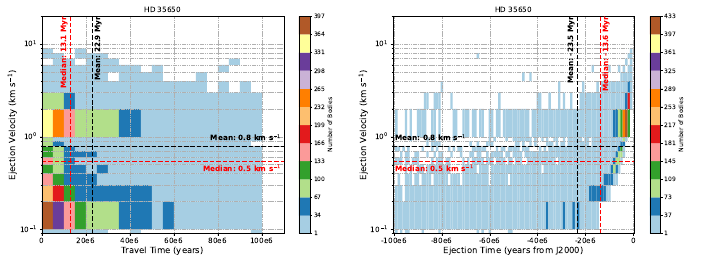}{\textit{HD 35650}}{HD-35650}
\appendixHeatplot{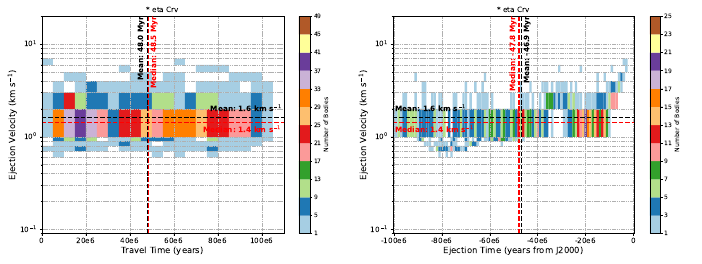}{\textit{* eta Crv}}{star-eta-Crv}
\appendixHeatplot{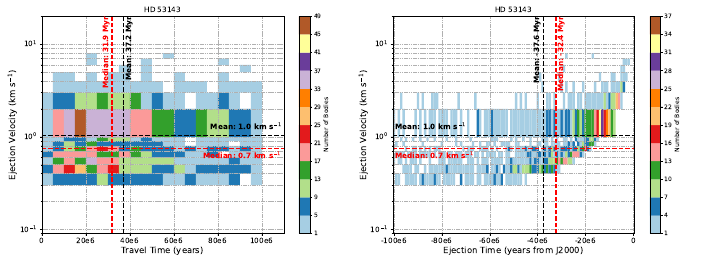}{\textit{HD 53143}}{HD-53143}
\appendixHeatplot{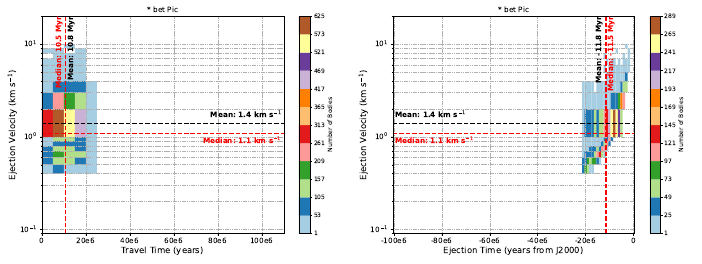}{\textit{* bet Pic}}{star-bet-Pic}

\clearpage

% \makeappendixsection{Radiant Heatmaps for All Systems}{append:A:Rad}
\subsection{Radiant Heatmaps for All Systems}\label{append:A:Rad}

% Helper macro for consistent formatting
\newcommand{\appendixRadiant}[3]{%
  \begin{figure*}[htb]
  % \begin{sidewaysfigure}[p]
    \centering
    \includegraphics[width=\textwidth]{figs_Catalogue/fig02set/#1}
    \caption{The heliocentric equatorial radiant for the close approaches from #2 at the time of their closest Solar approach (``Arrival Time''), with the current ``effective radiant'' of #2 corresponding to its apparent velocity relative to the Sun. The arrival directions of the three confirmed km-scale interstellar objects (ISOs), the direction of the local interstellar cloud (LIC) upstream direction of interstellar dust (ISD), the Solar apex with respect to the Local Standard of Rest (LSR) and within the Milky Way are also included, along with the direction towards the Galactic Center (see Table \ref{tab:ISO}). The heat density is on a log-scale for the number of simulated particles transferred to the Solar System.}
    \label{fig:radiant:#3}
  \end{figure*}
  % \end{sidewaysfigure}
}

% ------- Individual systems -------
\appendixRadiant{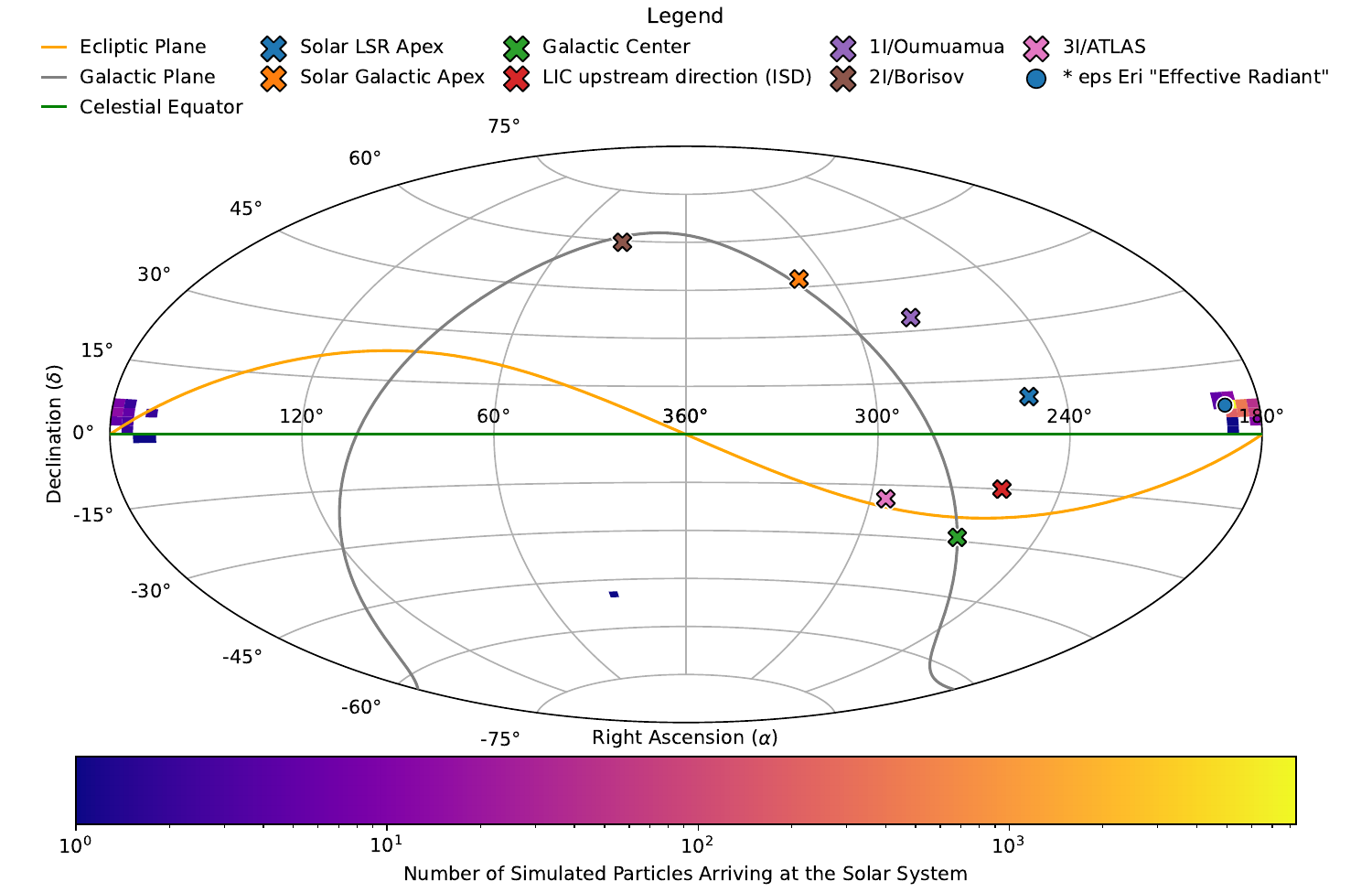}{\textit{* eps Eri}}{star-eps-Eri}
\appendixRadiant{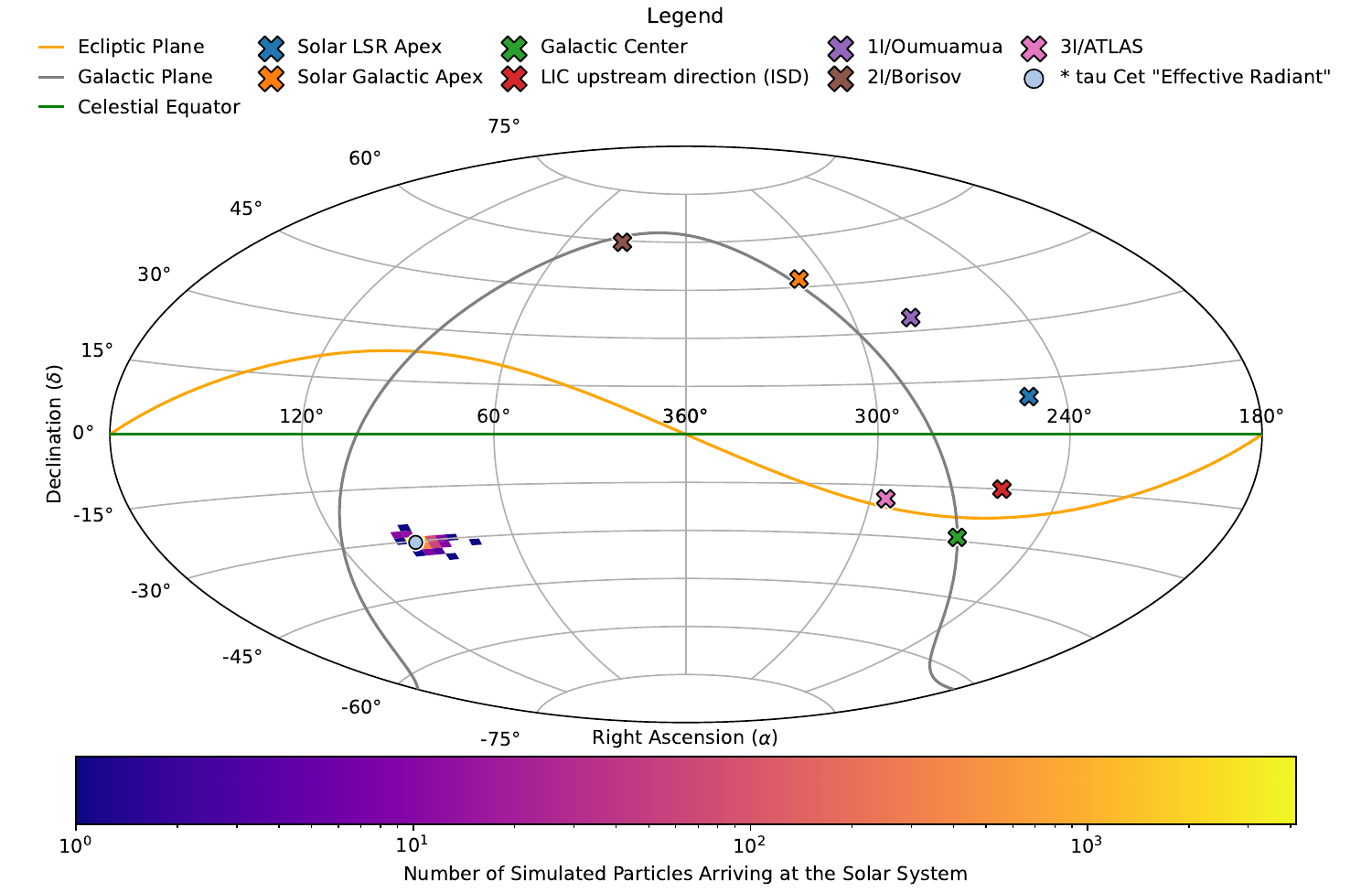}{\textit{* tau Cet}}{star-tau-Cet}
\appendixRadiant{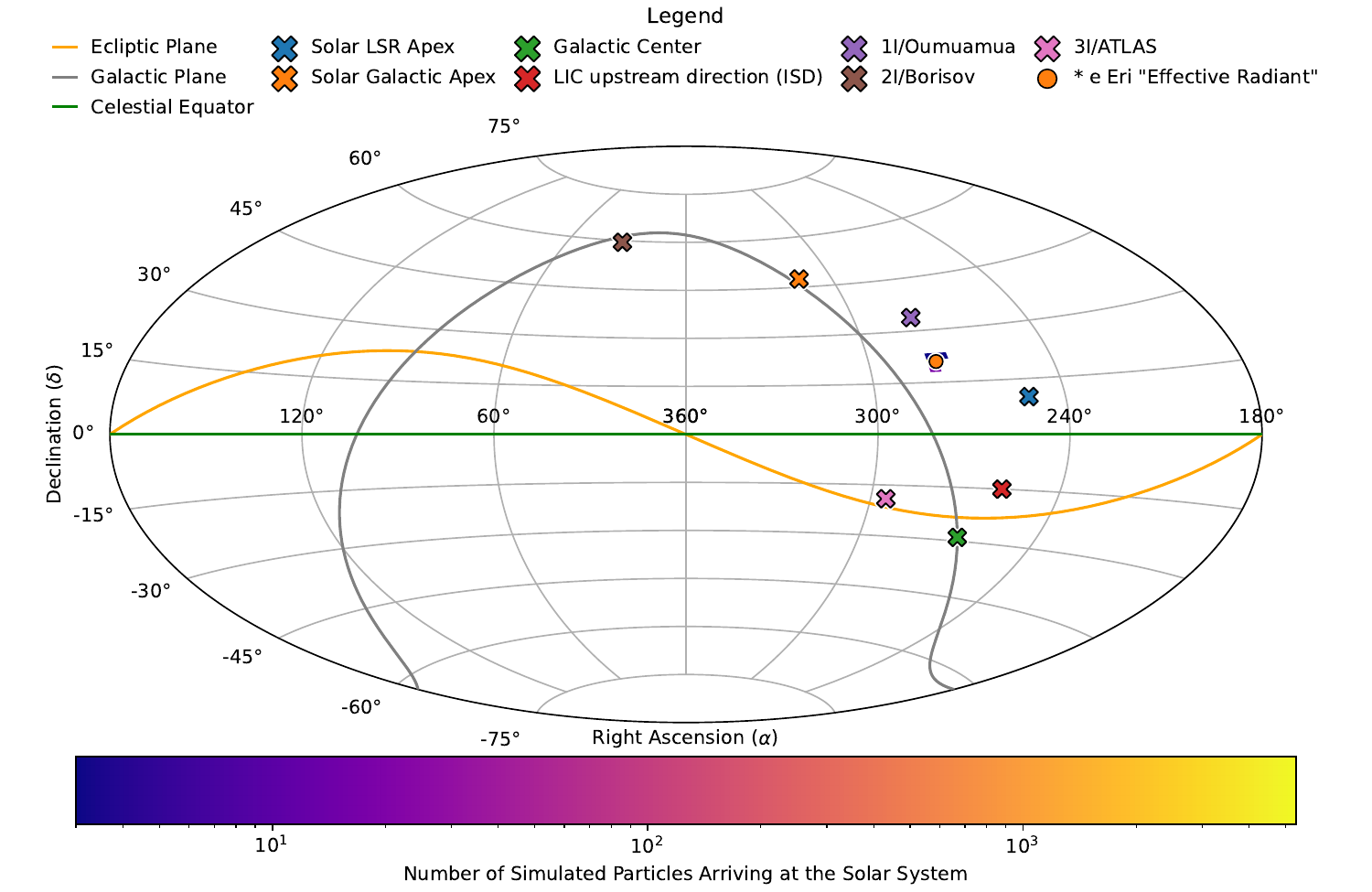}{\textit{* e Eri}}{star-e-Eri}
\appendixRadiant{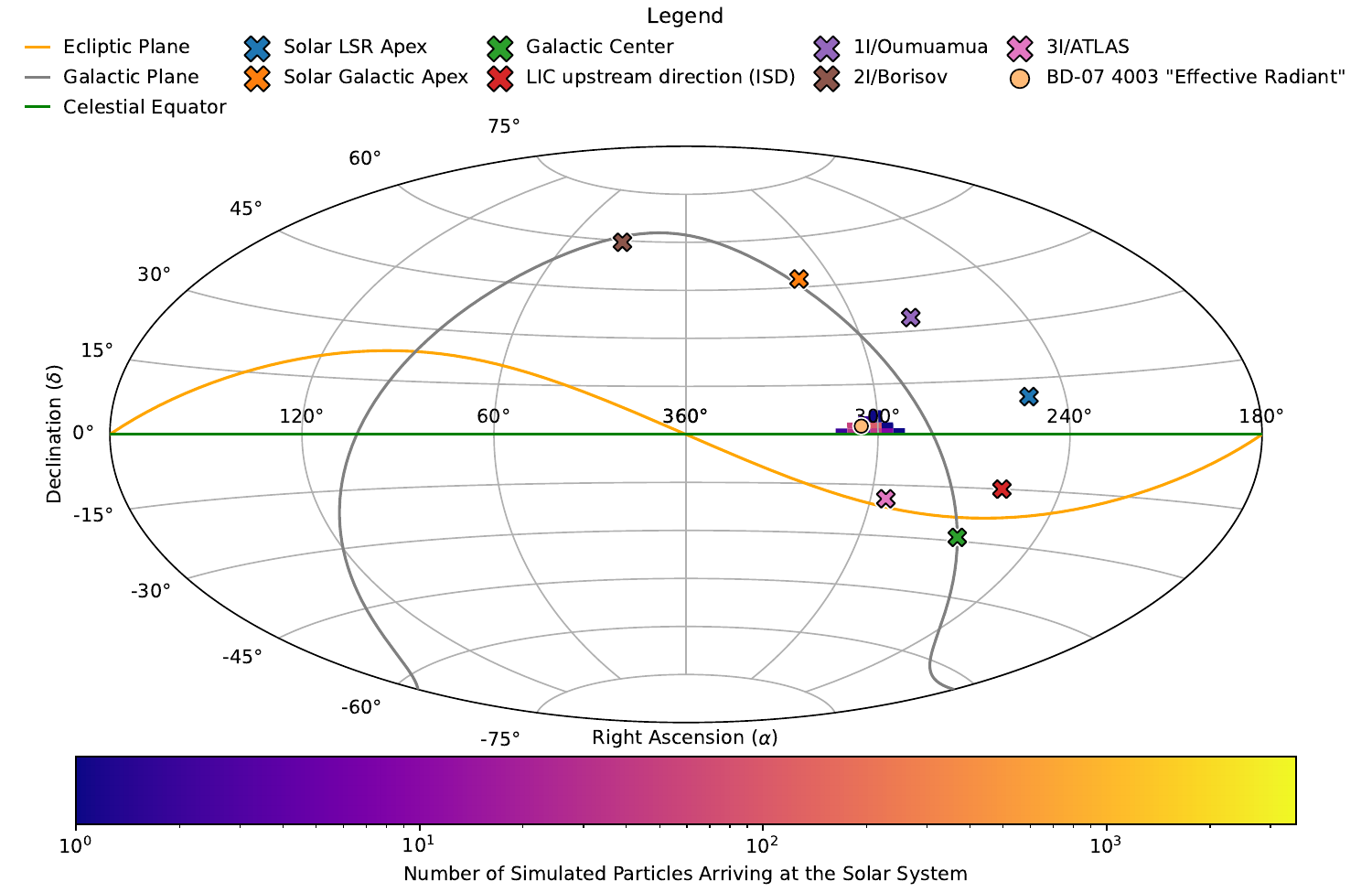}{\textit{BD-07 4003}}{BD-07-4003}
\appendixRadiant{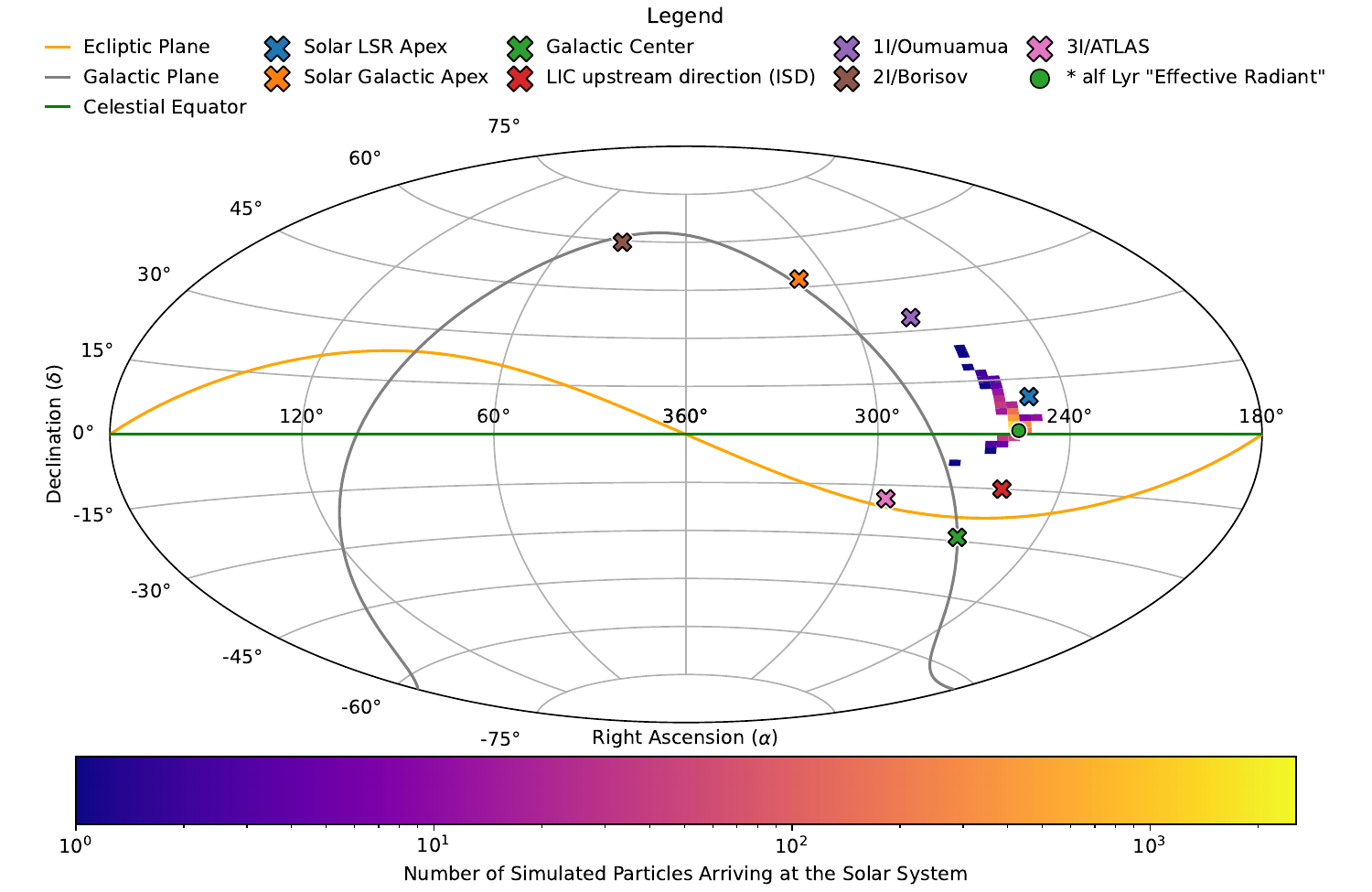}{\textit{* alf Lyr}}{star-alf-Lyr}
\appendixRadiant{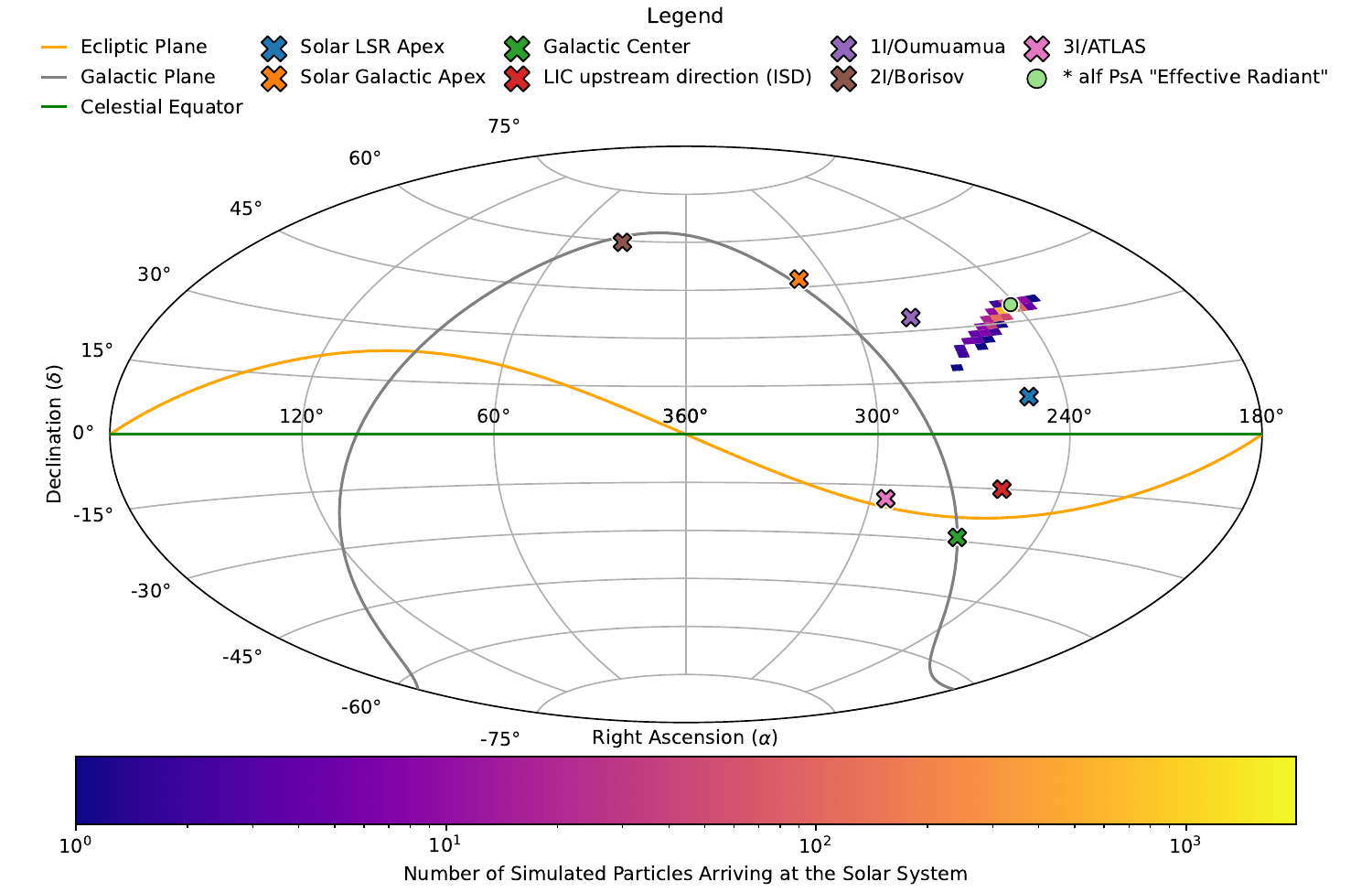}{\textit{* alf PsA}}{star-alf-PsA}
\appendixRadiant{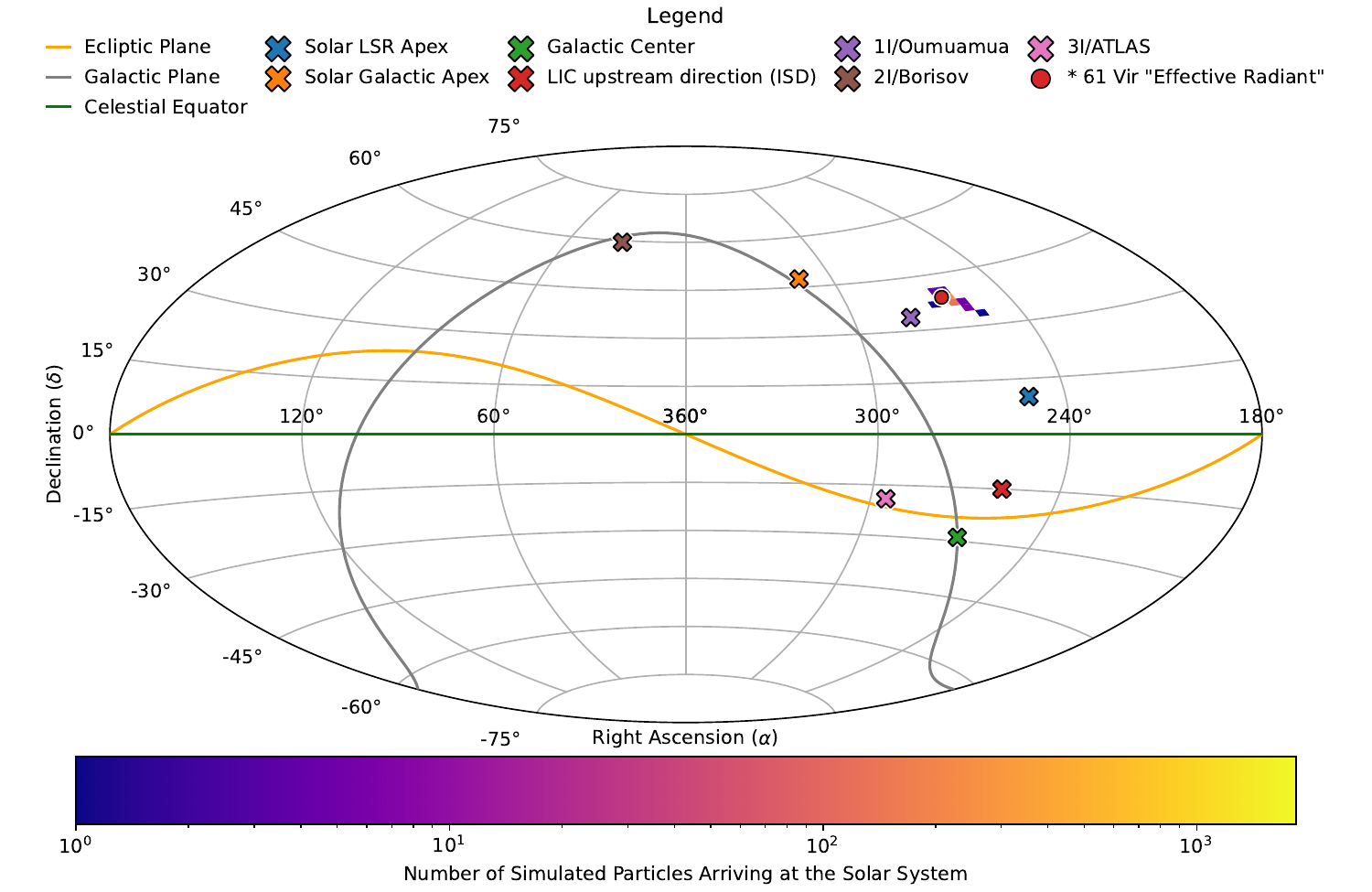}{\textit{* 61 Vir}}{star-61-Vir}
\appendixRadiant{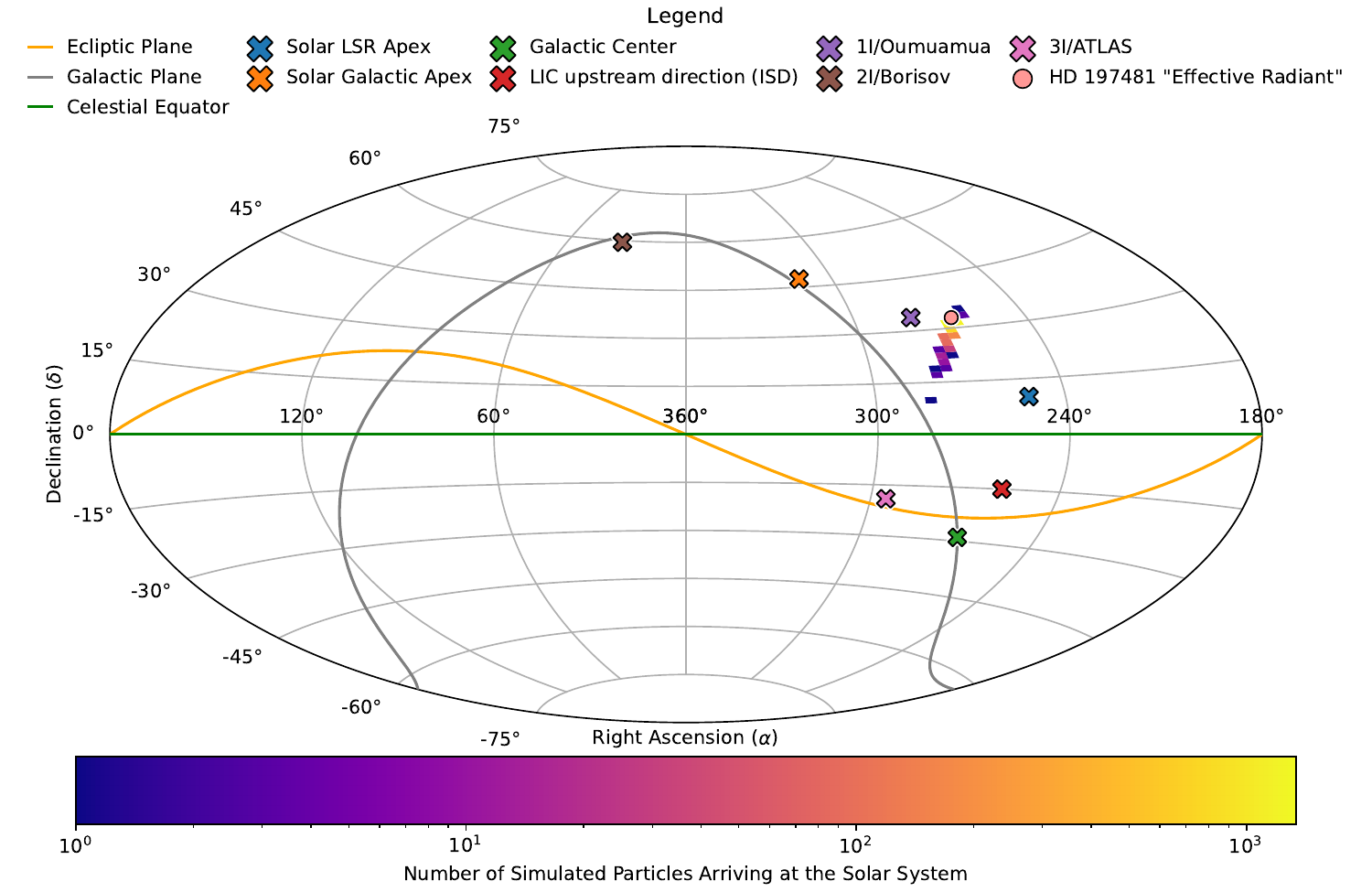}{\textit{HD 197481}}{HD-197481}
\appendixRadiant{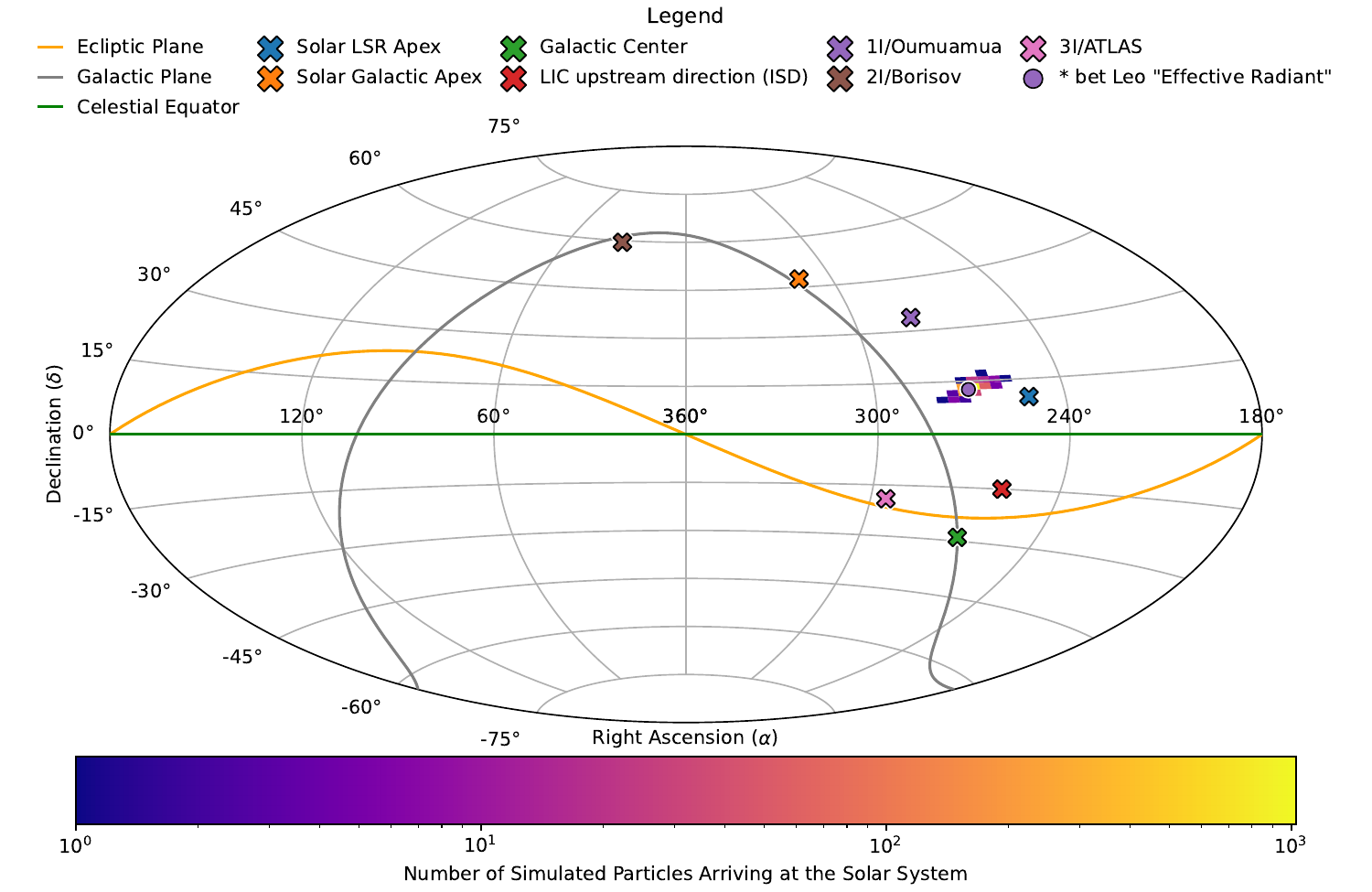}{\textit{* bet Leo}}{star-bet-Leo}
\appendixRadiant{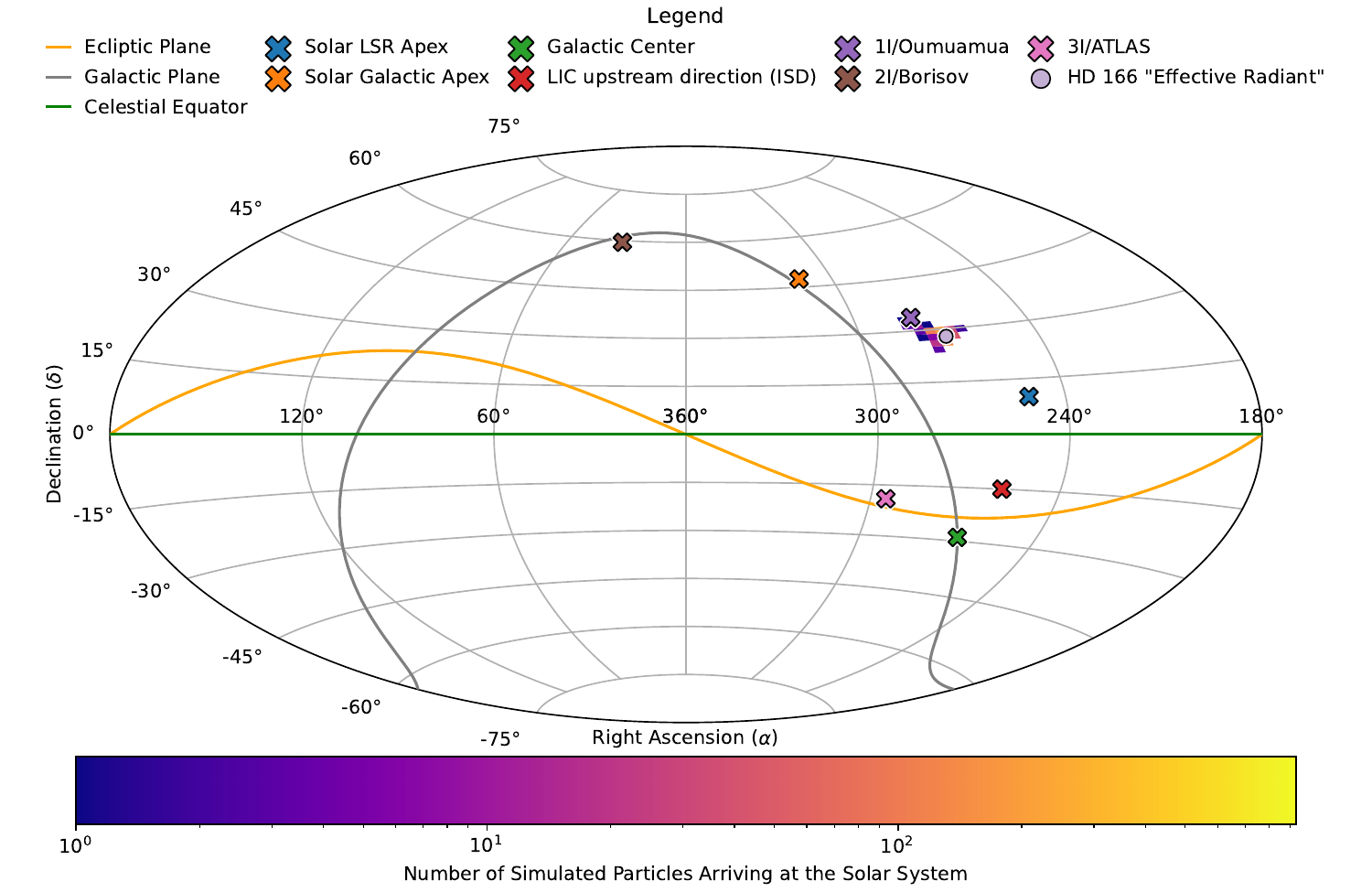}{\textit{HD 166}}{HD-166}
\appendixRadiant{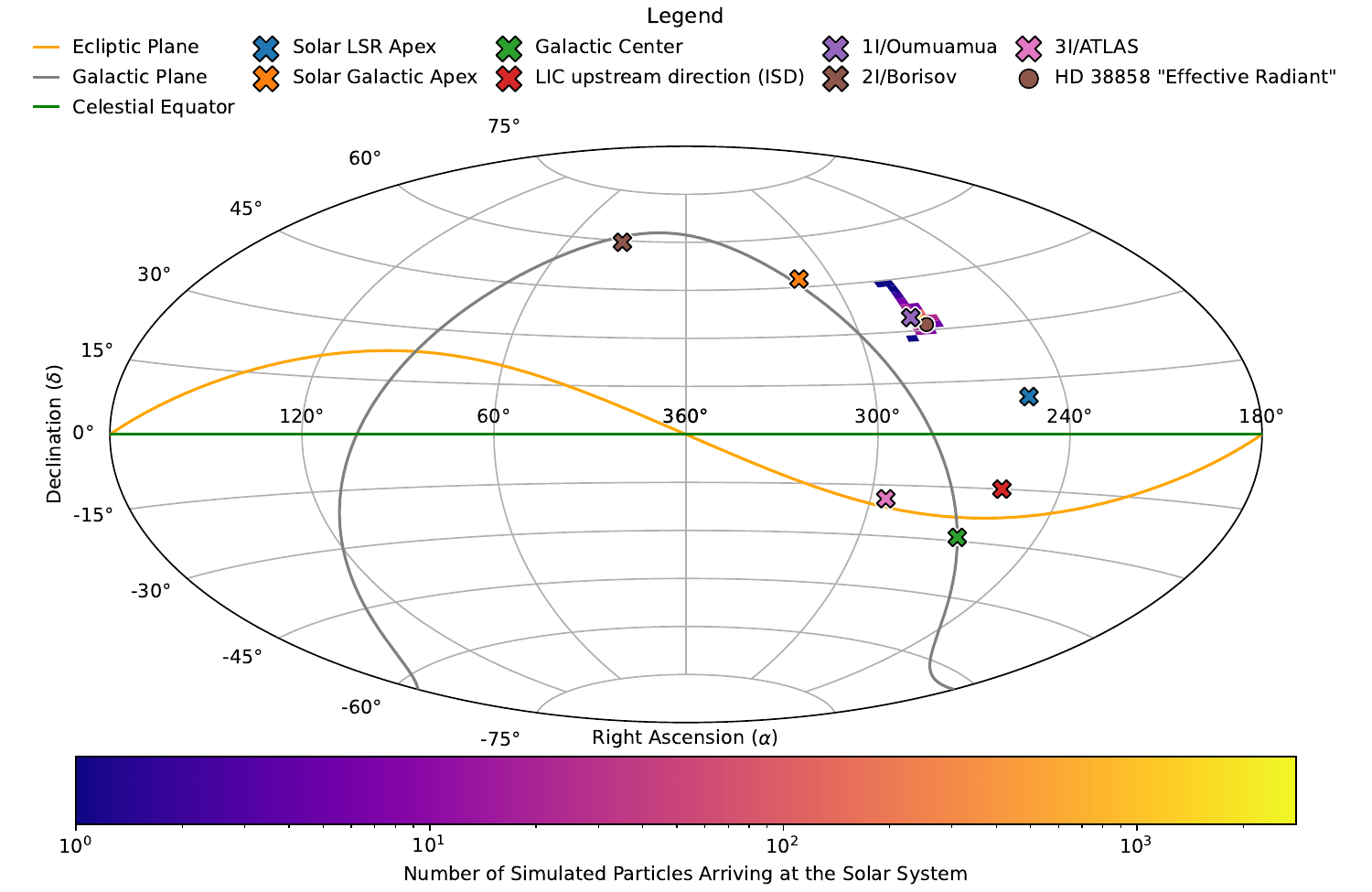}{\textit{HD 38858}}{HD-38858}
\appendixRadiant{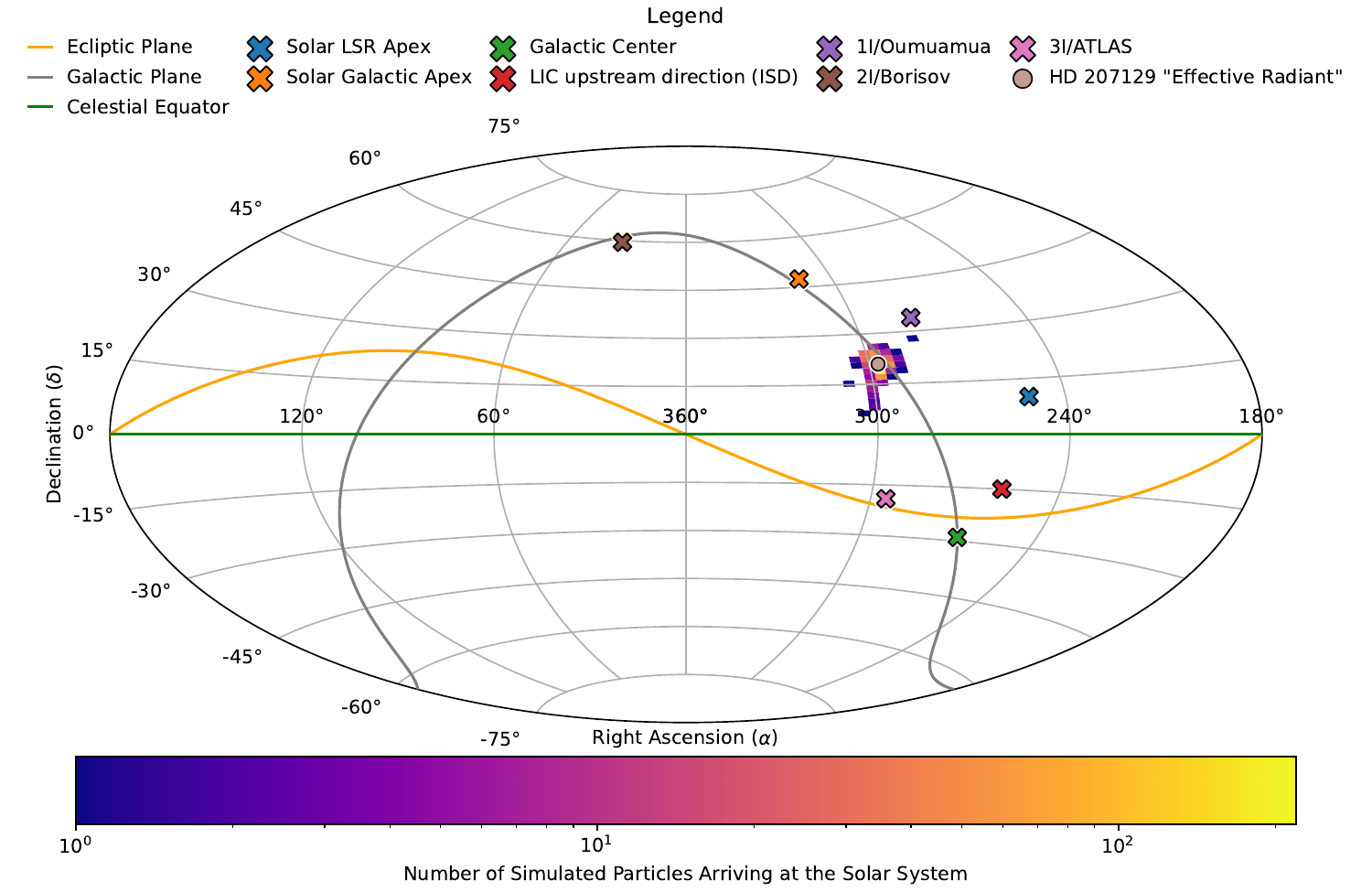}{\textit{HD 207129}}{HD-207129}
\appendixRadiant{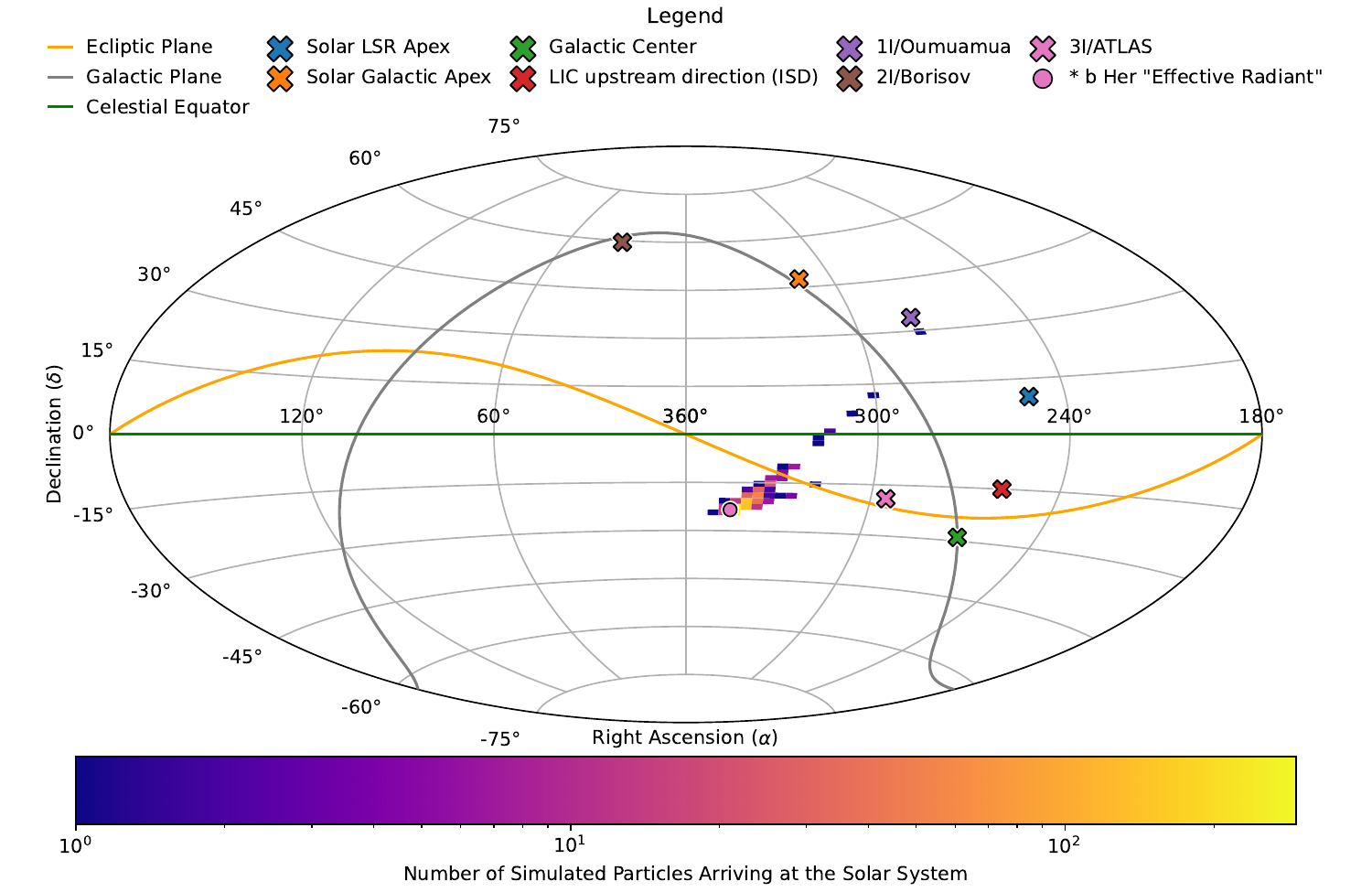}{\textit{* b Her}}{star-b-Her}
\appendixRadiant{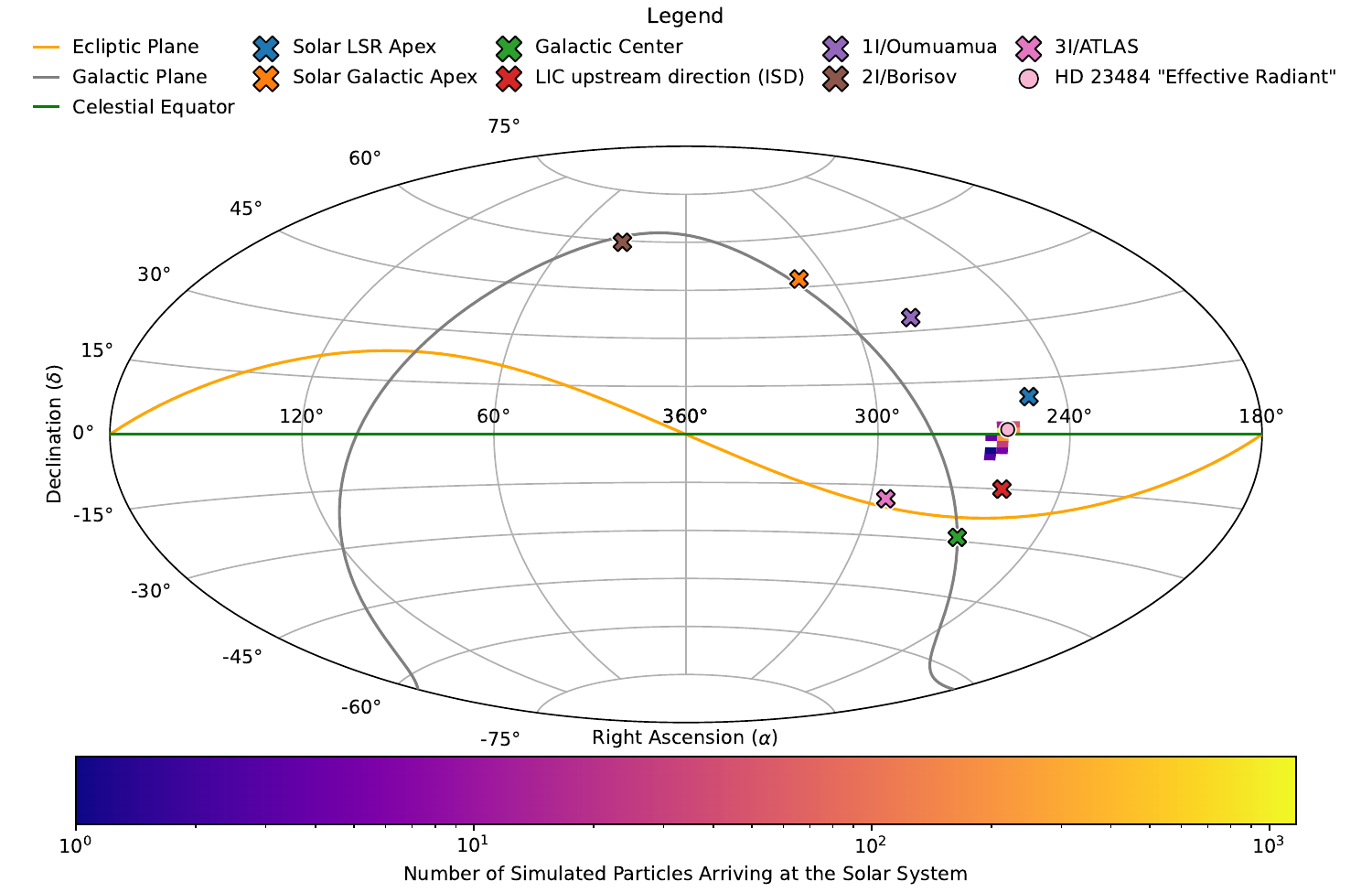}{\textit{HD 23484}}{HD-23484}
\appendixRadiant{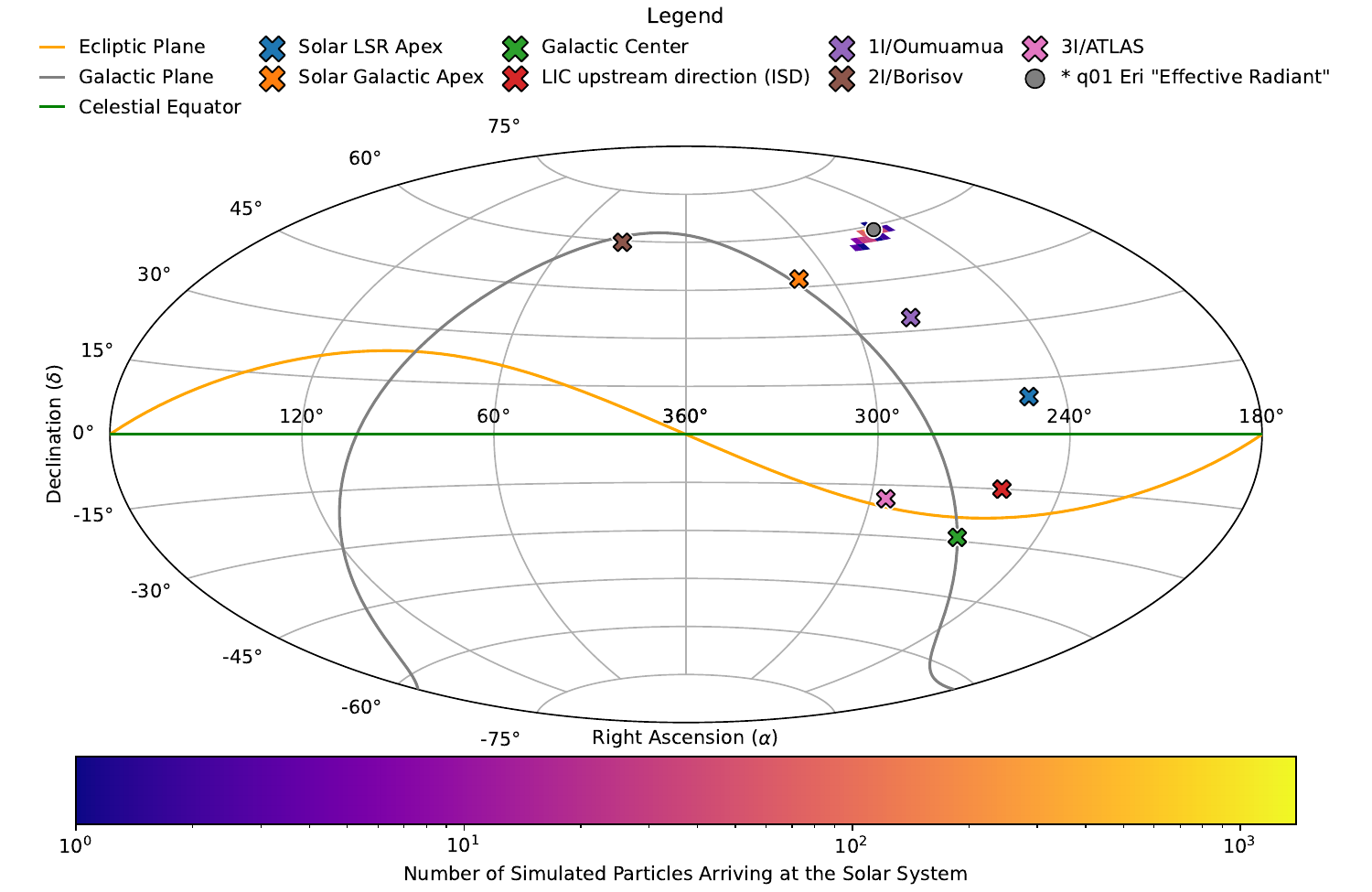}{\textit{* q01 Eri}}{star-q01-Eri}
\appendixRadiant{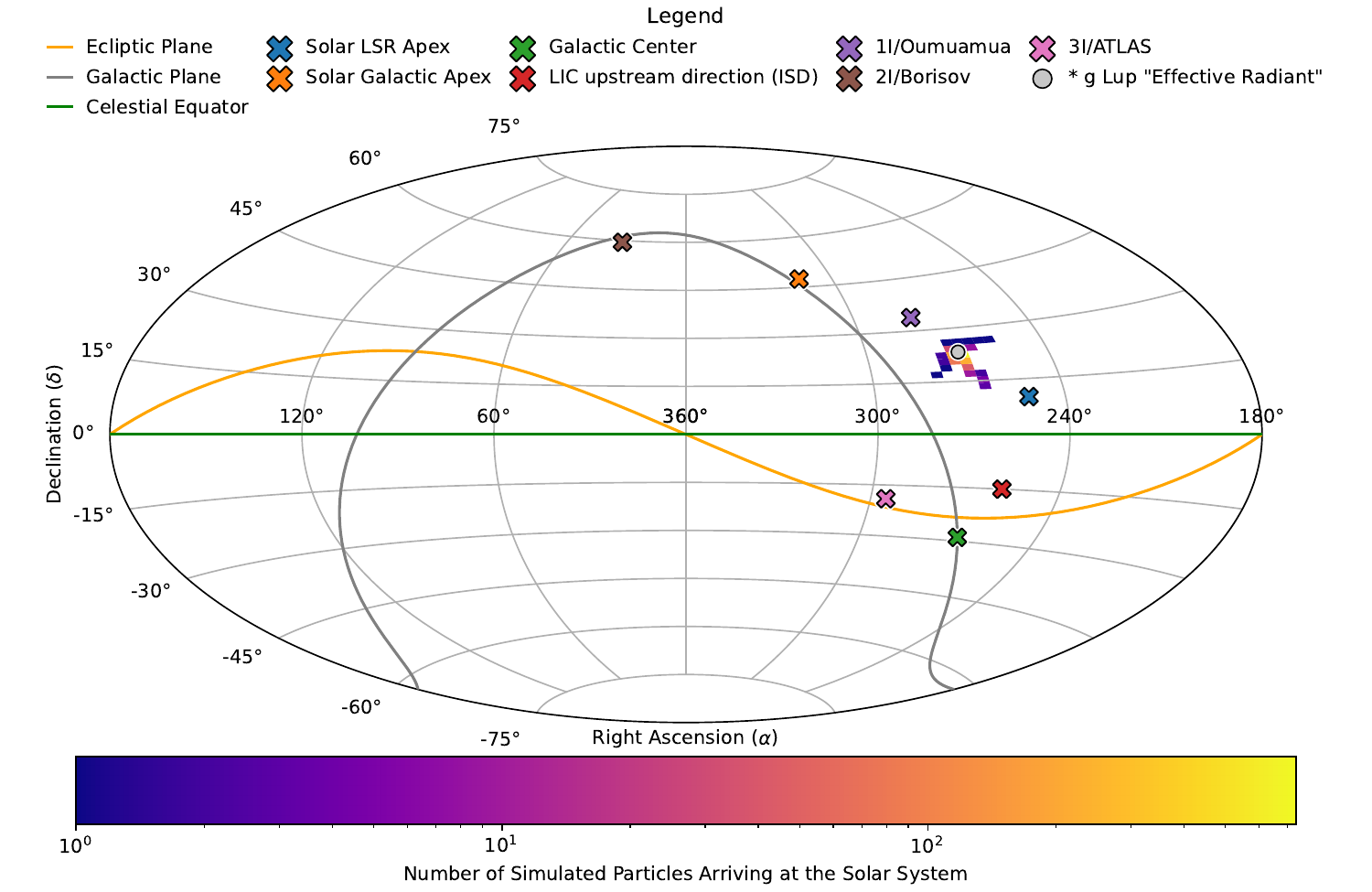}{\textit{* g Lup}}{star-g-Lup}
\appendixRadiant{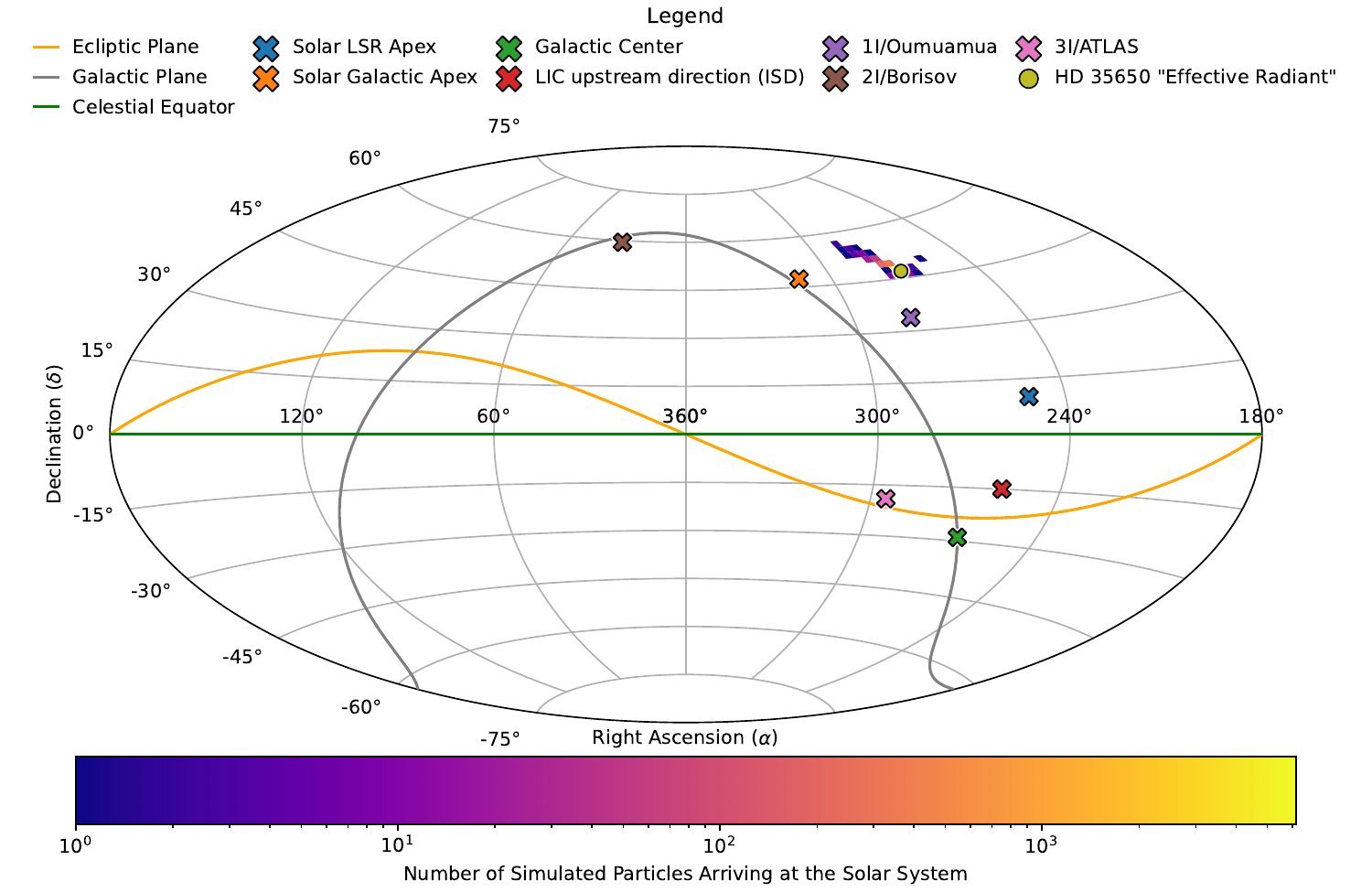}{\textit{HD 35650}}{HD-35650}
\appendixRadiant{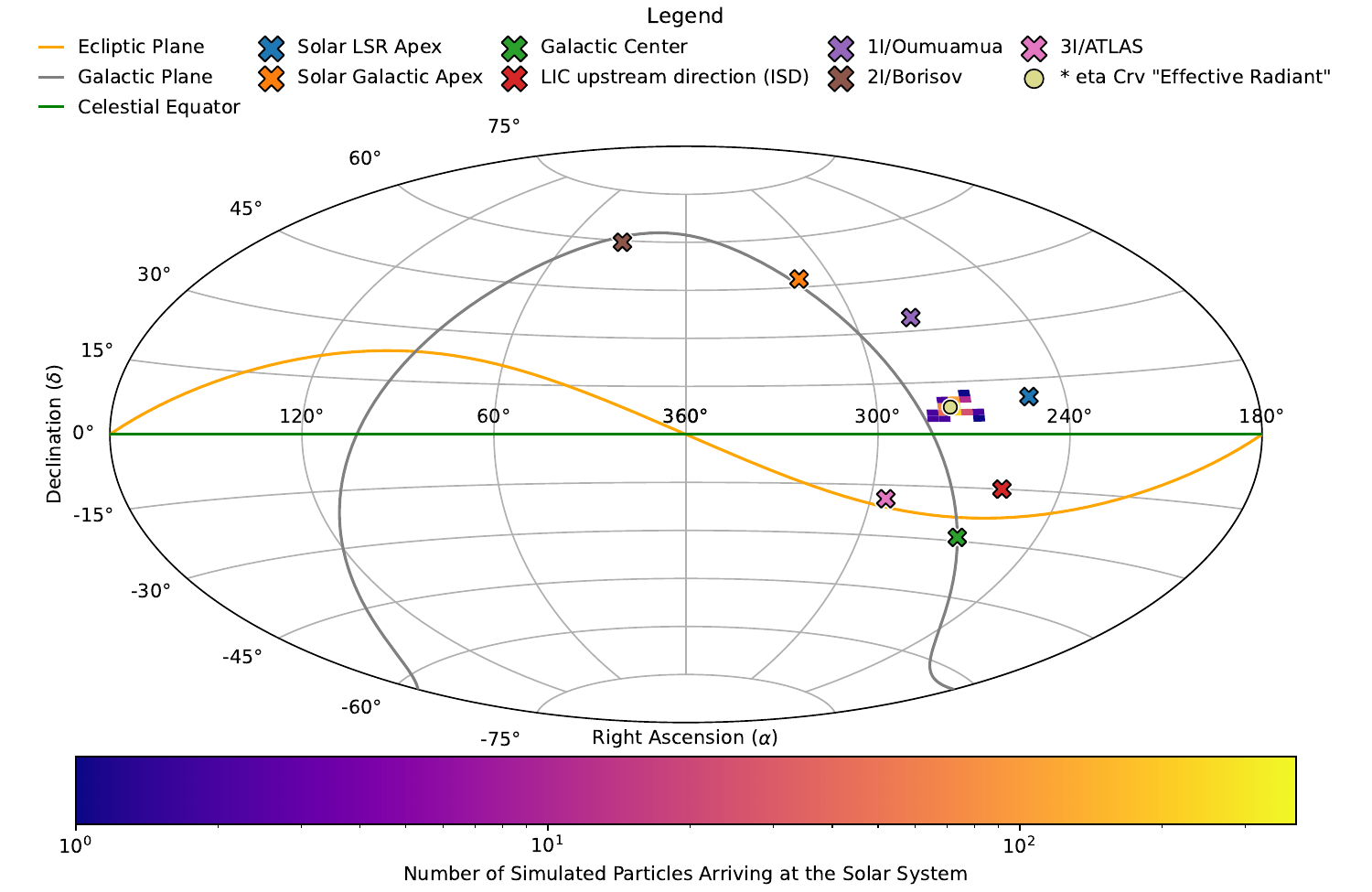}{\textit{* eta Crv}}{star-eta-Crv}
\appendixRadiant{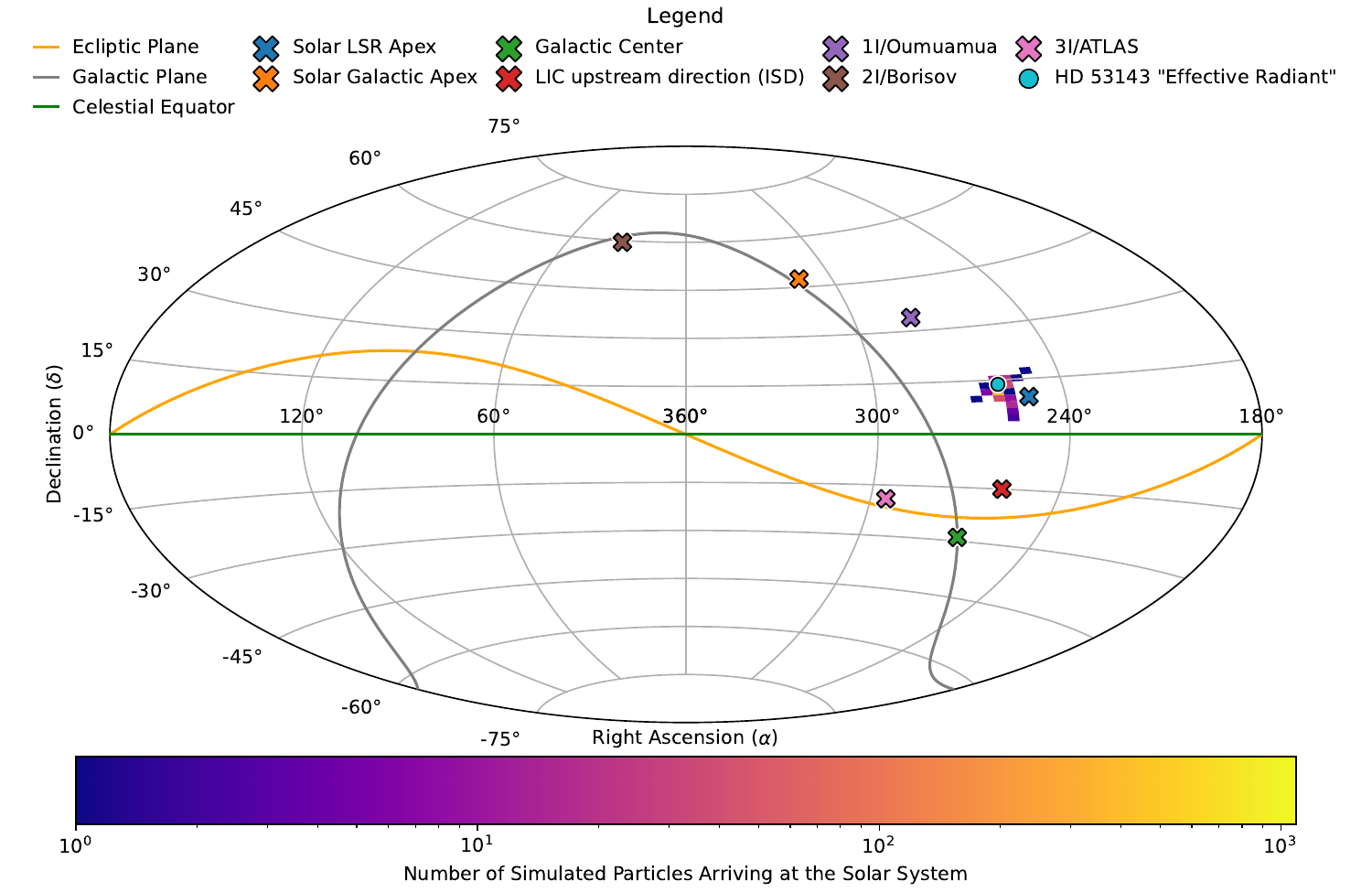}{\textit{HD 53143}}{HD-53143}
\appendixRadiant{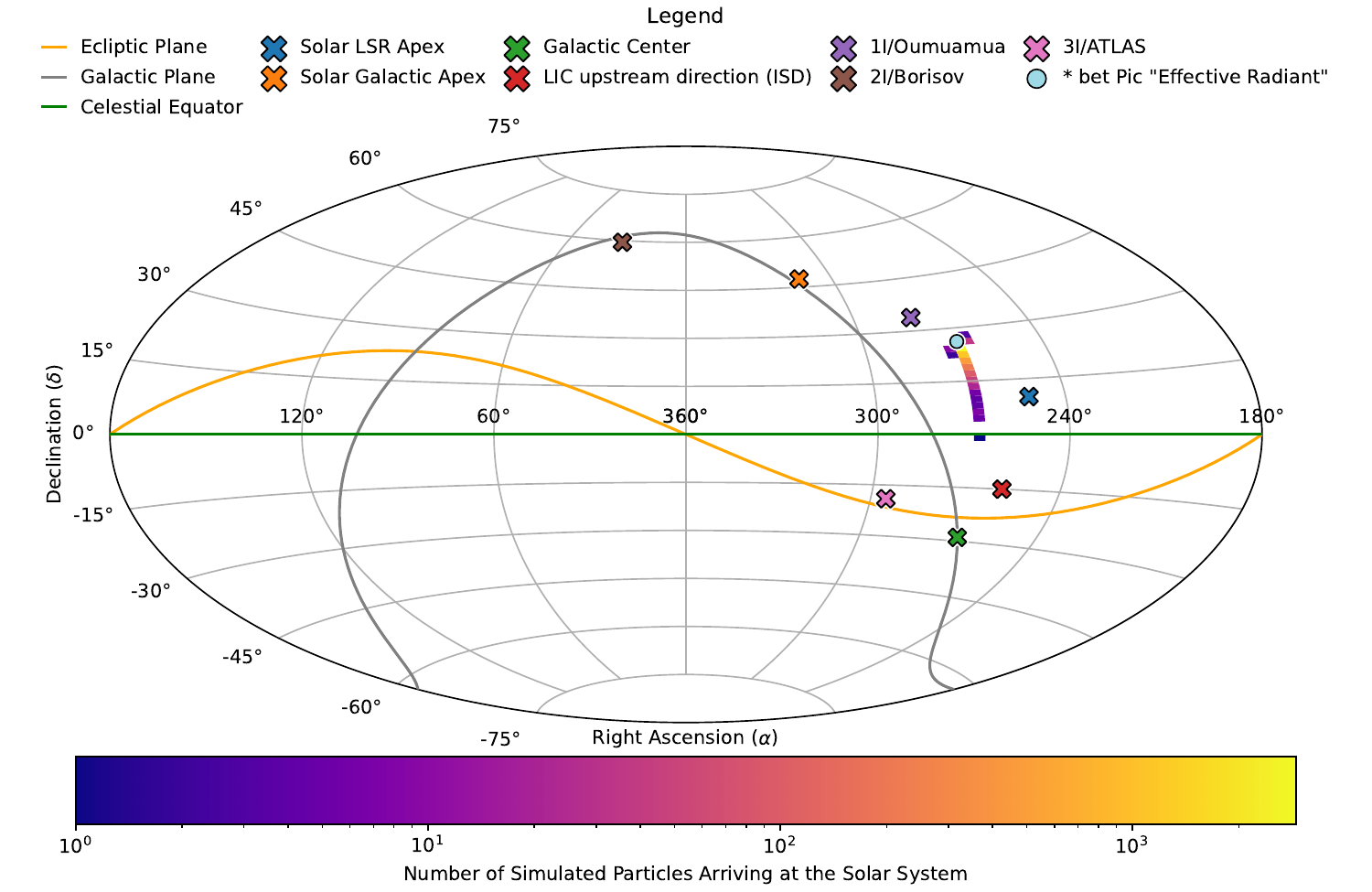}{\textit{* bet Pic}}{star-bet-Pic}

\clearpage

\subsection{Heliocentric Velocity and Eccentricity Plots for All Systems}\label{append:A:vel_e}
% \makeappendixsection{Heliocentric Velocity and Eccentricity Plots for All Systems}{append:A:vel_e}

% Helper macro for consistent formatting
\newcommand{\appendixVelE}[3]{%
  \begin{figure*}[htb]
    \centering
    \includegraphics[width=\textwidth]{figs_Catalogue/fig05set/#1}
    \caption{Two-panel plot for #2: These are histograms of all simulated particles arriving at the Solar System, weighted according to their contribution to the flux at Earth (see Section~\ref{sec:flux}) but with no size dependence. Heliocentric velocity is computed at 1 au ($v_{hel,1au}$) and eccentricities ($e$) with an assumed perihelion at 1 au. Note the log scale on the y-axis.}
    \label{fig:velE:#3}
  \end{figure*}
}

% ------- Individual systems -------
\appendixVelE{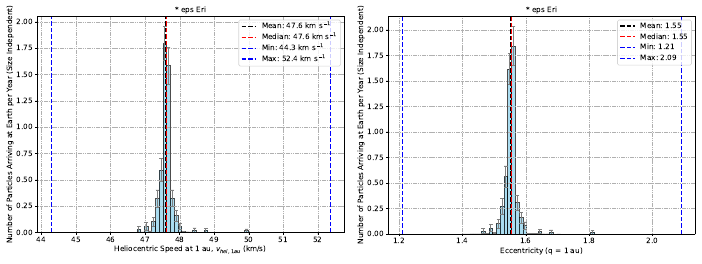}{\textit{* eps Eri}}{star-eps-Eri}
\appendixVelE{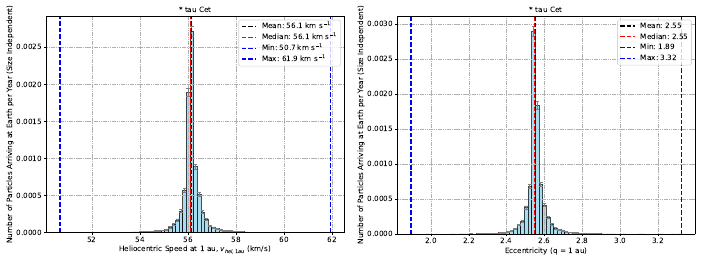}{\textit{* tau Cet}}{star-tau-Cet}
\appendixVelE{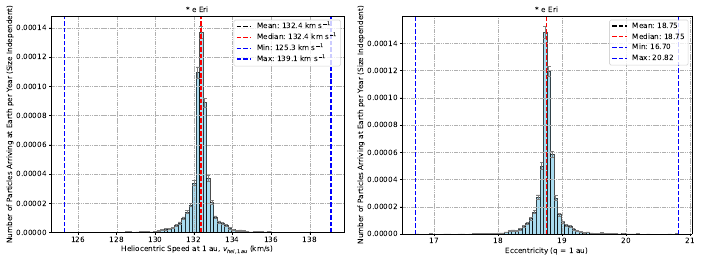}{\textit{* e Eri}}{star-e-Eri}
\appendixVelE{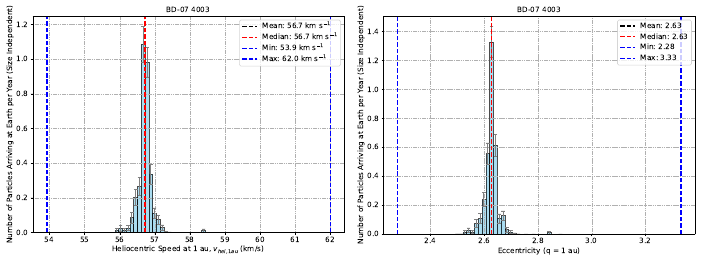}{\textit{BD-07 4003}}{BD-07-4003}
\appendixVelE{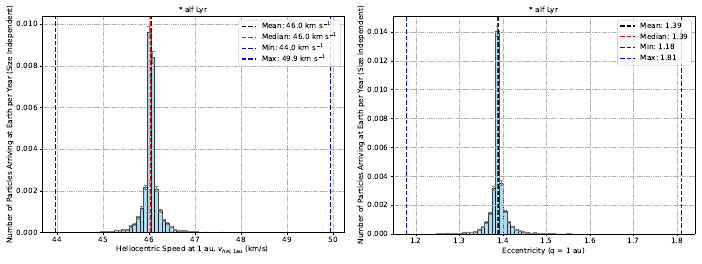}{\textit{* alf Lyr}}{star-alf-Lyr}
\appendixVelE{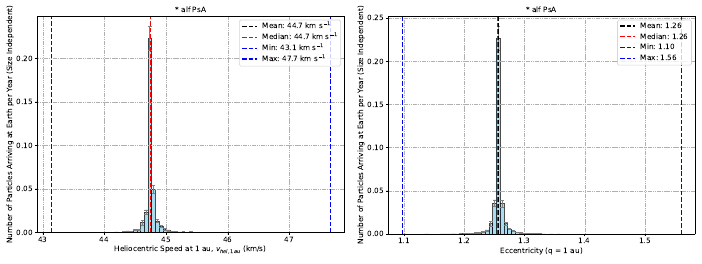}{\textit{* alf PsA}}{star-alf-PsA}
\appendixVelE{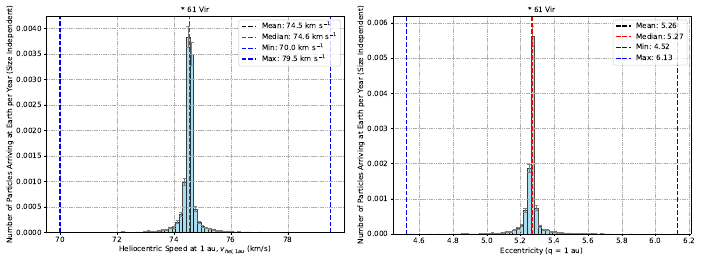}{\textit{* 61 Vir}}{star-61-Vir}
\appendixVelE{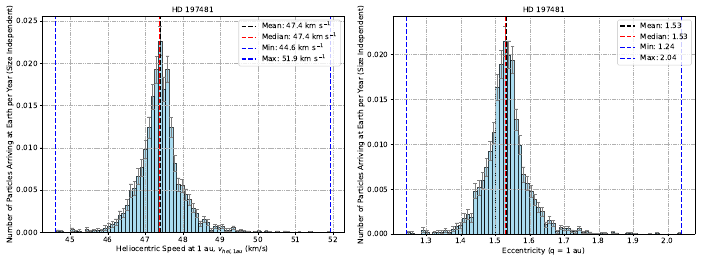}{\textit{HD 197481}}{HD-197481}
\appendixVelE{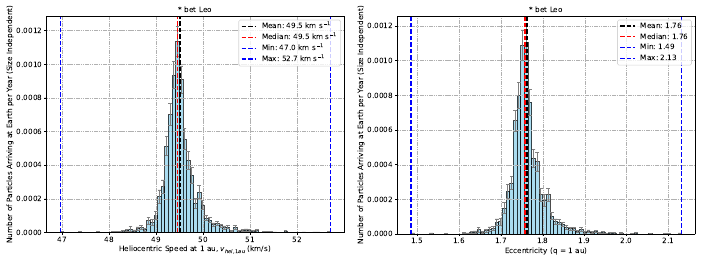}{\textit{* bet Leo}}{star-bet-Leo}
\appendixVelE{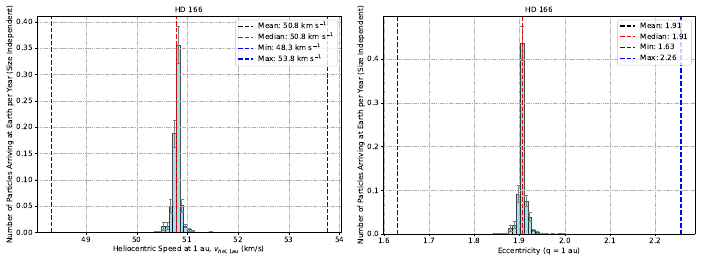}{\textit{HD 166}}{HD-166}
\appendixVelE{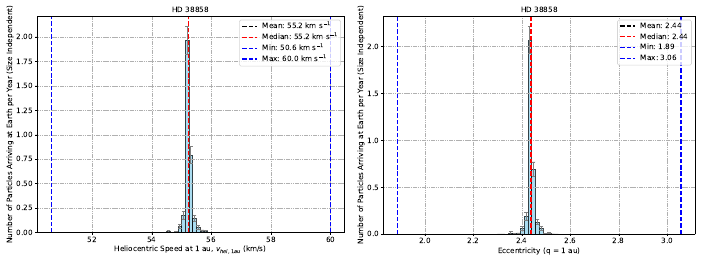}{\textit{HD 38858}}{HD-38858}
\appendixVelE{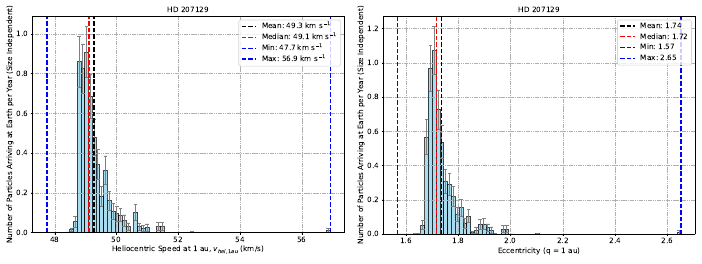}{\textit{HD 207129}}{HD-207129}
\appendixVelE{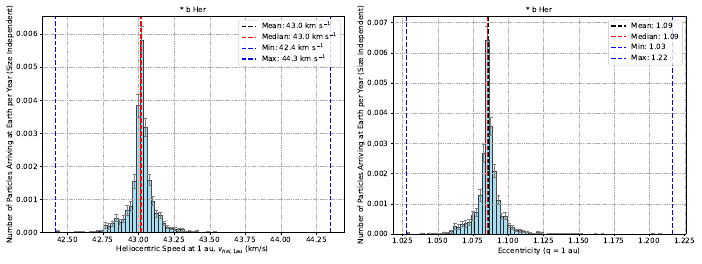}{\textit{* b Her}}{star-b-Her}
\appendixVelE{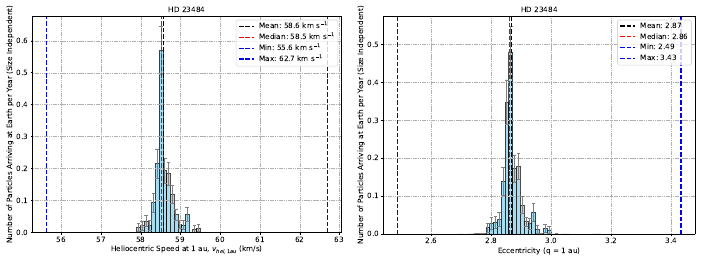}{\textit{HD 23484}}{HD-23484}
\appendixVelE{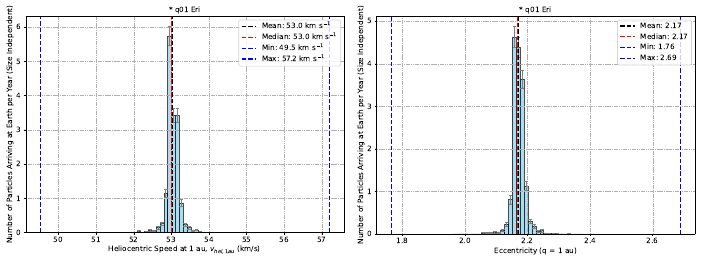}{\textit{* q01 Eri}}{star-q01-Eri}
\appendixVelE{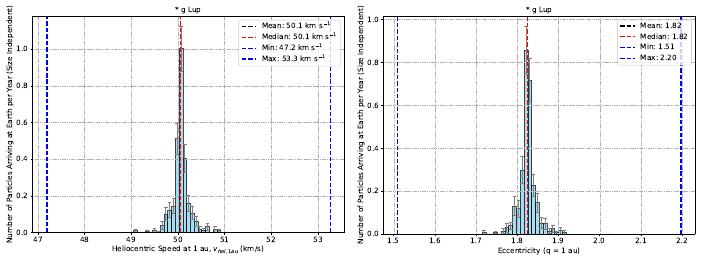}{\textit{* g Lup}}{star-g-Lup}
\appendixVelE{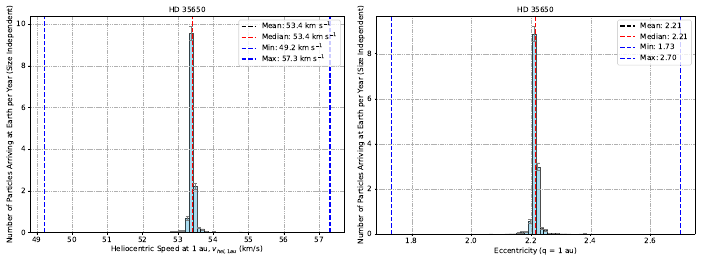}{\textit{HD 35650}}{HD-35650}
\appendixVelE{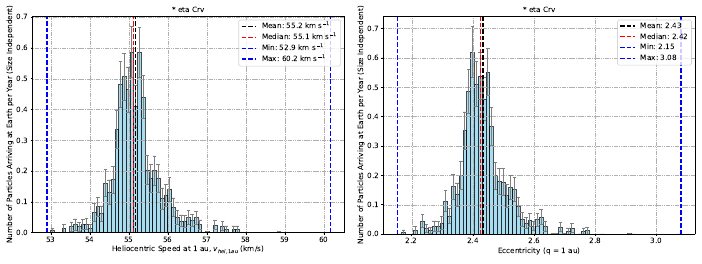}{\textit{* eta Crv}}{star-eta-Crv}
\appendixVelE{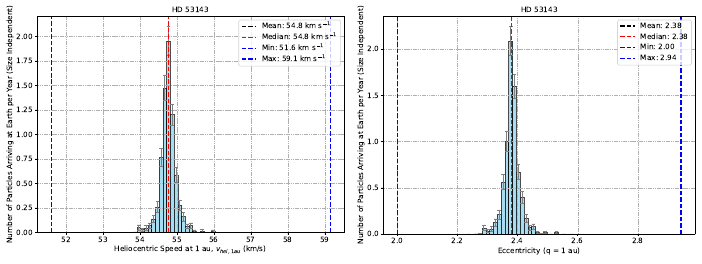}{\textit{HD 53143}}{HD-53143}
\appendixVelE{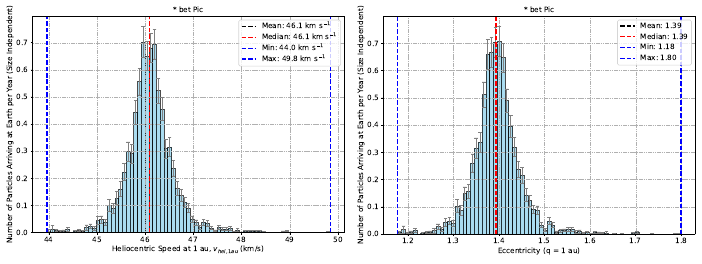}{\textit{* bet Pic}}{star-bet-Pic}

\clearpage

\section{Fluxes/Populations For a Critical Age of 3 Myr}\label{append:tcrit3}

The flux estimates in the main body of this paper assume a critical age ($t_{\mathrm{crit}}$; the age after planet formation where ejections could likely begin) of $10~\mathrm{Myr}$, values provided here are with the assumption $t_{\mathrm{crit}}=3~\mathrm{Myr}$.

\begin{deluxetable*}{cccccc}[htb!]
\tablecaption{Maximum and Current Flux into Earth's Atmosphere for Particles $\geq 200~\mu$m (assuming $t_{crit}=3$ Myr)\label{tab:200micron_ageDepend_tcrit3}}
\tablehead{
\colhead{Star Name} & \colhead{Time at Max} & \colhead{Max Flux} & \colhead{Error} & \colhead{Current Flux} & \colhead{Error} \\
\colhead{} & \colhead{} & \colhead{Earth} & \colhead{} & \colhead{Earth} & \colhead{} \\
\colhead{} & \colhead{(yr)} & \colhead{(yr$^{-1}$)} & \colhead{} & \colhead{(yr$^{-1}$)} & \colhead{}
}
\startdata
\protect* eps Eri & -1.1e+05 & 1.8e+05 & 2.6e+04 & 8.3e+04 & 1.8e+04 \\
\protect* tau Cet & 6.5e+04 & 9.5e+00 & 7.4e-02 & 7.8e+00 & 6.3e-02 \\
\protect* e Eri & -5.5e+04 & 1.4e+00 & 6.9e-03 & 1.1e+00 & 5.9e-03 \\
BD-07 4003 & -1.6e+05 & 5.2e+04 & 9.3e+03 & 3.1e+04 & 7.1e+03 \\
\protect* alf Lyr & 3.3e+05 & 3.2e+01 & 4.6e-01 & 6.9e+00 & 2.4e-01 \\
\protect* alf PsA & -6.2e+05 & 4.8e+02 & 3.1e+01 & 2.8e+01 & 5.9e+00 \\
\protect* 61 Vir & 1.6e+05 & 1.6e+01 & 6.2e-01 & 8.5e+00 & 4.2e-01 \\
HD 197481 & -8.0e+04 & 3.0e+02 & 8.2e+00 & 2.9e+02 & 7.7e+00 \\
\protect* bet Leo & 1.3e+06 & 3.6e+00 & 1.7e-01 & 1.2e+00 & 6.3e-02 \\
HD 166 & -6.6e+05 & 1.3e+03 & 1.4e+02 & 4.2e+01 & 1.7e+01 \\
HD 38858 & -3.2e+05 & 1.4e+05 & 1.6e+04 & 4.1e+02 & 1.6e+02 \\
HD 207129 & 3.3e+06 & 3.7e+04 & 5.1e+03 & 9.5e+03 & 5.1e+03 \\
\protect* b Her & -6.2e+05 & 1.8e+01 & 7.2e-01 & 2.6e+00 & 3.4e-01 \\
HD 23484 & -6.8e+05 & 3.0e+04 & 6.9e+03 & 1.6e+01 & 5.2e+00 \\
\protect* q01 Eri & -7.8e+05 & 8.0e+05 & 2.0e+04 & 7.6e+03 & 3.8e+03 \\
\protect* g Lup & 2.6e+05 & 4.5e+04 & 1.1e+04 & 9.8e+03 & 5.0e+03 \\
HD 35650 & -6.0e+05 & 7.3e+05 & 3.7e+04 & 3.3e+03 & 2.4e+03 \\
\protect* eta Crv & 2.0e+06 & 7.0e+04 & 5.1e+03 & 1.5e+04 & 4.9e+03 \\
HD 53143 & -3.9e+05 & 1.5e+05 & 1.6e+04 & 5.0e+04 & 8.5e+03 \\
\protect* bet Pic & -1.1e+06 & 5.1e+04 & 4.3e+03 & 4.4e+03 & 1.7e+03 \\
\tableline
\textbf{Total} & -6.0e+05 & 1.0e+06 & 4.3e+04 & 2.1e+05 & 2.3e+04 \\
\enddata
\end{deluxetable*}

\begin{deluxetable*}{cccccccccc}
\tablecaption{Maximum and Current Flux into the Oort Cloud and the Resulting Population Within the Inner Solar System for Particles $\geq 100$ m (assuming $t_{crit}=3$ Myr)\label{tab:100m_ageDepend_tcrit3}}
\tablehead{
\colhead{Star Name} & \colhead{Time at Max} & \colhead{Max Flux} & \colhead{Error} & \colhead{Max Population} & \colhead{Error} & \colhead{Current Flux} & \colhead{Error} & \colhead{Current Population} & \colhead{Error} \\
\colhead{} & \colhead{} & \colhead{Oort Cloud} & \colhead{} & \colhead{Inner Solar System} & \colhead{} & \colhead{Oort Cloud} & \colhead{} & \colhead{Inner Solar System} & \colhead{} \\
\colhead{} & \colhead{(yr)} & \colhead{(yr$^{-1}$)} & \colhead{} & \colhead{} & \colhead{} & \colhead{(yr$^{-1}$)} & \colhead{} & \colhead{} & \colhead{}
}
\startdata
\protect* eps Eri & -1.1e+05 & 1.2e+09 & 1.8e+08 & 3.5e+01 & 7.2e+00 & 5.6e+08 & 1.2e+08 & 1.6e+01 & 4.6e+00 \\
\protect* tau Cet & 6.5e+04 & 1.3e+05 & 1.1e+03 & 2.2e-03 & 2.1e-05 & 1.1e+05 & 8.9e+02 & 1.8e-03 & 1.7e-05 \\
\protect* e Eri & -5.5e+04 & 4.0e+04 & 1.9e+02 & 2.0e-04 & 1.0e-06 & 3.0e+04 & 1.6e+02 & 1.5e-04 & 8.5e-07 \\
BD-07 4003 & -1.6e+05 & 7.3e+08 & 1.3e+08 & 1.2e+01 & 2.4e+00 & 4.3e+08 & 1.0e+08 & 7.2e+00 & 1.9e+00 \\
\protect* alf Lyr & 3.3e+05 & 1.6e+05 & 2.4e+03 & 5.6e-03 & 1.3e-04 & 3.6e+04 & 1.2e+03 & 1.2e-03 & 6.5e-05 \\
\protect* alf PsA & -6.2e+05 & 1.7e+06 & 1.1e+05 & 7.2e-02 & 8.4e-03 & 9.7e+04 & 2.1e+04 & 4.1e-03 & 1.6e-03 \\
\protect* 61 Vir & 1.6e+05 & 3.4e+05 & 1.3e+04 & 3.5e-03 & 1.3e-04 & 1.8e+05 & 8.8e+03 & 1.9e-03 & 9.5e-05 \\
HD 197481 & -8.0e+04 & 2.0e+06 & 5.4e+04 & 5.7e-02 & 2.2e-03 & 1.9e+06 & 5.2e+04 & 5.5e-02 & 2.0e-03 \\
\protect* bet Leo & 1.3e+06 & 3.1e+04 & 1.6e+03 & 7.5e-04 & 4.8e-05 & 1.0e+04 & 5.6e+02 & 2.5e-04 & 1.7e-05 \\
HD 166 & -6.6e+05 & 1.3e+07 & 1.4e+06 & 2.8e-01 & 3.7e-02 & 4.0e+05 & 1.6e+05 & 9.0e-03 & 4.4e-03 \\
HD 38858 & -3.2e+05 & 1.8e+09 & 2.1e+08 & 3.1e+01 & 4.3e+00 & 5.4e+06 & 2.1e+06 & 9.5e-02 & 4.0e-02 \\
HD 207129 & 3.3e+06 & 3.0e+08 & 4.0e+07 & 7.6e+00 & 1.3e+00 & 8.3e+07 & 4.5e+07 & 2.0e+00 & 1.3e+00 \\
\protect* b Her & -6.2e+05 & 2.3e+04 & 9.0e+02 & 1.7e-03 & 2.2e-04 & 3.4e+03 & 4.3e+02 & 2.4e-04 & 1.1e-04 \\
HD 23484 & -6.8e+05 & 4.5e+08 & 1.1e+08 & 7.0e+00 & 1.8e+00 & 2.4e+05 & 8.0e+04 & 3.8e-03 & 1.3e-03 \\
\protect* q01 Eri & -7.8e+05 & 9.3e+09 & 2.3e+08 & 1.8e+02 & 5.4e+00 & 8.7e+07 & 4.3e+07 & 1.7e+00 & 1.0e+00 \\
\protect* g Lup & 2.6e+05 & 4.2e+08 & 1.0e+08 & 9.7e+00 & 2.9e+00 & 9.0e+07 & 4.6e+07 & 2.1e+00 & 1.3e+00 \\
HD 35650 & -6.0e+05 & 8.6e+09 & 4.5e+08 & 1.7e+02 & 1.0e+01 & 3.9e+07 & 2.9e+07 & 7.5e-01 & 6.0e-01 \\
\protect* eta Crv & 2.0e+06 & 9.0e+08 & 6.6e+07 & 1.6e+01 & 1.4e+00 & 2.0e+08 & 6.6e+07 & 3.5e+00 & 1.2e+00 \\
HD 53143 & -3.9e+05 & 2.0e+09 & 2.0e+08 & 3.5e+01 & 4.0e+00 & 6.4e+08 & 1.1e+08 & 1.2e+01 & 2.1e+00 \\
\protect* bet Pic & -1.1e+06 & 2.7e+08 & 2.2e+07 & 9.0e+00 & 1.1e+00 & 2.1e+07 & 7.2e+06 & 7.4e-01 & 4.1e-01 \\
\tableline
\textbf{Total} & -6.0e+05 & 1.2e+10 & 5.2e+08 & 2.4e+02 & 1.2e+01 & 2.2e+09 & 2.2e+08 & 4.6e+01 & 6.0e+00 \\
\enddata
\end{deluxetable*}

\clearpage

\end{document}